\documentclass[aps,pre,showkeys,
superscriptaddress,
longbibliography,
reprint,
onecolumn, 
nofootinbib,
floatfix]{revtex4-1}
\usepackage[T1]{fontenc}
\usepackage{amsmath}
\usepackage{amsfonts}
\usepackage{graphicx}
\usepackage{subcaption}
\usepackage{listings}
\usepackage{tkz-euclide}
\usetkzobj{all}
\usepackage[justification=centerlast]{caption}
\usepackage[colorlinks=true,breaklinks=true]{hyperref}

\begin{document}
%% \setulcolor{green}
%% \setstcolor{red}

\title{Noise induced unanimity and disorder in opinion formation}

\author{Agnieszka Kowalska-Stycze\'n}
\thanks{\includegraphics[scale=0.1]{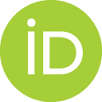}~\href{https://orcid.org/0000-0002-7404-9638}{0000-0002-7404-9638}}
\email{Agnieszka.Kowalska-Styczen@polsl.pl}
\affiliation{\href{http://www.polsl.pl/}{Silesian University of Technology},
Faculty of Organisation and Management,\\
ul. Roosevelta 26/28, 41-800 Zabrze, Poland}

\author{Krzysztof Malarz}
\thanks{\includegraphics[scale=0.1]{ORCID.png}~\href{https://orcid.org/0000-0001-9980-0363}{0000-0001-9980-0363}}
%% \homepage{http://home.agh.edu.pl/malarz/}
\email{malarz@agh.edu.pl}

\affiliation{\href{http://www.agh.edu.pl/}{AGH University of Science and Technology},
\href{http://www.pacs.agh.edu.pl/}{Faculty of Physics and Applied Computer Science},
al. Mickiewicza 30, 30-059 Krak\'ow, Poland}

\begin{abstract}
	We propose an opinion dynamics model based on Latan\'e's social impact theory. {} {Actors} in this model are heterogeneous and, in addition to opinions, are characterised by their varying levels of persuasion and support. The model is tested for two and three initial opinions randomly distributed among {} {actors}.
	We examine how the {noise (}randomness of behaviour{)} and the flow of information {} {among actors} affect the formation and spread of opinions.
Our main research involves the process of opinion
	{} 
formation and
	{}
{finding phases of the system in terms of parameters describing noise and flow of the information for two and three opinions available in the system.}
	The results show that opinion formation and spread are influenced by both {(i)} flow of information {} {among} {} {actors (effective range of} interactions {among actors} {}) and {(ii) noise (}randomness in adopting opinions{)}.
{The noise not only leads to opinions disorder but also it promotes consensus under certain conditions.}
\end{abstract}

\date{\today}


\maketitle

%% ###########################################################
\section{Introduction}
%% ###########################################################

Understanding how opinions are formed and spread in society is very important in studying consumer behaviour, organisational behaviour, predicting election results, and many others.
As pointed out by \citet{Acemoglu-2011}, we acquire our opinions and beliefs in the process of social learning, during which people get information and update their opinions as a result of their own experience, as well as observation of other people's activities and from their experience.
This process takes place in a social network consisting of friends, co-workers, family members and a certain group of leaders that we listen to and respect \citep{Acemoglu-2011,Jackson-2011}.
Units update and create their views by communicating with other people who belong to their social network.
It is communication that connects people and creates relationships  \citep{Duncan-1998}.

It should be noted that  people often copy the choices of others  \cite{Simon-1955,Bentley-2011}. 
This applies, for example, to the choice of names for children \citep{Kulakowski-2018,Krawczyk-2014}, a popular book, dishes ordered in a restaurant (instead of studying the menu, we look at what the others have ordered), and even ideological beliefs~\cite{Bentley-2011}.
This copying of opinions and behaviours often takes place in a network of informal contacts and it is based on social relations between people \citep{Guffy-2005} and plays an important role in forming opinions.
In addition, we are often dealing with unpredictability or indifference in opinion-forming or decision-making (despite the positive attitude towards the proposed actions).
This applies, among others, to electricity tariffs, eco-innovations or pro-environmental attitudes \citep{Kowalska-Pyzalska-2014a,Kowalska-Pyzalska-2014b,Byrka-2016}, as well as voting behaviour \citep{Stadelmann-2013}, in which human rationality is bounded.
One of the most active discussions in psychology of the opinion dynamics is also about the irrational processing of information \cite{Sobkowicz-2018}, therefore, this aspect should be taken into account in studying opinion formation.
Furthermore, individuals belong to many groups, or have many interactions outside the main group (a group of closest neighbours). Such a connection with people from other groups (neighbourhoods) increases the information advantage  \citep{Apolloni-2011},  and can be interesting in disseminating information.

We therefore propose a model of forming an opinion based on the social impact theory formulated by \citet{Latane-1981}, in which we take into account the randomness of the actors' behaviour by introducing a {noise (}social temperature{)}, as well as interactions with agents; not only close neighbours, but in the whole network by $\alpha$ parameter (scale the distance function). Our agents are heterogeneous through a different level of persuasion intensity and support intensity, as well as the possibility of having different opinions.

Recently, the multi-choice opinion dynamics model \citep{1902.03454,ThesisBancerowski} based on Latan\'e theory \citep{Latane-1981,Latane-1976,Darley1968,Latane-1981a,Nowak-1990} was proposed.
In this model, it is possible to test the diffusion of opinions in case there are more than two opinions available in the system.
The earlier attempts to modelling multiple-choice of opinions include among others multi-state and discrete-state opinions models~\cite{PhysRevE.91.032808,PhysRevE.74.056108,Martins_2019,000432967700004,000316891200004,Kulakowski2010,Gekle-2005,000238502800018,000223811700008,A1991FU10100002,A1990EP96200025}
or discrete vector-like variables~\cite{Axelrode-1997,Weimer2019,ISI:000317452800014,Sznajd-2005a}.
The rest of huge literature (see papers by~\textcite{Sirbu2017,RevModPhys.81.591,Stauffer2009,Anderson1992,GalamReview} for reviews) is devoted to the systems with binary opinions (see for example Refs.~\onlinecite{Kulakowski2008,Slanina-2008,Sznajd-2005,Sznajd-2000}) or 
the continuous space of opinions (see for example Refs.~\onlinecite{Sobkowicz-2018,Gargiulo-2017,Mathias-2016,Kulakowski2014,Kulakowski2012,Gronek2011,Kulakowski2009469,Deffuant-2006,Hegselmann-2002,Deffuant-2000,000411951900001,Malarz2006b,000413837700014,000414818100049,000409101900001,000399951700004,000399951000005,000396054300030}).

{There are also many examples showing that randomness is useful and beneficial.
Random noise facilitates the dynamics and reduces relaxation times in the models of social influence} \cite{PhysRevE.79.046108}. 
{In addition, noise plays a beneficial role in developing cooperation} \cite{PhysRevE.75.045101}, 
{in the application of social and financial strategies} \cite{Biondo2013} {and in addressing the coordination problem of human groups} \cite{Shirado2017}.

In this paper we study how opinions are formed and how they spread in the community.
Agent based model with lattice fully populated by actors has been adopted, where each of the network nodes refers to one person.
We take into account the flow of information in the community ({effective range of actors} interactions {}) and {noise (}randomness of human behaviour{)}.
{We show with computer simulation, that small level of noise induces unaminity of opinion.
Unfortunately, this beneficial noise role was overlooked in Ref.}~\onlinecite{1902.03454} {due to unreasonable extrapolation of results for middle noise level towards noiseless system.}

%% ###########################################################
\section{\label{S:model}Model}
%% ###########################################################

To study the diffusion of opinions, the theory of social influence introduced by \citet{Latane-1981} in the dynamic manner proposed by \citet{Nowak-1990}---as implemented by \citet{1902.03454}---has been used.

Social influence is a process that results in a change in the behaviour, opinion or feelings of a human being as a result of what other people do, think or feel. The essence of social influence is of course not only exerting social influence, but also succumbing to it, which will be taken into account in the used model by means of appropriate parameters (intensity of persuasion and intensity of support).
The \citet{Latane-1981} theory rely on three experimentally proven \cite{Latane-1976,Darley1968,Latane-1981a} assumptions:
\begin{description}
\item[social force principle] it says that social impact $I$ {(details are given in description of Eq.}~\eqref{eq:wplyw}{)} on $i$-th actors is a function of the product of strength $S$, immediacy $J$, and the number of sources $N$;
The strength of influence is the intensity, power or importance of the source of influence. This concept may reflect socio-economical status of the one that affects our opinion, his/her age, prestige or position in the society.
The immediacy determines the relationship between the source and the goal of influence. This may mean closeness in the social relationship, lack of communication barriers and ease of communication among actors;
\item[psycho-social law] it states that each next actor $j$ sharing the same opinion as actor $i$ exerts the lower impact on the $i$-th actor;  
\item[division of impact theory] it is based on the bystander effect and is observed as the errors of reacting to crisis events, along with an increase in the number of witnesses to this event.
\end{description}
Based on these assumptions \citet{Nowak-1990} proposed computerised model of opinion dynamics based on Latan\'e \cite{Latane-1976,Darley1968,Latane-1981a} social impact theory (see Ref.~\onlinecite{ARCPIX253} for review).

{Every agent $i$ is characterised by the following parameters:}
\begin{description}
\item[{opinion $\xi_i$}] {the current opinion supported by agent $i$,}
\item[{intensity of support $s_i$}] {the strength of the agent $i$ influence on other agents, which determines the ability of this agent to convince other agents not to change {} {their} opinion if this opinion is identical to his/her opinion ($0\le s_i \le 1$),}
\item[{intensity of persuasion $p_i$}] {the strength of agent $i$ influence on agents, which determines the ability of this agent to convince other agents to accept his/her opinions ($0\le p_i \le 1$).}
\end{description}

{Each agent is influenced by all other agents on the network. The strength of this influence decreases as the distance between agents increases. In the presented model a cellular automaton was used, which consists of a square grid of $L^2$ cells, where exactly one agent is assigned to each cell. The distance $d_{ij}$ between agents $i$ and $j$ is calculated as the Euclidean distance between cells.}

To take into account the varied flow of information in the community, we use the $\alpha$ parameter, which was adapted to scale the distance function.
Parameter $\alpha$ {} {adjusts} the influence of close and distant neighbours in the community.
Small $\alpha$ values mean good communication between agents and good access to information, because it 
{allows for} an exchange of information with a large number of agents in the network.
The larger values of $\alpha$, weaker the communication
{among} the groups of agents, {weaker effective} exchange of information and {weaker} access to information, because the exchange of information takes place only in the closest neighbourhood {of actors, although we still keep long-range interactions among actors}.

Here we are on a position to recapitulate the formal model composition as proposed in Ref.~\onlinecite{1902.03454}.

%% ===============================================================
\subsection{\label{S:formalism}Formal model description}
%% ===============================================================

Actors occupy the nodes of the square lattice with linear size $L$.
Every actor $1\le i\le L^2$ is characterised by his/her discrete opinion $\xi_i\in\{\Xi_1,\Xi_2,\cdots,\Xi_K\}$, where $K$ is the number of opinions available in the system.
Additionally, we assign random real value $p_i\in[0,1]$ and $s_i\in[0,1]$ describing actor's persuasiveness and his/her supportiveness, respectively.

The system evolution depends on the social temperature $T$.
If $T=0$, then a lack of {noise} {} is assumed, and the {} {actor} $i$ adopts an opinion $\Xi_k$ that has the most {impact} on it:
\begin{equation}
\label{eq:T=0}
\xi_i(t+1)=\Xi_k \iff I_{i,k}(t)=\max(I_{i,1}(t), I_{i,2}(t),\cdots,I_{i,K}(t)),
\end{equation}
where $k$ is the label of this opinion which believers exert the largest social impact on $i$-th actor and $I_{i,k}$ are the social influence on actor $i$ exerted by actors sharing opinion $\Xi_k$.

The social impact on actor $i$ from actors $j$ sharing opinion of actor $i$ ($\xi_j=\xi_i$) is calculated as
\begin{widetext}
\begin{subequations}
\label{eq:wplyw}
\begin{equation}
\label{eq:wplyw_ta_sama}
I_{i,k}(t) = 4\mathcal{J}_s\left(\sum_{j=1}^{N}\dfrac{q(s_j)}{g(d_{i,j})} \delta(\Xi_k,\xi_j(t))\delta(\xi_j(t),\xi_i(t)) \right)
\end{equation}
while $K-1$ social impacts on actor $i$ from all other actors having $K-1$ different opinions ($\xi_j\ne\xi_i$) is given as
\begin{equation}
\label{eq:wplyw_inna}
I_{i,k}(t) = 4\mathcal{J}_p\left(\sum_{j=1}^{N}\dfrac{q(p_j)}{g(d_{i,j})} \delta(\Xi_k,\xi_j(t))[1-\delta(\xi_j(t),\xi_i(t))] \right),
\end{equation}
\end{subequations}
where $1\le k\le K$ enumerates the opinions and Kronecker's delta $\delta(x,y)=1$ if $x=y$ and zero otherwise \citep{1902.03454}.
\end{widetext}

As in Ref.~\onlinecite{1902.03454} we assume identity function for scaling functions $\mathcal{J}_S(x)\equiv x$, $\mathcal{J}_P(x)\equiv x$, $q(x)\equiv x$.
The distance scaling function should be an increasing function of its argument.
Here, we assume the distance scaling function as
\begin{equation}
\label{eq:g}
g(x)=1+x^\alpha,
\end{equation}
what ensures non-zero values $g(0)=1$ of denominator for self-supportiveness in Eq.~\eqref{eq:wplyw_ta_sama}.

The exponent $\alpha$ is an arbitrary quantity which characterise the long-range interaction among actors.
For small values of $\alpha$ (for instance for $\alpha=2$) we assume good communication among actors, good access to information in the society and effective exchange of information.
In contrary, for larger values of $\alpha$ (for instance for $\alpha=6$) discussion and information exchange takes place only in the actors' nearest neighbourhood.

For $T>0$, the larger the social temperature $T$ {(noise)}, the more often the opinions, that do not have the greatest impact are selected.
As it was shown in Ref.~\onlinecite{1902.03454} in the modelled system the phase transition occurs: below critical temperature $T\ll T_c$ the ordered phase is observed with domination of one of the available opinion, while for $T\gg T_c$ all opinions become equally supported by {} {actors}.
Critical temperatures $T_c$ (but for homogeneous society with $\forall i: s_i=p_i=0.5$) are $T_c = 6.1$ and $T_c = 4.7$, for two and for three opinions, respectively \cite{1902.03454}.
In this article, simulations were carried out for $T \le 5$ to take into account different levels of {noise} {}, reaching the critical level at which agents more often take random opinions than guided by the opinion of their neighbours.

For finite values of social temperature $T>0$ we apply the Boltzmann choice
\begin{subequations}
\label{eq:T=T}
\begin{equation}
        p_{i,k}(t) = \exp\left(\dfrac{I_{i,k}(t)}{T}\right),
\label{eq:E_ik}
\end{equation}
which yields probabilities
\begin{equation}
        P_{i,k}(t) = \frac{p_{i,k}(t)}{\sum_{j=1}^{K}p_{i,j}(t)}
\label{eq:P_ik}
\end{equation}
of choosing by $i$-th actor in the next time step $k$-th opinion:
\begin{equation}
        \xi_i(t+1)=\Xi_k, \text{ with probability } P_{i,k}(t).
\end{equation}
\end{subequations}
The form of dependence~\eqref{eq:E_ik} in statistics and economy is called logit function~\citep{Anderson1992,Byrka-2016}.

Both, for $T=0$ and $T>0$ the calculated social impacts $I_{i,k}(t)$ influence the $i$-th actor opinion $\xi_i(t+1)$ at the subsequent time step.
Newly evaluated opinions are applied synchronously to all actors.
The simulations takes one {}{thousand} time steps which ensures reaching a plateau in time evolution of several observables. {}

The simulations are carried out on square lattice of linear size $L=41$ with open boundary conditions.
{To check the system behaviour, also simulations for $L=21$ and $L=61$ were carried out.}
We assume random values of supportiveness $s_i$  and persuasiveness $p_i$ for all actors.
The studies for homogeneous society, i.e. with $\forall i:  p_i=s_i=0.5$ were carried out in Ref.~\onlinecite{1902.03454}.
{The results are averaged over one hundred independent simulations with various initial distribution of opinions and actors persuasiveness and actors supportiveness.}

The example of social impact calculations for a small system (with nine actors and three opinions) is given in Appendix \ref{A:small}.
The model implementation in Fortran95 \citep{Fortran95} is attached as Listing{s}~\ref{lst:T0} {and} \ref{lst:TT} in Appendix \ref{A:code}.

%% ###############################################################
\section{\label{S:results}Results}
%% ###############################################################

%% ===============================================================
\subsection{\label{S:resultsspo}Spatial distribution of opinions}
%% ===============================================================

We start presentation of our results by showing the spatial distribution of opinions for $K=2$, 3 (various numbers of opinions available in the system), for $\alpha=${2}, 3, 6 (various {levels of} flow of information), for $T = 0$, 1, 3, 5 (various levels of {noise} {}). 

%% ---------------------------------------------------------------
\subsubsection{\label{S:resultsK2}$K=2$}
%% ---------------------------------------------------------------

%% ===============================================================
\begin{figure*}
%% ---------------------------------------------------------------
\begin{subfigure}[b]{.28\textwidth}
\caption{\label{fig:rev2K2T0a2xi}{$T=0$, $\alpha=2$}}
\includegraphics[width=\textwidth]{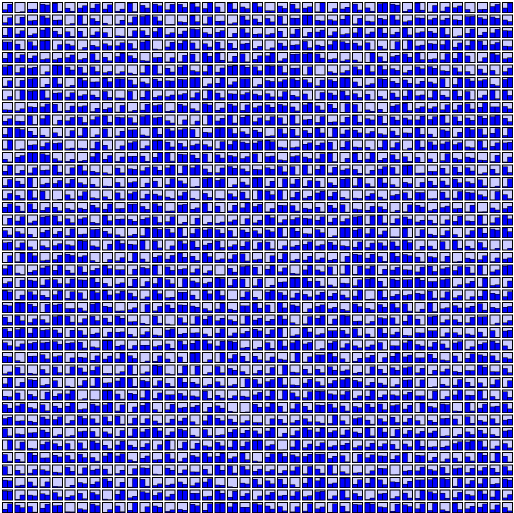}
\end{subfigure}
%% ---------------------------------------------------------------
\begin{subfigure}[b]{.28\textwidth}
\caption{\label{fig:rev2K2T0a3xi}{$T=0$, $\alpha=3$}}
\includegraphics[width=\textwidth]{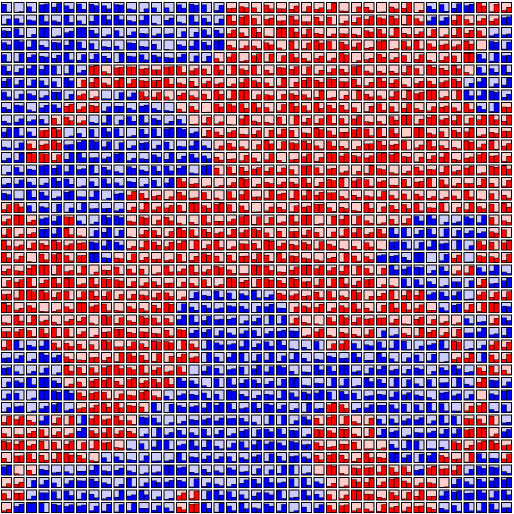}
\end{subfigure}
%% ---------------------------------------------------------------
\begin{subfigure}[b]{.28\textwidth}
\caption{\label{fig:rev2K2T0a6xi}{$T=0$, $\alpha=6$}}
\includegraphics[width=\textwidth]{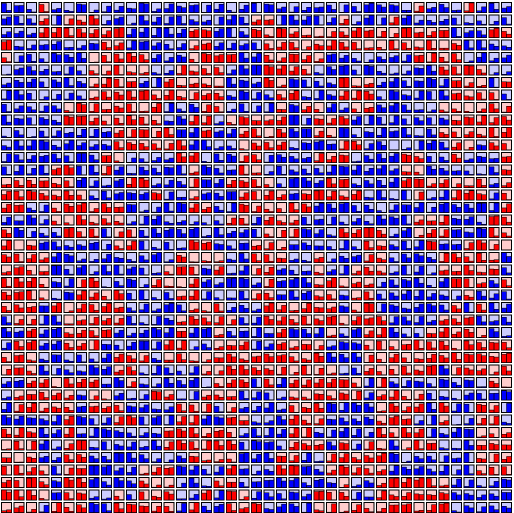}
\end{subfigure}
%% ---------------------------------------------------------------
\begin{subfigure}[b]{.28\textwidth}
\caption{\label{fig:rev2K2T1a2xi}{$T=1$, $\alpha=2$}}
\includegraphics[width=\textwidth]{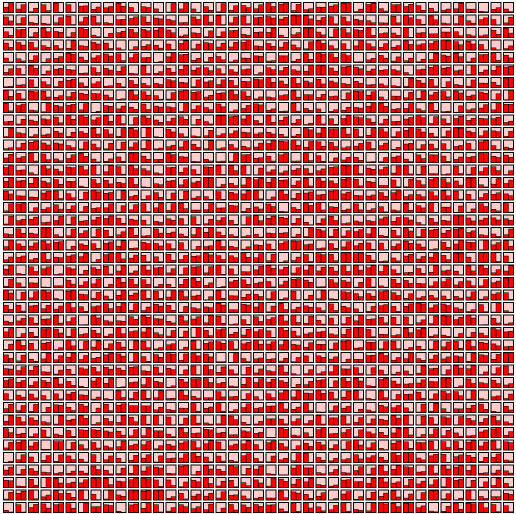}
\end{subfigure}
%% ---------------------------------------------------------------
\begin{subfigure}[b]{.28\textwidth}
\caption{\label{fig:rev2K2T1a3xi}{$T=1$, $\alpha=3$}}
\includegraphics[width=\textwidth]{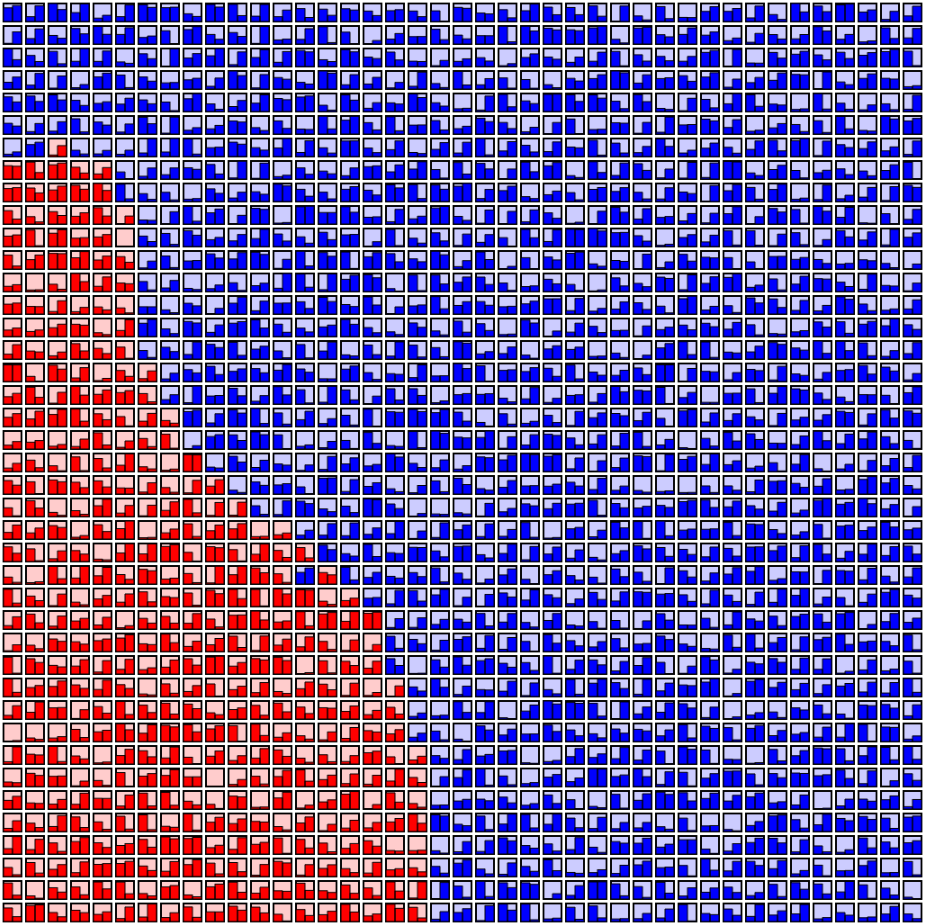}
\end{subfigure}
%% ---------------------------------------------------------------
\begin{subfigure}[b]{.28\textwidth}
\caption{\label{fig:rev2K2T1a6xi}{$T=1$, $\alpha=6$}}
\includegraphics[width=\textwidth]{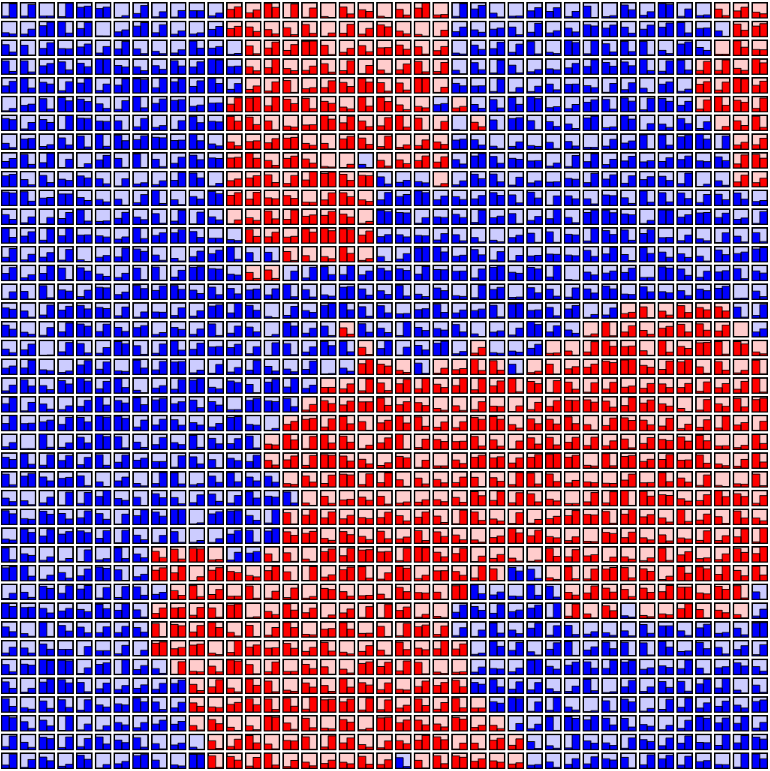}
\end{subfigure}
%% ---------------------------------------------------------------
\begin{subfigure}[b]{.28\textwidth}
\caption{\label{fig:rev2K2T3a2xi}{$T=3$, $\alpha=2$}}
\includegraphics[width=\textwidth]{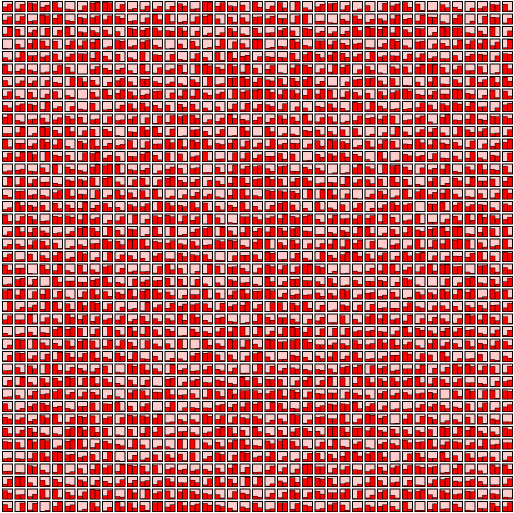}
\end{subfigure}
%% ---------------------------------------------------------------
\begin{subfigure}[b]{.28\textwidth}
\caption{\label{fig:rev2K2T3a3xi}{$T=3$, $\alpha=3$}}
\includegraphics[width=\textwidth]{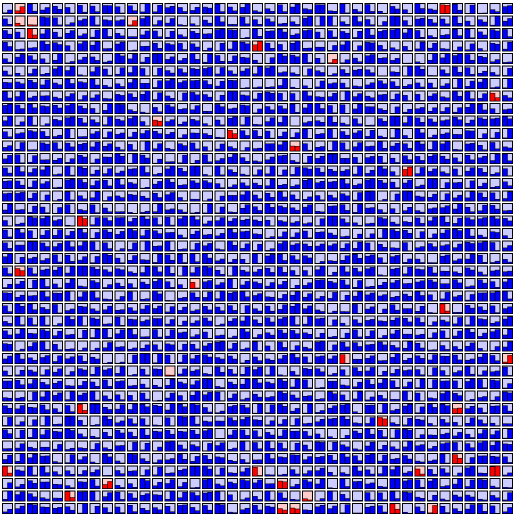}
\end{subfigure}
%% ---------------------------------------------------------------
\begin{subfigure}[b]{.28\textwidth}
\caption{\label{fig:rev2K2T3a6xi}{$T=3$, $\alpha=6$}}
\includegraphics[width=\textwidth]{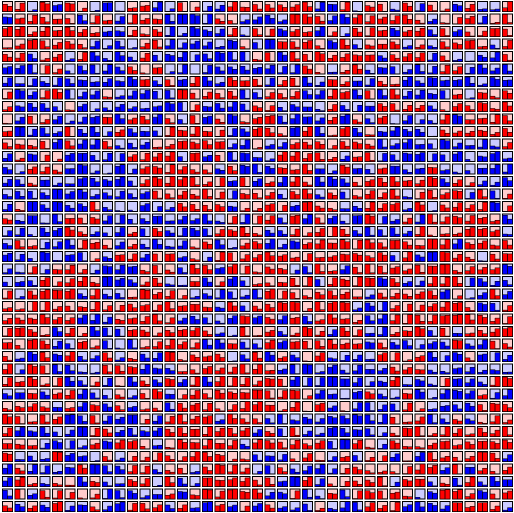}
\end{subfigure}
%% ---------------------------------------------------------------
\begin{subfigure}[b]{.28\textwidth}
\caption{\label{fig:rev2K2T5a2xi}{$T=5$, $\alpha=2$}}
\includegraphics[width=\textwidth]{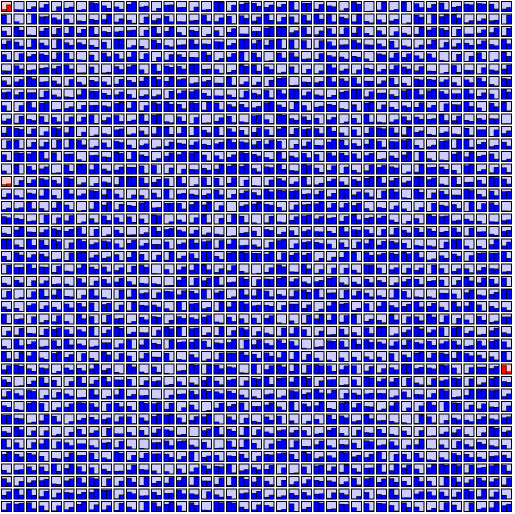}
\end{subfigure}
%% ---------------------------------------------------------------
\begin{subfigure}[b]{.28\textwidth}
\caption{\label{fig:rev2K2T5a3xi}{$T=5$, $\alpha=3$}}
\includegraphics[width=\textwidth]{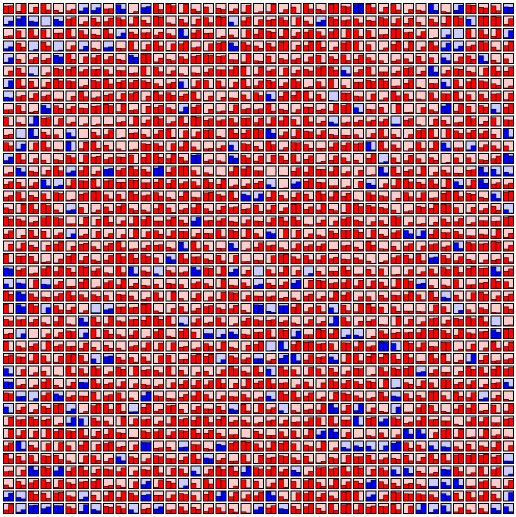}
\end{subfigure}
%% ---------------------------------------------------------------
\begin{subfigure}[b]{.28\textwidth}
\caption{\label{fig:rev2K2T5a6xi}{$T=5$, $\alpha=6$}}
\includegraphics[width=\textwidth]{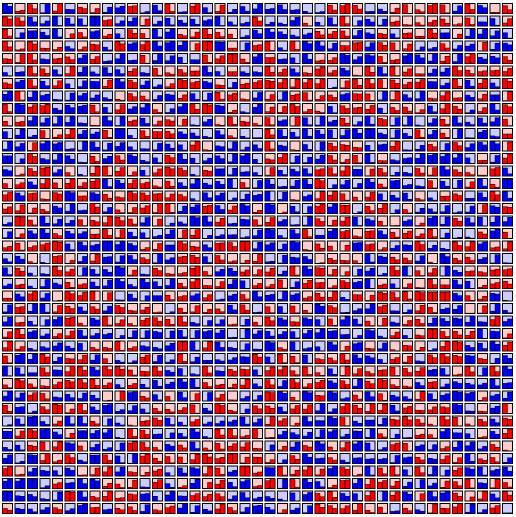}
\end{subfigure}
%% ---------------------------------------------------------------
\caption{\label{fig:rev2K2xi}{{Spatial distribution of opinions $\xi$ for various values of social temperature $T$ and exponent $\alpha$ after $t_{\max}=10^3$ time steps. $L^2=1681$, $K=2$. Pictures are produced with application}~\cite{Bancerowski-app}.}}
\end{figure*}
%% ===============================================================

In Fig.~\ref{fig:rev2K2xi} the simulation results for $K=2$, $\alpha=${2}, 3, 6 and $T=0$, 1, 3, 5 are presented.

{Both, $\alpha$ (information flow) and $T$ (noise) influence opinion formation and  the spatial distribution of opinions. For $\alpha=2$ the consensus takes place (all actors adopt one of two opinions, except of few actors for large $T$ values).
Interesting phenomena for $\alpha=3$ are observed, where the frozen initial system goes into a polarized phase of two large clusters for $T=1$ and into a consensus phase for $T=3$ (one cluster with single actors with a different opinion) before disordering for $T=5$.
For $\alpha=6$ many clusters are visible, although the introduction of noise ($T=1$) results in a more ordered system than for other $T$ values.
To sum up, the increase of $\alpha$ and $T$ generally causes greater disorder in the system (many clusters with both opinions), but for some values of these parameters their increase leads to order (consensus)---see Fig.}~\ref{fig:rev2K2T3a3xi}.

{In general, we can observe four main types of structures in the formation of opinions for 
$K = 2$:}
\begin{itemize}
\item {formation of single cluster, when all agents adopt one opinion and consensus takes place (Figs.}~\ref{fig:rev2K2T0a2xi}, \ref{fig:rev2K2T1a2xi}, \ref{fig:rev2K2T3a2xi}, \ref{fig:rev2K2T5a2xi}),
\item {the majority of agents with the same opinion and single agents with opposing opinions scattered across the lattice (Fig.}~\ref{fig:rev2K2T3a3xi}),
\item {formation of several large clusters of agents with different opinions---polarisation of the group opinion (Fig.}~\ref{fig:rev2K2T1a3xi}),
\item {formation of plenty small clusters with both opinions (e.g. Figs.}~\ref{fig:rev2K2T0a3xi}, \ref{fig:rev2K2T1a6xi}).
\end{itemize}

%% ---------------------------------------------------------------
\subsubsection{\label{S:resultsK3}$K=3$}
%% ---------------------------------------------------------------

%% ===============================================================
\begin{figure*}
%% ---------------------------------------------------------------
\begin{subfigure}[b]{.28\textwidth}
\caption{\label{fig:rev2K3T0a2xi}{$T=0$, $\alpha=2$}}
\includegraphics[width=\textwidth]{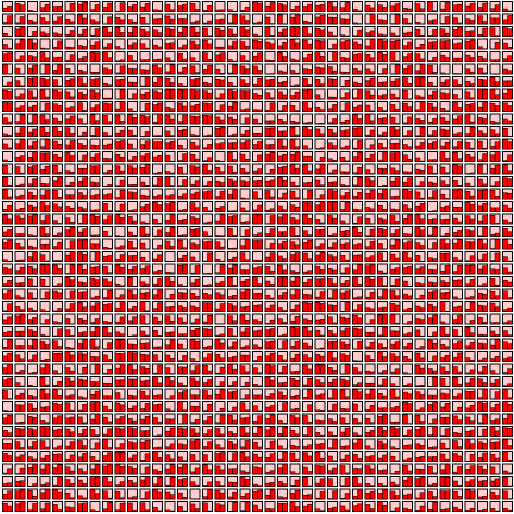}
\end{subfigure}
%% ---------------------------------------------------------------
\begin{subfigure}[b]{.28\textwidth}
\caption{\label{fig:rev2K3T0a3xi}{$T=0$, $\alpha=3$}}
\includegraphics[width=\textwidth]{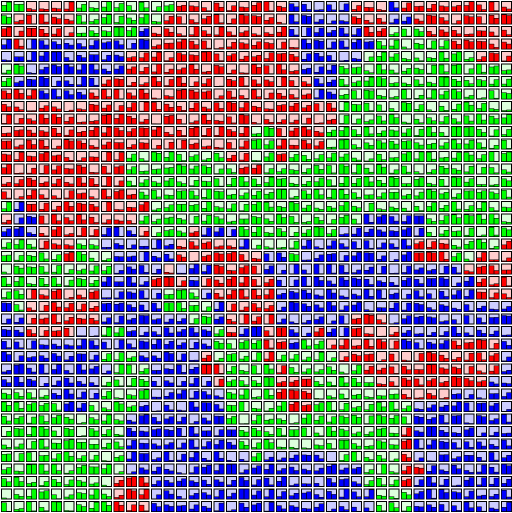}
\end{subfigure}
%% ---------------------------------------------------------------
\begin{subfigure}[b]{.28\textwidth}
\caption{\label{fig:rev2K3T0a6xi}{$T=0$, $\alpha=6$}}
\includegraphics[width=\textwidth]{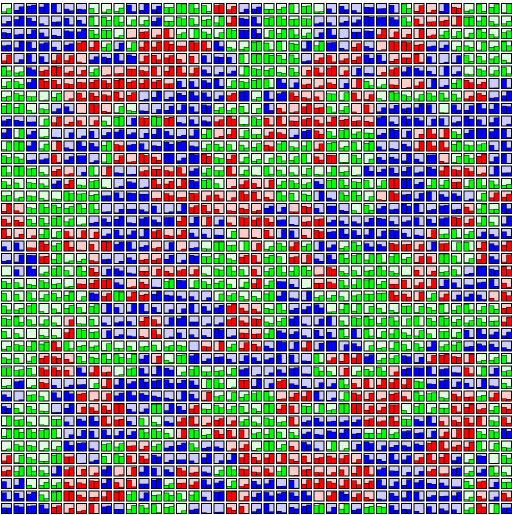}
\end{subfigure}
%% ---------------------------------------------------------------
\begin{subfigure}[b]{.28\textwidth}
\caption{\label{fig:rev2K3T1a2xi}{$T=1$, $\alpha=2$}}
\includegraphics[width=\textwidth]{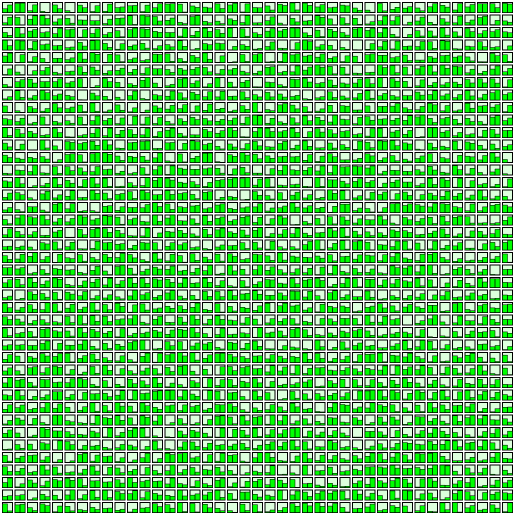}
\end{subfigure}
%% ---------------------------------------------------------------
\begin{subfigure}[b]{.28\textwidth}
\caption{\label{fig:rev2K3T1a3xi}{$T=1$, $\alpha=3$}}
\includegraphics[width=\textwidth]{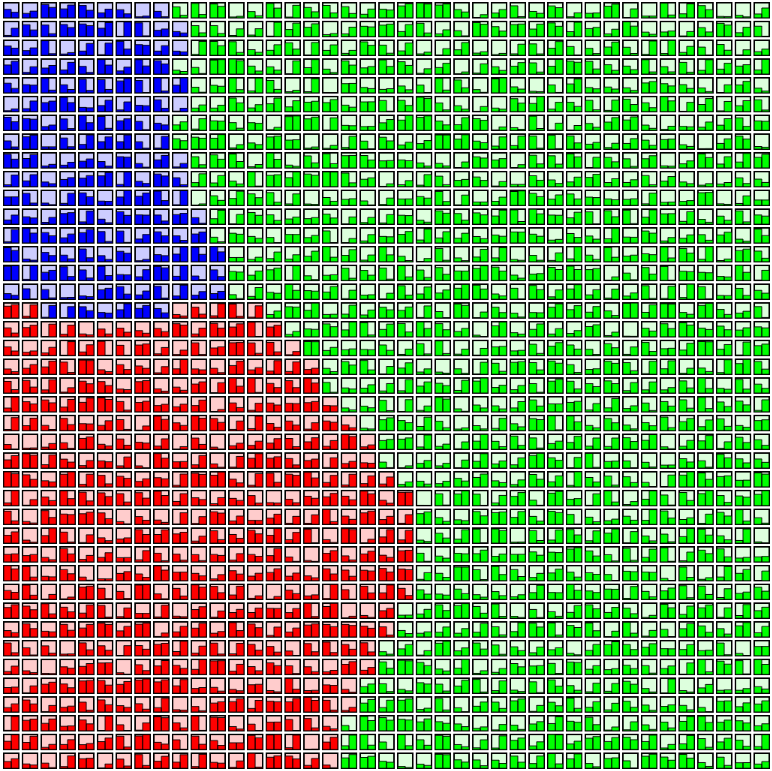}
\end{subfigure}
%% ---------------------------------------------------------------
\begin{subfigure}[b]{.28\textwidth}
\caption{\label{fig:rev2K3T1a6xi}{$T=1$, $\alpha=6$}}
\includegraphics[width=\textwidth]{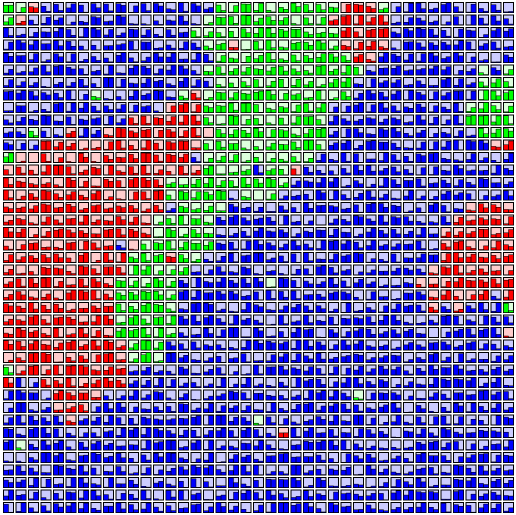}
\end{subfigure}
%% ---------------------------------------------------------------
\begin{subfigure}[b]{.28\textwidth}
\caption{\label{fig:rev2K3T3a2xi}{$T=3$, $\alpha=2$}}
\includegraphics[width=\textwidth]{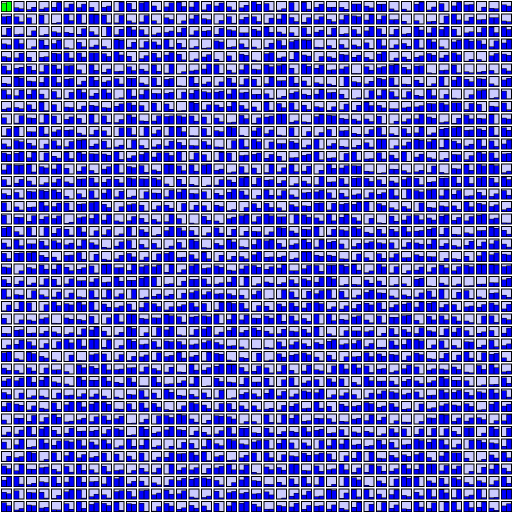}
\end{subfigure}
%% ---------------------------------------------------------------
\begin{subfigure}[b]{.28\textwidth}
\caption{\label{fig:rev2K3T3a3xi}{$T=3$, $\alpha=3$}}
\includegraphics[width=\textwidth]{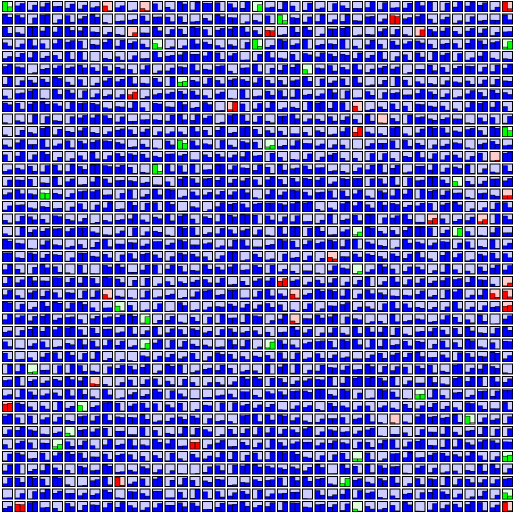}
\end{subfigure}
%% ---------------------------------------------------------------
\begin{subfigure}[b]{.28\textwidth}
\caption{\label{fig:rev2K3T3a6xi}{$T=3$, $\alpha=6$}}
\includegraphics[width=\textwidth]{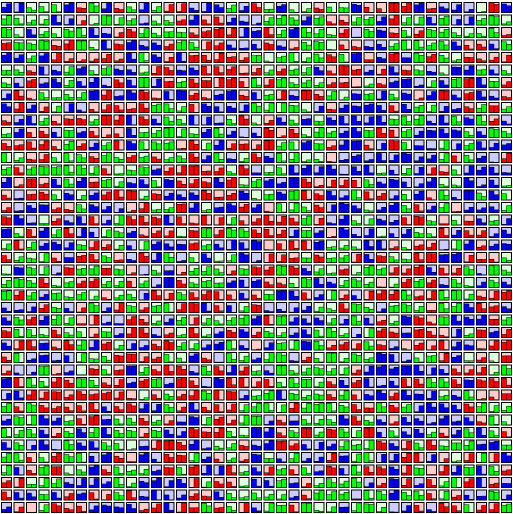}
\end{subfigure}
%% ---------------------------------------------------------------
\begin{subfigure}[b]{.28\textwidth}
\caption{\label{fig:rev2K3T5a2xi}{$T=5$, $\alpha=2$}}
\includegraphics[width=\textwidth]{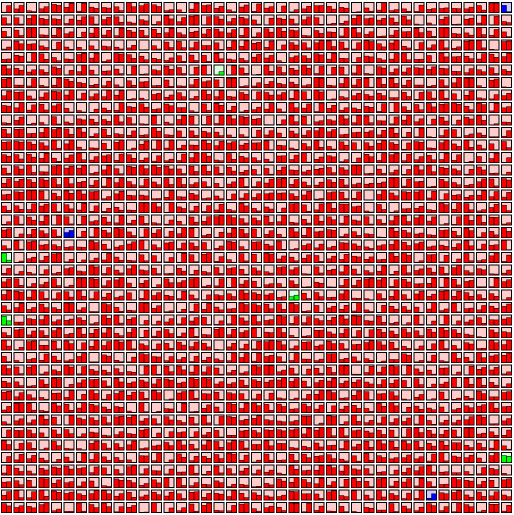}
\end{subfigure}
%% ---------------------------------------------------------------
\begin{subfigure}[b]{.28\textwidth}
\caption{\label{fig:rev2K3T5a3xi}{$T=5$, $\alpha=3$}}
\includegraphics[width=\textwidth]{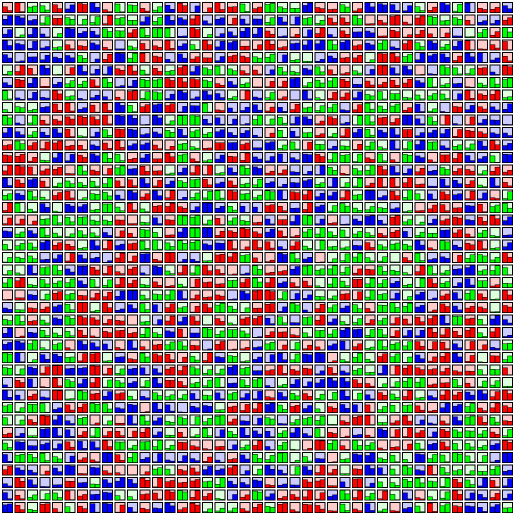}
\end{subfigure}
%% ---------------------------------------------------------------
\begin{subfigure}[b]{.28\textwidth}
\caption{\label{fig:rev2K3T5a6xi}{$T=5$, $\alpha=6$}}
\includegraphics[width=\textwidth]{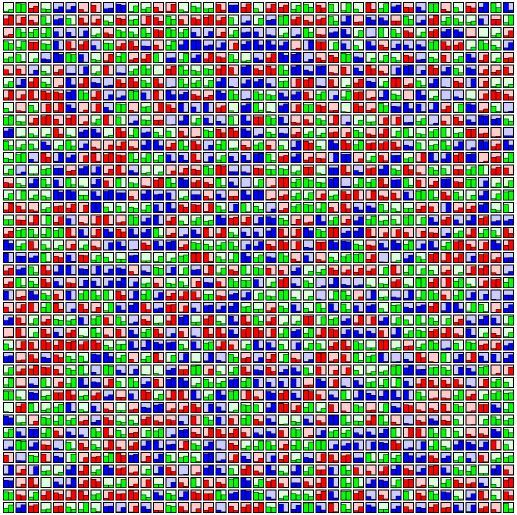}
\end{subfigure}
%% ---------------------------------------------------------------
\caption{\label{fig:rev2K3xi}{{Spatial distribution of opinions $\xi$ for various values of social temperature $T$ and exponent $\alpha$ after $t_{\max}=10^3$ time steps. $L^2=1681$, $K=3$. Pictures are produced with application}~\cite{Bancerowski-app}.}}
\end{figure*}
%% ===============================================================

The simulation results for three opinions among {} {actors} (where $K=3$, $\alpha=${2}, 3, 6 and $T=0$, 1, 3, 5)  are presented in Fig.~\ref{fig:rev2K3xi}.

{}
Similarly to $K=2$, the formation of opinions (the formation of clusters of opinion) depends on the level of {noise} {} {and} the {effective range of interactions among actros} {}.
{}
{For $\alpha=2$, one cluster is formed---consensus take place. For $K=3$ the consensus among actors with three different opinions is also possible for $\alpha=3$. In other cases, many clusters with three opinions or disordered system state are visible.}

{}

{In general, we can observe four types of structures in the formation of opinions for $K=3$, after thousand time steps:}
\begin{itemize}
\item {formation of single cluster, when all agents adopt one opinion and consensus takes place (Fig.}~\ref{fig:rev2K3T0a2xi}, \ref{fig:rev2K3T1a2xi}, \ref{fig:rev2K3T3a2xi}, \ref{fig:rev2K3T5a2xi}),
\item {the majority of agents with the same opinion and single agents with opposing opinions scattered across the lattice (Fig.}~\ref{fig:rev2K3T3a3xi}),
\item {formation of clusters with all possible opinions---polarisation of the group opinion (Fig.}~\ref{fig:rev2K3T1a3xi}),
\item {formation of plenty small clusters with all opinions (e.g. Fig.}~\ref{fig:rev2K3T0a3xi}, \ref{fig:rev2K3T0a6xi}, \ref{fig:rev2K3T1a6xi}).
\end{itemize}

%% ===============================================================
\subsection{\label{S:num_clu}Clustering of opinions}
%% ===============================================================

\begin{figure*}
%% \psfrag{s}{$\mathcal{S}$}
%% \psfrag{H(s)}{$H(\mathcal{S})$}
%% ---------------------------------------------------------------
\begin{subfigure}[b]{.24\textwidth} \caption{\label{fig:hist210} $\alpha=1$, $T=0$}
\includegraphics[width=.99\textwidth]{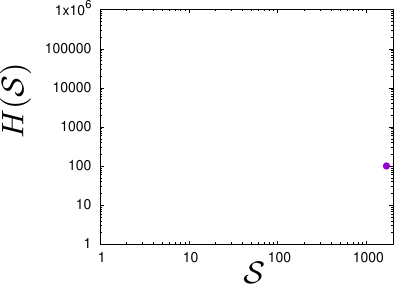}
\end{subfigure}
\begin{subfigure}[b]{.24\textwidth} \caption{\label{fig:hist220} $\alpha=2$, $T=0$}
\includegraphics[width=.99\textwidth]{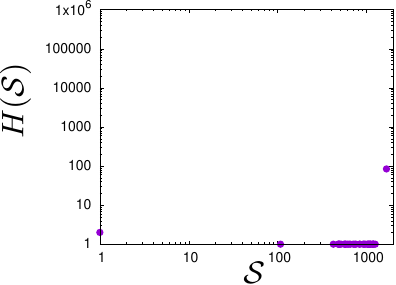}
\end{subfigure}
\begin{subfigure}[b]{.24\textwidth} \caption{\label{fig:hist230} $\alpha=3$, $T=0$}
\includegraphics[width=.99\textwidth]{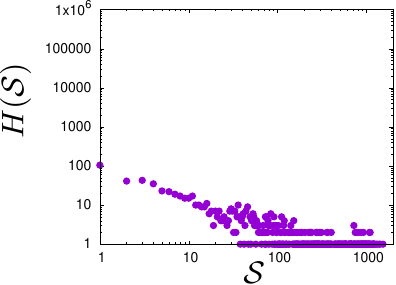}
\end{subfigure}
\begin{subfigure}[b]{.24\textwidth} \caption{\label{fig:hist260} $\alpha=6$, $T=0$}
\includegraphics[width=.99\textwidth]{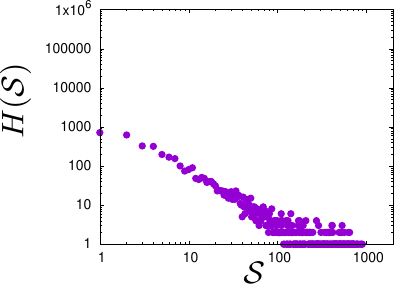}
\end{subfigure}
%% ---------------------------------------------------------------
\begin{subfigure}[b]{.24\textwidth} \caption{\label{fig:hist211} $\alpha=1$, $T=1$}
\includegraphics[width=.99\textwidth]{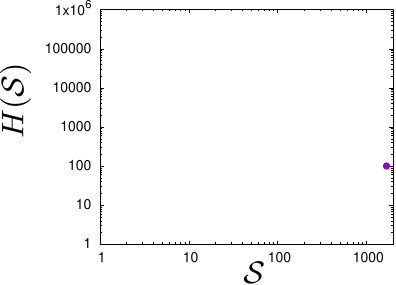}
\end{subfigure}
\begin{subfigure}[b]{.24\textwidth} \caption{\label{fig:hist221} $\alpha=2$, $T=1$}
\includegraphics[width=.99\textwidth]{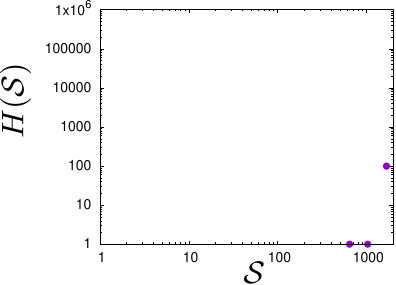}
\end{subfigure}
\begin{subfigure}[b]{.24\textwidth} \caption{\label{fig:hist231} $\alpha=3$, $T=1$}
\includegraphics[width=.99\textwidth]{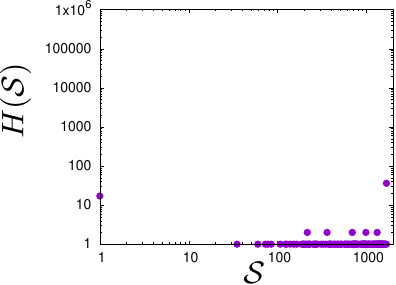}
\end{subfigure}
\begin{subfigure}[b]{.24\textwidth} \caption{\label{fig:hist261} $\alpha=6$, $T=1$}
\includegraphics[width=.99\textwidth]{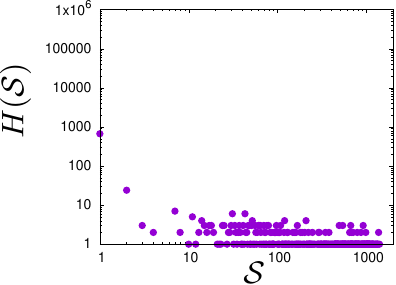}
\end{subfigure}
%% ---------------------------------------------------------------
\begin{subfigure}[b]{.24\textwidth} \caption{\label{fig:hist212} $\alpha=1$, $T=2$}
\includegraphics[width=.99\textwidth]{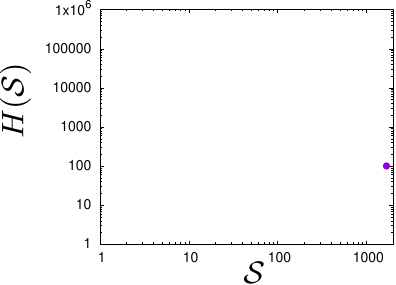}
\end{subfigure}
\begin{subfigure}[b]{.24\textwidth} \caption{\label{fig:hist222} $\alpha=2$, $T=2$}
\includegraphics[width=.99\textwidth]{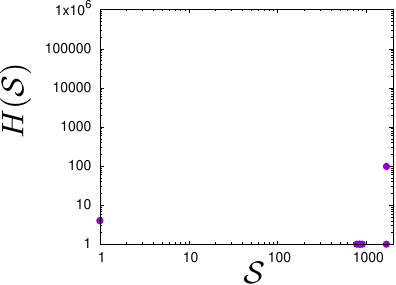}
\end{subfigure}
\begin{subfigure}[b]{.24\textwidth} \caption{\label{fig:hist232} $\alpha=3$, $T=2$}
\includegraphics[width=.99\textwidth]{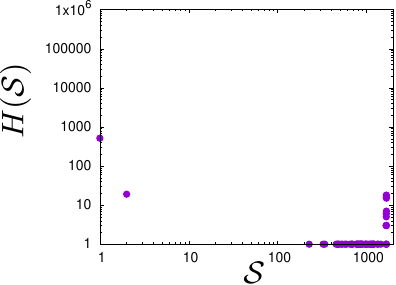}
\end{subfigure}
\begin{subfigure}[b]{.24\textwidth} \caption{\label{fig:hist262} $\alpha=6$, $T=2$}
\includegraphics[width=.99\textwidth]{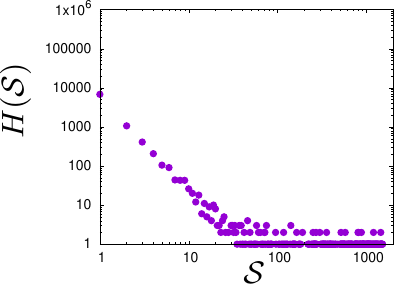}
\end{subfigure}
%% ---------------------------------------------------------------
\begin{subfigure}[b]{.24\textwidth} \caption{\label{fig:hist213} $\alpha=1$, $T=3$}
\includegraphics[width=.99\textwidth]{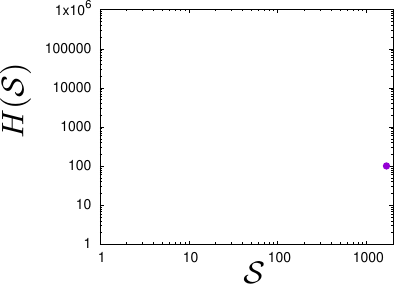}
\end{subfigure}
\begin{subfigure}[b]{.24\textwidth} \caption{\label{fig:hist223} $\alpha=2$, $T=3$}
\includegraphics[width=.99\textwidth]{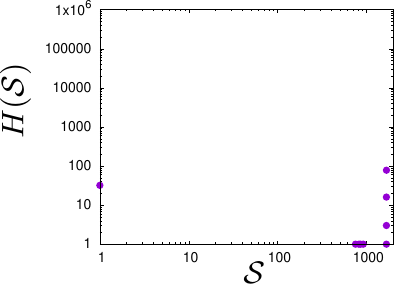}
\end{subfigure}
\begin{subfigure}[b]{.24\textwidth} \caption{\label{fig:hist233} $\alpha=3$, $T=3$}
\includegraphics[width=.99\textwidth]{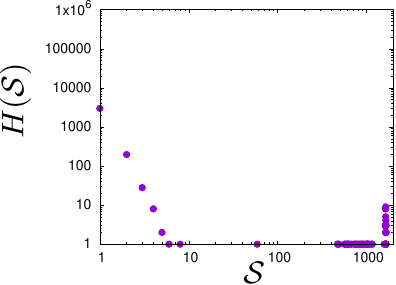}
\end{subfigure}
\begin{subfigure}[b]{.24\textwidth} \caption{\label{fig:hist263} $\alpha=6$, $T=3$}
\includegraphics[width=.99\textwidth]{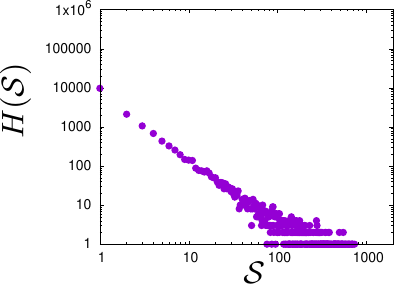}
\end{subfigure}
%% ---------------------------------------------------------------
\begin{subfigure}[b]{.24\textwidth} \caption{\label{fig:hist214} $\alpha=1$, $T=4$}
\includegraphics[width=.99\textwidth]{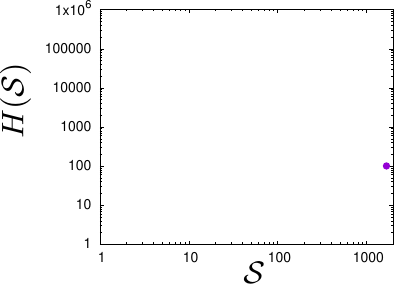}
\end{subfigure}
\begin{subfigure}[b]{.24\textwidth} \caption{\label{fig:hist224} $\alpha=2$, $T=4$}
\includegraphics[width=.99\textwidth]{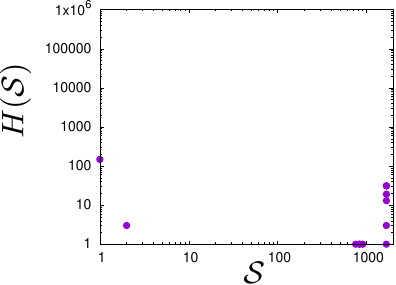}
\end{subfigure}
\begin{subfigure}[b]{.24\textwidth} \caption{\label{fig:hist234} $\alpha=3$, $T=4$}
\includegraphics[width=.99\textwidth]{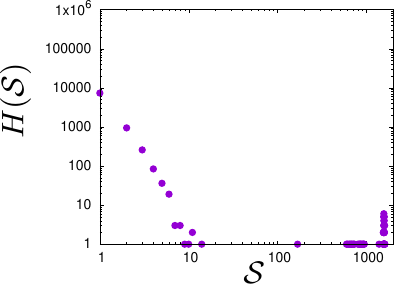}
\end{subfigure}
\begin{subfigure}[b]{.24\textwidth} \caption{\label{fig:hist264} $\alpha=6$, $T=4$}
\includegraphics[width=.99\textwidth]{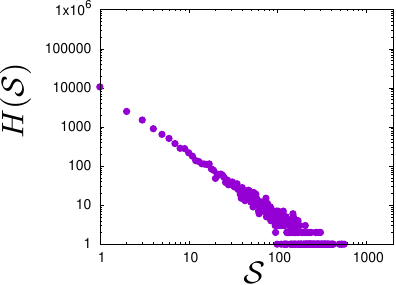}
\end{subfigure}
%% ---------------------------------------------------------------
\begin{subfigure}[b]{.24\textwidth} \caption{\label{fig:hist215} $\alpha=1$, $T=5$}
\includegraphics[width=.99\textwidth]{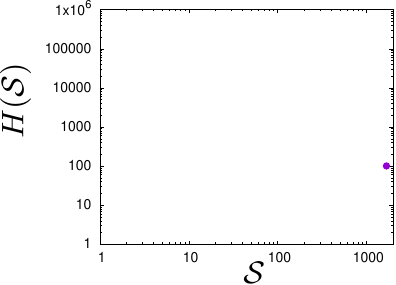}
\end{subfigure}
\begin{subfigure}[b]{.24\textwidth} \caption{\label{fig:hist225} $\alpha=2$, $T=5$}
\includegraphics[width=.99\textwidth]{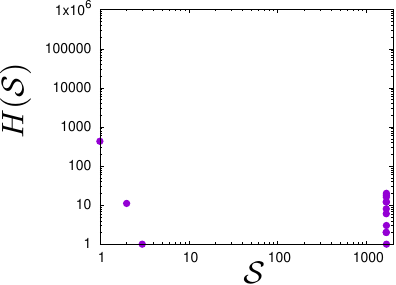}
\end{subfigure}
\begin{subfigure}[b]{.24\textwidth} \caption{\label{fig:hist235} $\alpha=3$, $T=5$}
\includegraphics[width=.99\textwidth]{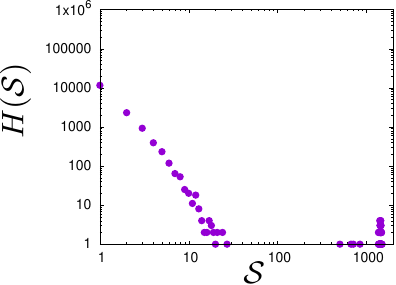}
\end{subfigure}
\begin{subfigure}[b]{.24\textwidth} \caption{\label{fig:hist265} $\alpha=6$, $T=5$}
\includegraphics[width=.99\textwidth]{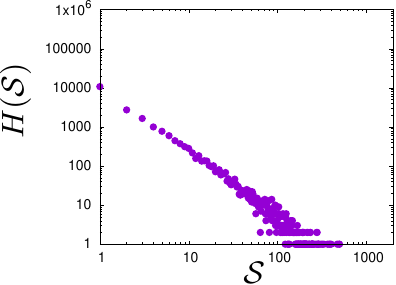}
\end{subfigure}
%% ---------------------------------------------------------------
\caption{\label{fig:histK2}{Histograms $H(\mathcal{S})$ of cluster sizes $\mathcal{S}$ for various values of social temperature $T$ and exponent $\alpha$. $L=41$, $K=2$. {The results are gathered from hundred runings with different initial conditions after $t_{\max}=1000$ time steps.}}}
\end{figure*}

\begin{figure*}
%% \psfrag{s}{$\mathcal{S}$}
%% \psfrag{H(s)}{$H(\mathcal{S})$}
\begin{subfigure}[b]{.24\textwidth} \caption{\label{fig:hist310} $\alpha=1$, $T=0$}
\includegraphics[width=.99\textwidth]{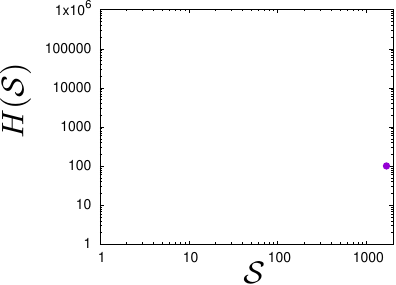}
\end{subfigure}
\begin{subfigure}[b]{.24\textwidth} \caption{\label{fig:hist320} $\alpha=2$, $T=0$}
\includegraphics[width=.99\textwidth]{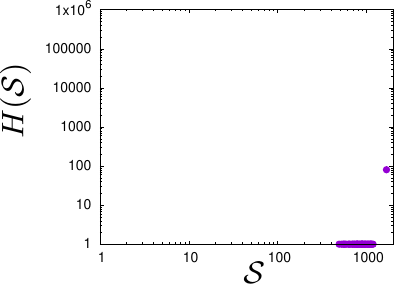}
\end{subfigure}
\begin{subfigure}[b]{.24\textwidth} \caption{\label{fig:hist330} $\alpha=3$, $T=0$}
\includegraphics[width=.99\textwidth]{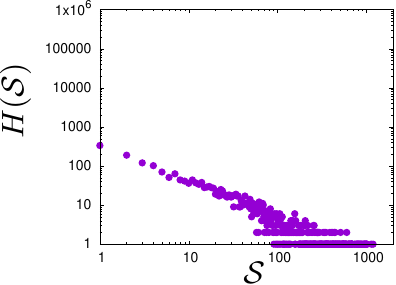}
\end{subfigure}
\begin{subfigure}[b]{.24\textwidth} \caption{\label{fig:hist360} $\alpha=6$, $T=0$}
\includegraphics[width=.99\textwidth]{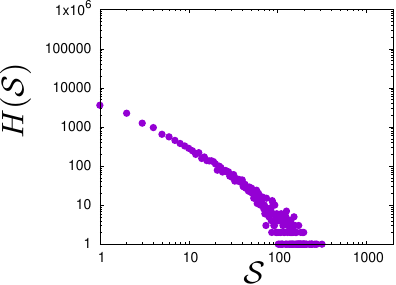}
\end{subfigure}
%% ---------------------------------------------------------------
\begin{subfigure}[b]{.24\textwidth} \caption{\label{fig:hist311} $\alpha=1$, $T=1$}
\includegraphics[width=.99\textwidth]{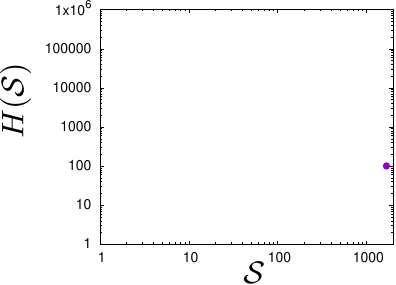}
\end{subfigure}
\begin{subfigure}[b]{.24\textwidth} \caption{\label{fig:hist321} $\alpha=2$, $T=1$}
\includegraphics[width=.99\textwidth]{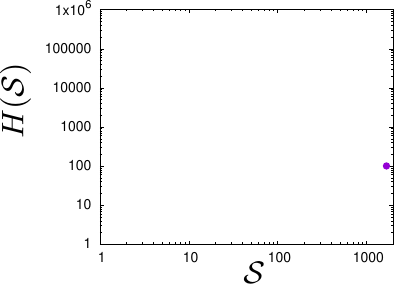}
\end{subfigure}
\begin{subfigure}[b]{.24\textwidth} \caption{\label{fig:hist331} $\alpha=3$, $T=1$}
\includegraphics[width=.99\textwidth]{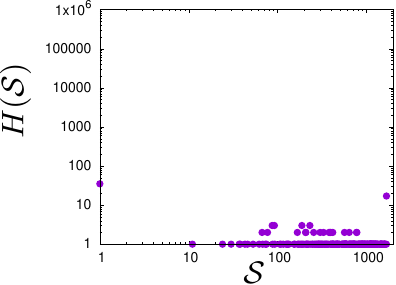}
\end{subfigure}
\begin{subfigure}[b]{.24\textwidth} \caption{\label{fig:hist361} $\alpha=6$, $T=1$}
\includegraphics[width=.99\textwidth]{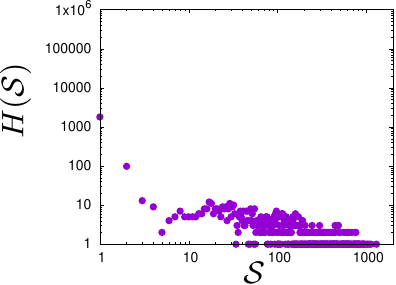}
\end{subfigure}
%% ---------------------------------------------------------------
\begin{subfigure}[b]{.24\textwidth} \caption{\label{fig:hist312} $\alpha=1$, $T=2$}
\includegraphics[width=.99\textwidth]{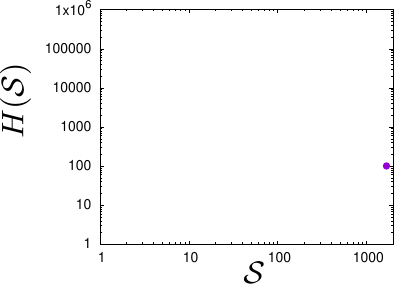}
\end{subfigure}
\begin{subfigure}[b]{.24\textwidth} \caption{\label{fig:hist322} $\alpha=2$, $T=2$}
\includegraphics[width=.99\textwidth]{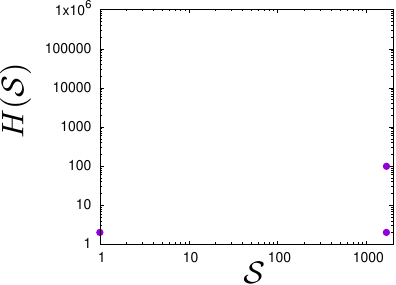}
\end{subfigure}
\begin{subfigure}[b]{.24\textwidth} \caption{\label{fig:hist332} $\alpha=3$, $T=2$}
\includegraphics[width=.99\textwidth]{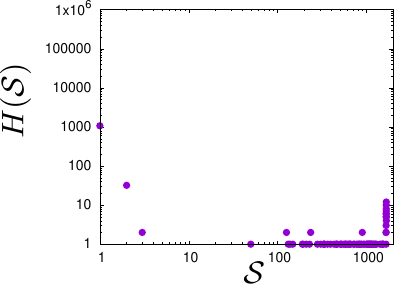}
\end{subfigure}
\begin{subfigure}[b]{.24\textwidth} \caption{\label{fig:hist362} $\alpha=6$, $T=2$}
\includegraphics[width=.99\textwidth]{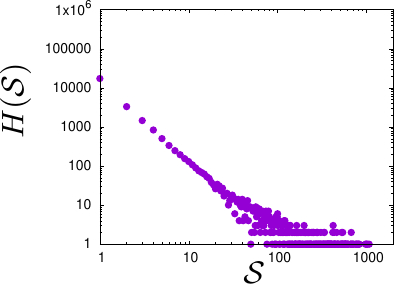}
\end{subfigure}
%% ---------------------------------------------------------------
\begin{subfigure}[b]{.24\textwidth} \caption{\label{fig:hist313} $\alpha=1$, $T=3$}
\includegraphics[width=.99\textwidth]{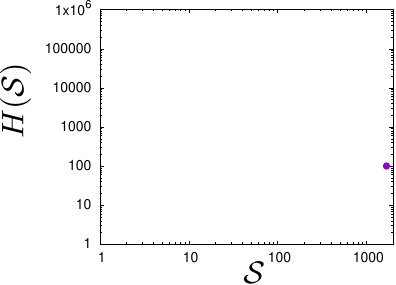}
\end{subfigure}
\begin{subfigure}[b]{.24\textwidth} \caption{\label{fig:hist323} $\alpha=2$, $T=3$}
\includegraphics[width=.99\textwidth]{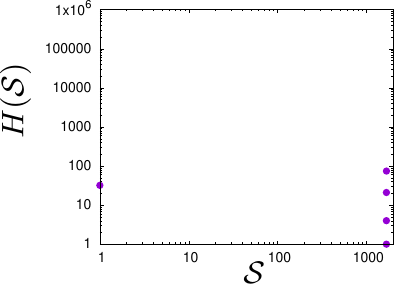}
\end{subfigure}
\begin{subfigure}[b]{.24\textwidth} \caption{\label{fig:hist333} $\alpha=3$, $T=3$}
\includegraphics[width=.99\textwidth]{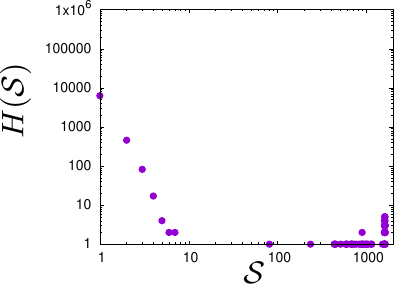}
\end{subfigure}
\begin{subfigure}[b]{.24\textwidth} \caption{\label{fig:hist363} $\alpha=6$, $T=3$}
\includegraphics[width=.99\textwidth]{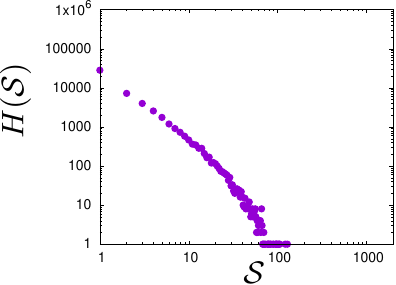}
\end{subfigure}
%% ---------------------------------------------------------------
\begin{subfigure}[b]{.24\textwidth} \caption{\label{fig:hist314} $\alpha=1$, $T=4$}
\includegraphics[width=.99\textwidth]{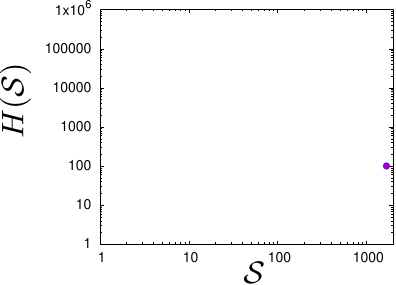}
\end{subfigure}
\begin{subfigure}[b]{.24\textwidth} \caption{\label{fig:hist324} $\alpha=2$, $T=4$}
\includegraphics[width=.99\textwidth]{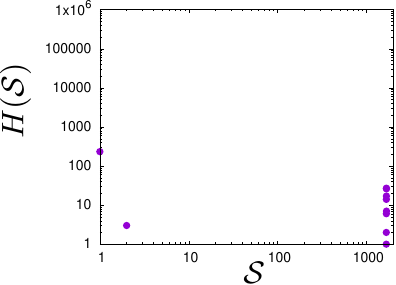}
\end{subfigure}
\begin{subfigure}[b]{.24\textwidth} \caption{\label{fig:hist334} $\alpha=3$, $T=4$}
\includegraphics[width=.99\textwidth]{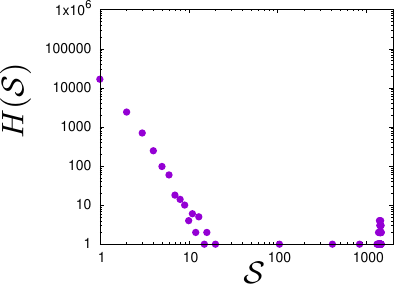}
\end{subfigure}
\begin{subfigure}[b]{.24\textwidth} \caption{\label{fig:hist364} $\alpha=6$, $T=4$}
\includegraphics[width=.99\textwidth]{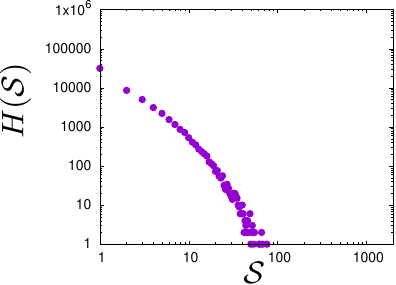}
\end{subfigure}
%% ---------------------------------------------------------------
\begin{subfigure}[b]{.24\textwidth} \caption{\label{fig:hist315} $\alpha=1$, $T=5$}
\includegraphics[width=.99\textwidth]{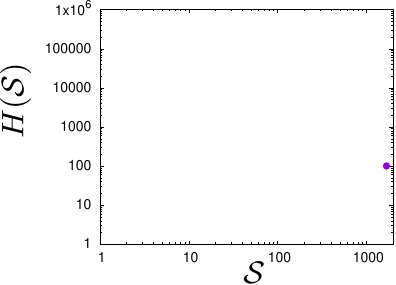}
\end{subfigure}
\begin{subfigure}[b]{.24\textwidth} \caption{\label{fig:hist325} $\alpha=2$, $T=5$}
\includegraphics[width=.99\textwidth]{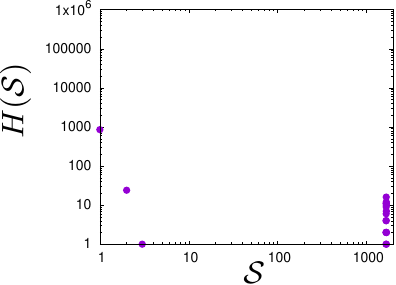}
\end{subfigure}
\begin{subfigure}[b]{.24\textwidth} \caption{\label{fig:hist335} $\alpha=3$, $T=5$}
\includegraphics[width=.99\textwidth]{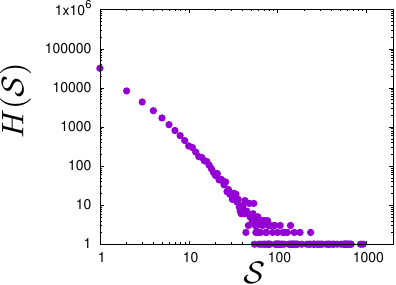}
\end{subfigure}
\begin{subfigure}[b]{.24\textwidth} \caption{\label{fig:hist365} $\alpha=6$, $T=5$}
\includegraphics[width=.99\textwidth]{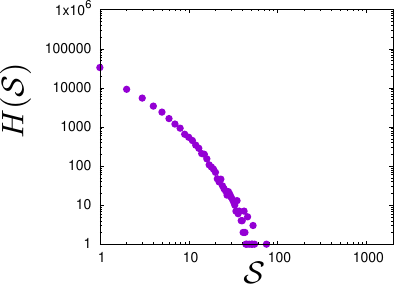}
\end{subfigure}
%% ---------------------------------------------------------------
\caption{\label{fig:histK3}{Histogram $H(\mathcal{S})$ of cluster sizes $\mathcal{S}$ for various values of social temperature $T$ and exponent $\alpha$. $L=41$, $K=3$. {The results are gathered from hundred runings with different initial conditions after $t_{\max}=1000$ time steps.}}}
\end{figure*}

{In Figs.}~\ref{fig:rev2K2xi} ~{and} \ref{fig:rev2K3xi}{---discussed in the previous section---different phases of the system depending on the parameters $\alpha$ and $T$ were presented. In order to get a better look at influence of $T$ and $\alpha$ on system behaviour, we anlyzed the histograms $H(\mathcal{S})$ of cluster sizes $\mathcal{S}$ after thousand steps of simulation gathered from hundred simulations (see Figs.}~\ref{fig:histK2}, \ref{fig:histK3}).

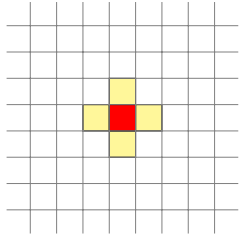
\begin{figure}[b]
\begin{tikzpicture}[scale=0.35]
\filldraw[fill=yellow!50!white, draw=black] ( 0, 1) rectangle ( 1, 2);
\filldraw[fill=yellow!50!white, draw=black] ( 0,-1) rectangle ( 1, 0);
\filldraw[fill=yellow!50!white, draw=black] (-1, 0) rectangle ( 0, 1);
\filldraw[fill=yellow!50!white, draw=black] ( 1, 0) rectangle ( 2, 1);
\filldraw[fill=red, draw=black]             ( 0, 0) rectangle ( 1, 1);
\draw[step=10mm,gray,very thin] (-3.9,-3.9) grid (4.9,4.9);
\end{tikzpicture}
\caption{\label{F:vN} The actors with identical opinions $\Xi_k$ belong to the common cluster if they are in von Neumann neighbourhood.}
\end{figure}

We apply the Hoshen--Kopelman algorithm \cite{Hoshen1976a} for clusters detection.
In Hoshen--Kopelman algorithm each actor is labelled in such way, that actors with the same opinions and in the same cluster have identical labels.
The algorithm allows for cluster detection in multi-dimensional space and for complex neighbourhoods \cite{1803.09504,Malarz2015,Kurzawski2012,Majewski2007,Galam2005a}, here however, we assume the simplest case, i.e. square lattice with von Neumann neighbourhood (see Fig.~\ref{F:vN}).

{To better explain the phenomena observed in Figs}.~\ref{fig:histK2} {and} \ref{fig:histK3}, {the following parameters describing the number and size of clusters were selected:}
\begin{itemize}
\item {average largest cluster size $\langle\mathcal{S}_{\max}\rangle$,}
\item {average cluster number $\langle n_c\rangle$,}
\item {average number of small clusters $\langle n_s\rangle$.}
\end{itemize}

{The example of clusters}
\begin{itemize}
\item {size distribution counting (for preparation histograms $H(\mathcal{S})$ for Figs.}~\ref{fig:histK2} {and} \ref{fig:histK3}),
\item {average largest value $\langle \mathcal{S}_{\max}\rangle$ (for Figs.} \ref{fig:rev2SmaxK2}, \ref{fig:rev2SmaxK3}),
\item {and average numbers $\langle n_c\rangle$ (for Figs.}~\ref{fig:rev2aveclunumK2}, \ref{fig:rev2aveclunumK3})  {and $\langle n_s\rangle$ (for Figs.}~\ref{fig:rev2aveclusizK2}, \ref{fig:rev2aveclusizK3})
\end{itemize}
{are provided in Appendix} \ref{B:small}.

{On the left panels of Fig.}~\ref{fig:rev2K2} {(for $K=2$) and Fig.}~\ref{fig:rev2K3} {(for $K=3$) the `heat maps' and numerical values of $\langle\mathcal{S}_{\max}\rangle$, $\langle n_c\rangle$ and $\langle n_s\rangle$ for different values of $\alpha$ and $T$ are presented. 
The supplementary `contour maps', presenting the same data but allowing for better visualisation of non-monotous character of these data, are included on the right panels of these figures. 
On left panels average values of the larges cluster sizes $\langle\mathcal{S}_{\max}\rangle$ are expressed as a fraction of the number of sites ($L^2$).
The results were averaged over hundred simulations and gathered after $t_{\max}=1000$ time steps.}

{As can be seen in Figs.}~\ref{fig:histK2}, \ref{fig:histK3}, {the shapes of histograms are very similar to each other for $K= 2$ (Fig.}~\ref{fig:histK2}) {and $K=3$ (Fig.}~\ref{fig:histK3}).
{In both cases, there are clear differences in the shape of histograms due to values of $\alpha$ (information flow, effective range of interaction among actors):}
\begin{itemize}
\item {For $\alpha=1$ and $\alpha=2$---where the effective range of interaction is the largest---the unaminity of opinions is achieved (in the case of $\alpha=2$, single agents and rarely small clusters with opposite opinions may appear).}
\item {The most interesting are the histograms for $\alpha=3$ (see figures forming columns from}~\ref{fig:hist230} {to} \ref{fig:hist235} {and from} \ref{fig:hist330} {to} \ref{fig:hist335}{), because the difference in the shape of histograms due to noise ($T$) is also visible.}
{For $\alpha=3$, various phases in the system behaviour are observed.
The system from the disordered state, with the growth of $T$ is increasingly ordered. 
As $T$ increases, more and more clusters appear close to the maximum cluster size in this lattice and more and more small clusters consisting of single agents.}
\item {For $\alpha=6$, histograms have similar shapes for all $T$ values and they indicate the phase of system disorder (a large number of clusters), apart from slight orderliness for $T=1$ (see Figs.}~\ref{fig:hist261} {and} \ref{fig:hist361}).
\end{itemize}
{In Figs.}~\ref{fig:histK2} {and} \ref{fig:histK3}, {we can also notice that for $\alpha>2$, the greater the randomness in adopting opinions by agents (greater $T$), the more single clusters containing one agent and several agents appear.}

%% ===============================================================
%% ===============================================================

%% ===============================================================
%% ===============================================================

%% ===============================================================
%% ===============================================================

%% ---------------------------------------------------------------
\subsubsection{$K=2$}
%% ---------------------------------------------------------------

\begin{figure*}
%% \psfrag{a}{$\alpha$}
%% \psfrag{T}{$T$}
%% ---------------------------------------------------------------
\begin{subfigure}[b]{\textwidth}
\caption{\label{fig:rev2SmaxK2}{Average largest cluster size $\langle\mathcal{S}_{\max}\rangle$ (in left panel normalized to $L^2$).}}
\includegraphics[width=.45\textwidth]{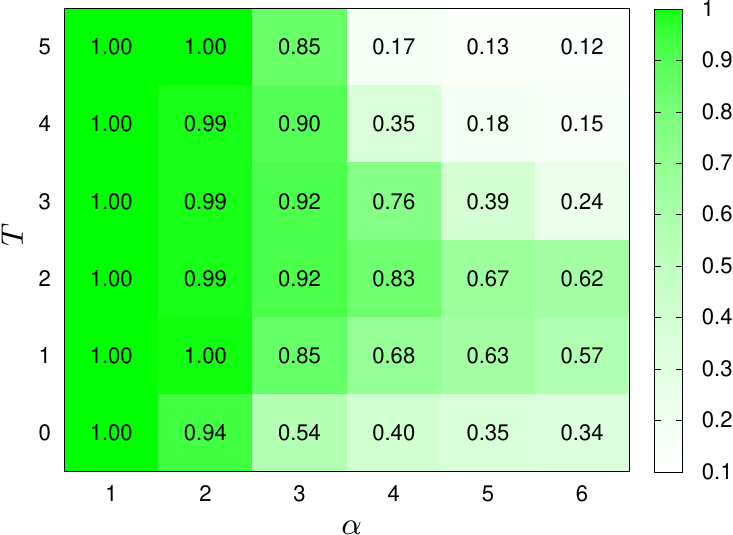} \hfill
\includegraphics[width=.45\textwidth]{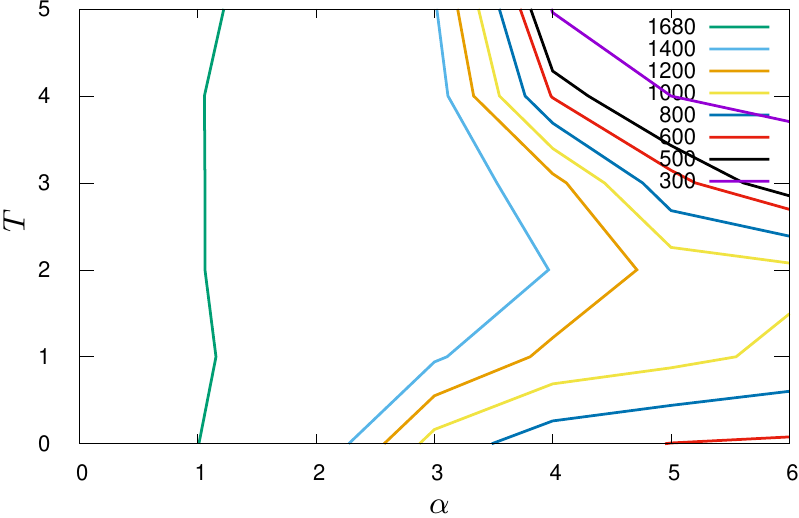}
\end{subfigure}
%% ---------------------------------------------------------------
\begin{subfigure}[b]{\textwidth}
\caption{\label{fig:rev2aveclunumK2}{Average number of clusters $\langle n_c\rangle$.}}
\includegraphics[width=.45\textwidth]{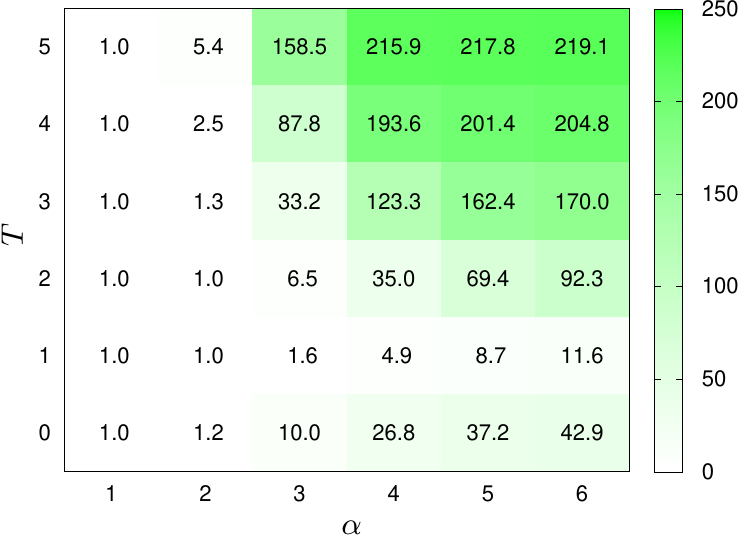} \hfill
\includegraphics[width=.45\textwidth]{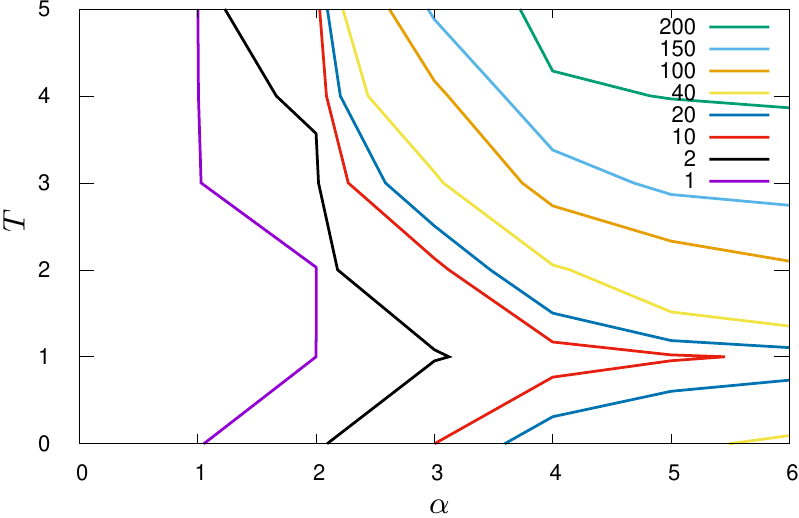} 
\end{subfigure}
%% ---------------------------------------------------------------
\begin{subfigure}[b]{\textwidth}
\caption{\label{fig:rev2aveclusizK2}{Average number of small cluster $\langle n_s\rangle$.}}
\includegraphics[width=.45\textwidth]{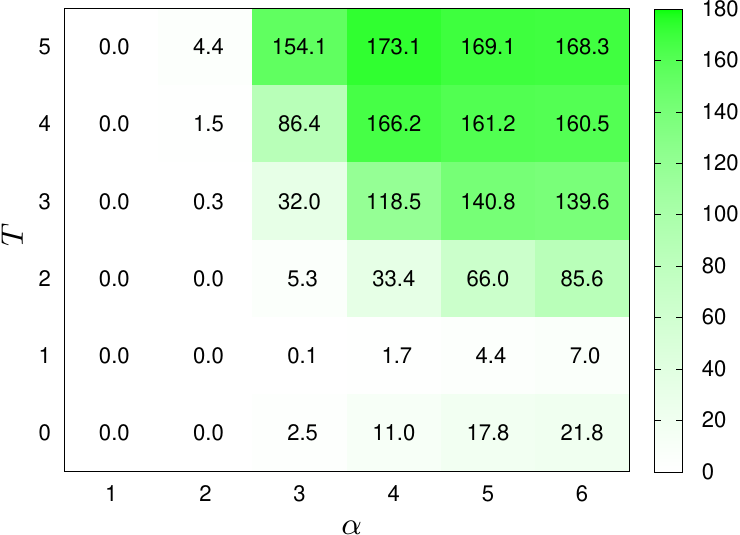} \hfill
\includegraphics[width=.45\textwidth]{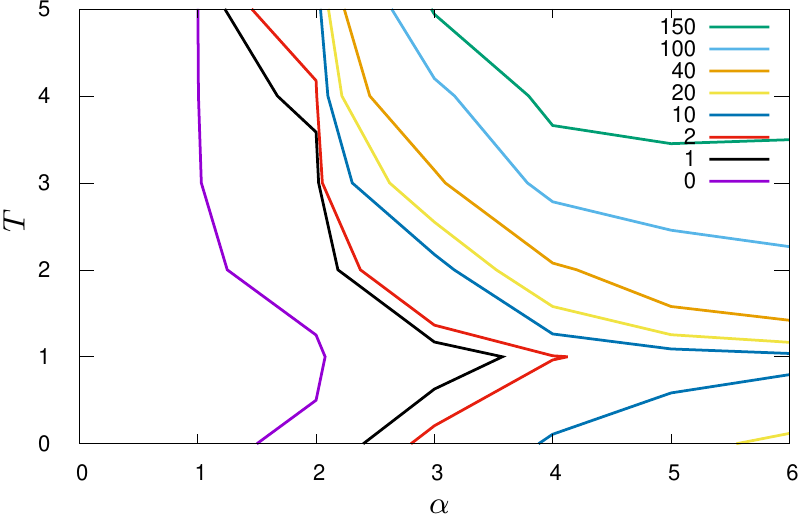}
\end{subfigure}
%% ---------------------------------------------------------------
\caption{\label{fig:rev2K2}{{Average (a) largest cluster size $\langle\mathcal{S}_{\max}\rangle$, (b) number of clusters $\langle n_c\rangle$ and (c)  average number of small clusters $\langle n_s\rangle$ for various values of noise level $T$ and exponent $\alpha$. $L^2=1681$, $K=2$. The results are averaged over hundred runings with different initial conditions and measured after $t_{\max}=10^3$ time steps.}}}
\end{figure*}

{As can be seen in Fig.}~\ref{fig:rev2SmaxK2}, {the average size of the maximum cluster $\langle\mathcal{S}_{\max}\rangle$ decreases with $\alpha$ for fixed $T$ values.
The appearance of noise in the system ($T=1$) slightly organizes the system in relation to the noiseless situation with $T=0$ (see Fig.}~\ref{fig:rev2SmaxK2}).
{Indeed, like in earlier studies} \cite{PhysRevE.75.045101,Shirado2017}{, small level of noise brought more order to the system.}
{In addition, the introduction of noise ($T$) in the adoption of opinions causes an increase in $\langle\mathcal{S}_{\max}\rangle$, and then its decrease, which is especially visible for $\alpha>2$.
The presense of noise $T>0$ results in greater orderliness of the system but only up to certain $T$ values.}
{Particularly interesting are the large values of $\langle\mathcal{S}_{\max}\rangle$ for $\alpha=3$ and $T=2$, 3, 4.
This issue will be disscused below.}

{Figure}~\ref{fig:rev2aveclunumK2} {shows the simulation results of the average number of clusters $\langle n_c\rangle$.
As can be seen, for fixed noise level $T$, the average number of clusters $\langle n_c\rangle$ increases with $\alpha$.
This figure also shows that for $\alpha\ge 3$ the average number of clusters increases with $T$.
In addition, comparing the results for all $\alpha$ values and for the noiseless system ($T=0$) with the results for small noise level ($T=1$), it can be seen that the introduction of noise results in system ordering.}

{The average number of small clusters $\langle n_s\rangle$ increases with $\alpha$ for fixed values of $T$ (see Fig.}~\ref{fig:rev2aveclusizK2}).
{For $\alpha=1$ there are no small clusters, and for $\alpha=2$ there are few of them. 
The increase in the number of small clusters due to $T$ and $\alpha$ is well visible for $\alpha\ge 2$, i.e. when the range of interaction between agents is small.}

{Considering all results presented in Fig.}~\ref{fig:rev2K2} {we can notice, that:} 
\begin{itemize}
\item {For $\alpha\le 2$ we observe the system phase in which there is a consensus (one large cluster representing actors sharing one of the available opinions).
The average size of the largest cluster is equal to the size of the lattice.
		For $\alpha=2$ and $T>2$, the average number of clusters is larger than one, but these higher $\langle n_c\rangle$ values only mean the appearance of clusters consisting of single agents, as $|\langle n_c\rangle-\langle n_s\rangle|\approx 1$, which is shown in Figs.}~\ref{fig:rev2aveclunumK2} and \ref{fig:rev2aveclusizK2} {and also visible in Figs.}~\ref{fig:hist223}, \ref{fig:hist224}, {and} \ref{fig:hist225}.

\item {An interesting phenomenon can be observed for $\alpha=3$, where the frozen noisless system goes into a more ordered phase for $T=1$ and $T=2$, and into a consensus phase for $T=3$ before disordering at $T=5$ (see. Fig.}~\ref{fig:rev2K2}).
{Polarization of opinion is observed for $T=1$ (about 60\% of simulations end with polarization and about 40\% of them end with consensus).
The average size of the largest cluster contains 83.1\% of all agents on the lattice.
The average number of clusters is very small and it equals 1.9, and the average number of small clusters is 0.17, which means that the simulations mostly end in a state containing two clusters, one of which is definitely larger than the other. This is also visible in Figs.}~\ref{fig:rev2K2xi} {and} \ref{fig:hist231}. 
{For $T=2$, more and more system ordering is observed (more and more simulations end with unanimity).}

{The second phase in the case of $\alpha = 3$ takes place for $T=3$.
Despite the high values of noise $T$, the size of the maximum cluster $\langle\mathcal{S}_{\max}\rangle$ is still very large and contains about 90\% of all actors (Fig.}~\ref{fig:rev2SmaxK2}), {and the average number of clusters is definitely larger than for $T\le 2$ (see Fig.}~\ref{fig:rev2aveclunumK2}).
{An increase in noise level surprisingly causes a kind of order.
In this case, one of the opinions dominates, but representatives of the opposite opinion appear in the form of small individual clusters.
The average number of small clusters $\langle n_s\rangle$ is 32 for $T=3$.
This number when compared to the average number of clusters $\langle n_c\rangle$ in Fig.}~\ref{fig:rev2aveclunumK2} {(33.2 for $T=3$) indicates the emergence of one large cluster and individual small clusters (the average number of small clusters $\langle n_s\rangle$ differs by approximately one from the average number of clusters $\langle n_c\rangle$).
This means, that system evolution toward unaminity of opinion dominates system dynamics for $T=3$.}

\item {For $T=4$, the size of the largest cluster $\langle\mathcal{S}_{\max}\rangle$ is still very large (and reaches ca. 90\% of all actors), but although $\langle n_c\rangle$ differs from $\langle n_s\rangle$ by about one, both the number of clusters and the number of small clusters are definitely larger than in the case of $T=3$ (Fig.}~\ref{fig:rev2K2}{). So, for $T=4$, the system goes into a disordered phase.}

{In the case of $T=5$, the system enters a phase of disorder, because there are definitely more clusters with opposite opinions and they have a larger size than for $T=3$ and $T=4$, which is also visible in Fig.}~\ref{fig:hist235} {and Fig.}~\ref{fig:rev2K2xi}. 

{For $\alpha=4$ and $\alpha=5$, when the noise level goes to $T=2$, a slight increase of noise induces more order (see Figs.}~\ref{fig:rev2K2} {and} \ref{fig:histK2}).
{This can be seen in the average size of the largest cluster ($\langle\mathcal{S}_{\max}\rangle$ for $T=2$ is larger than for $T=1$) and by comparing the average number of clusters $\langle n_c\rangle$ with the average number of small clusters $\langle n_s\rangle$ (the difference is smaller than for $T=1$)---Fig.}~\ref{fig:rev2K2}. 

{When the interaction effectively takes place only among the nearest neighbors (for $\alpha=6$), this effect vanishes.}
\end{itemize}

%% ===============================================================
%% ===============================================================

{To check the system behaviour for $K=2$, simulations for $L=21$ and $L=61$ were also carried out. 
The simulations for the smaller and larger network of agents showed results consistent with the presented for networks with size $L=41$. 
Slight differences were observed for $\alpha=3$, where for $T=1$ and $T=2$ after 1000 steps of simulation, in the case of a smaller network, consensus is more often observed, and for a larger network less often.}

{To sum up, three main types of structures in the formation of opinions for $K=2$ are observed:}
\begin{itemize}
\item {formation of single cluster, when all agents adopt one opinion and consensus takes place (for $\alpha \le 2$, $T=1$, 2, 3, 4, 5 and $\alpha=3$, $T=2$, 3, 4),}
\item {greater orderliness---polarization of opinions in clusters ($\alpha=3$, $T=1$, $T=2$ and for $\alpha=3$, 4, and $T=2$),}
\item {formation of plenty small clusters with both opinions---disorder (for other values of $\alpha$ and $T$).}
\end{itemize}

%% ---------------------------------------------------------------
\subsubsection{$K=3$}
%% ---------------------------------------------------------------

\begin{figure*}
%% \psfrag{a}{$\alpha$}
%% \psfrag{T}{$T$}
%% ---------------------------------------------------------------
\begin{subfigure}[b]{\textwidth}
\caption{\label{fig:rev2SmaxK3}{Average largest cluster size $\langle\mathcal{S}_{\max}\rangle$ (in left panel normalized to $L^2$).}}
\includegraphics[width=.45\textwidth]{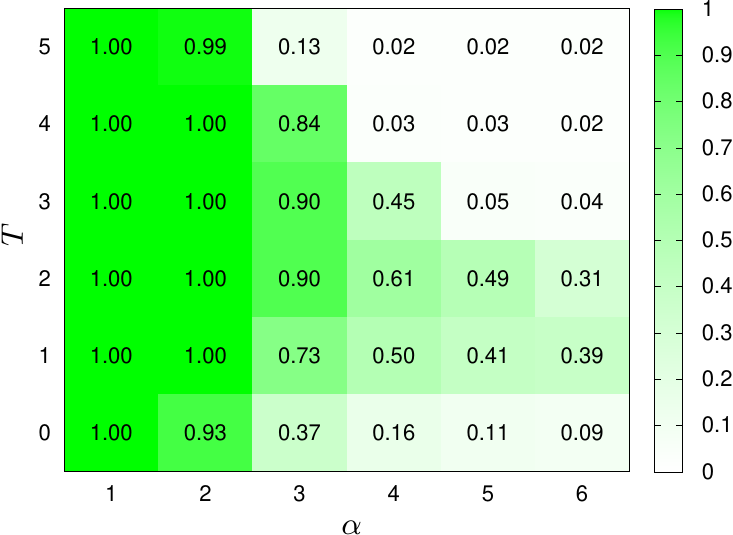} \hfill
\includegraphics[width=.45\textwidth]{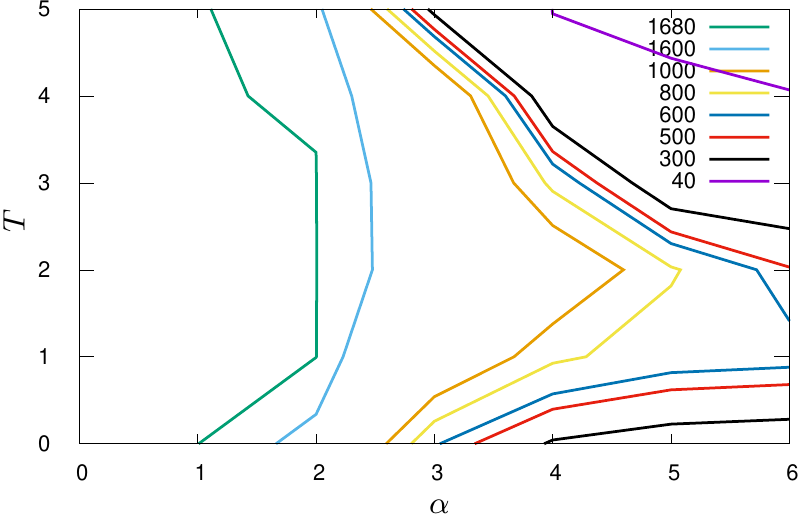}
\end{subfigure}
%% ---------------------------------------------------------------
\begin{subfigure}[b]{\textwidth}
\caption{\label{fig:rev2aveclunumK3}{Average number of clusters $\langle n_c\rangle$.}}
\includegraphics[width=.45\textwidth]{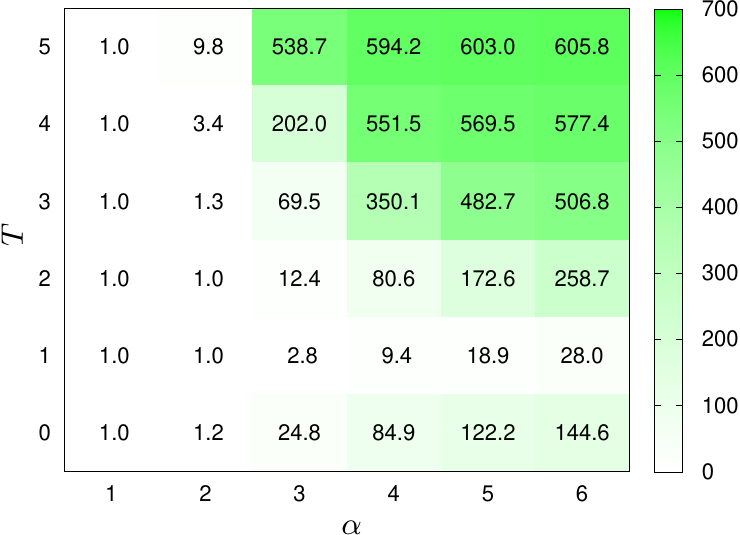} \hfill
\includegraphics[width=.45\textwidth]{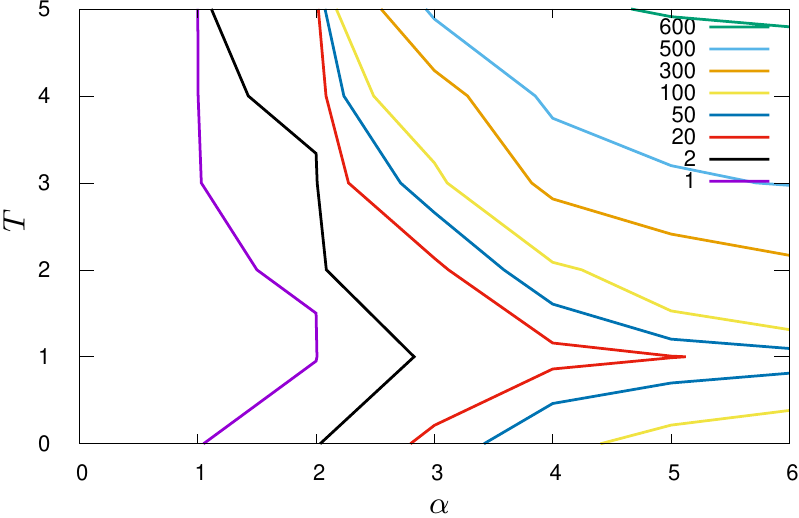}
\end{subfigure}
%% ---------------------------------------------------------------
\begin{subfigure}[b]{\textwidth}
\caption{\label{fig:rev2aveclusizK3}{Average number of small cluster $\langle n_s\rangle$.}}
\includegraphics[width=.45\textwidth]{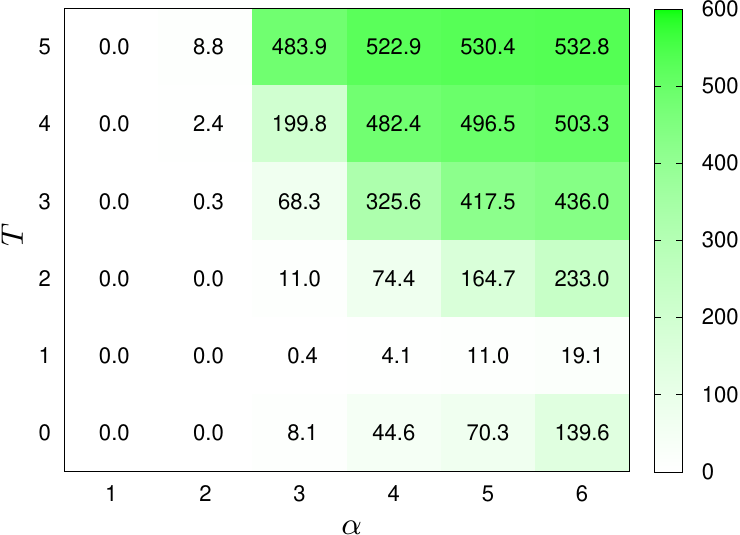} \hfill
\includegraphics[width=.45\textwidth]{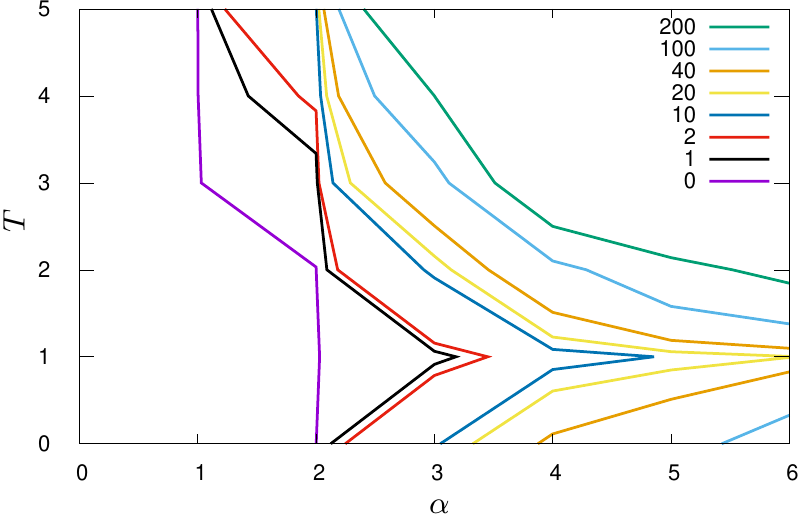}
\end{subfigure}
%% ---------------------------------------------------------------
\caption{\label{fig:rev2K3}{{Average (a) largest cluster size $\langle\mathcal{S}_{\max}\rangle$, (b) number of clusters $\langle n_c\rangle$ and (c)  average number of small clusters $\langle n_s\rangle$ for various values of noise level $T$ and exponent $\alpha$. $L^2=1681$, $K=3$. The results are averaged over hundred runings with different initial conditions and measured after $t_{\max}=10^3$ time steps.}}}
\end{figure*}

{In Fig.}~\ref{fig:rev2K3} {the simulation results for $K = 3$ are presented.
The average size of the largest cluster $\langle\mathcal{S}_{\max}\rangle$ (Fig.}~\ref{fig:rev2SmaxK3}) {decreases with $\alpha$ for fixed values of the noise level $T$, as it was observed for $K=2$.
For $\alpha=1$ and $\alpha=2$, $\langle\mathcal{S}_{\max}\rangle$ is equal to the number of all actors on the lattice. 
For $\alpha>2$, the average size of the largest cluster $\langle\mathcal{S}_{\max}\rangle$ increases up to a certain value of $T$, and then decreases.
This inflection point is nearly $T=2$.}

{In the case of the average number of clusters $\langle n_c\rangle$---which is presented in Fig.}~\ref{fig:rev2aveclunumK3}---{this number is the smallest for $T = 1$, i.e. as soon as noise is introduced to the system. 
Similarly to the case of $K=2$, for fixed noise level $T$, the average number of clusters $\langle n_c\rangle$ increases with $\alpha$.
Fig.}~\ref{fig:rev2aveclunumK3} {also shows, that for $\alpha>2$ the average number of clusters $\langle n_c\rangle$ increases with $T$ for all $T>1$.}

%% ===============================================================
%% ===============================================================

{The average number of small clusters $\langle n_s\rangle$, similarly to the case of $K=2$, increases with the increase $T$ and $\alpha$ (Fig.}~\ref{fig:rev2aveclusizK3}).
{For $\alpha=1$ there are no small clusters, and for $\alpha=2$ they appear only for $T>3$.
The increase in the number of small clusters $\langle n_s\rangle$ with increase of $T$ and $\alpha$ is clearly visible for $\alpha>2$.}

{As the simulation results suggest---also in the case of $K=3$ opinions available in the system---various phases in the system behaviour can be observed.
These phases are induced by interplay of noise level $T$ and effective range of interaction among actors $\alpha$.
First of all, for $\alpha = 1$ and all $T$ values, a large cluster with one opinion is formed (consensus takes place).
In this case, $\langle\mathcal{S}_{\max}\rangle$ are equal to the number of all agents on the lattice, and the average number of clusters $\langle n_c\rangle$ is one (see Figs.}~\ref{fig:rev2SmaxK3}, \ref{fig:rev2aveclunumK3}{---left panels).}
{Single cluster is also created in the case of $\alpha=2$.
As can be seen in Figs.}~\ref{fig:rev2SmaxK3} {and} \ref{fig:rev2aveclunumK3} {the average largest cluster size is close to the system size $\langle\mathcal{S}_{\max}\rangle\approx L^2$, and the average number of clusters is close to one $\langle n_c\rangle\approx 1$.
The exceptions occur for large values of the noise level $T\ge 4$, but this is associated with the appearance of single small and short-living clusters, which is typical for high randomness in actors behaviour.
In this situation the average number of small clusters differs by approximately one from the average number of clusters $|\langle n_c\rangle-\langle n_s\rangle|\le 1$. 
These results are confirmed by the data in Figs.}~\ref{fig:rev2aveclunumK3} {and} \ref{fig:rev2aveclusizK3}.

{As in the case of two opinions ($K=2$), interesting phenomena are visible for $\alpha=3$.
For $T=1$, polarization of opinions is observed.
In 80\% of cases of the final system state two or three clusters of opinions are created, one of which is larger than the others ($\mathcal{S}_{\max}/L^2=0.73$, see Fig.}~\ref{fig:rev2K3T1a3xi}).
{For $T=2$, as in the case of two opinions, more ordering is observed than for $T=1$. 
Consequently, by increasing noise level, for $T=3$ a consensus takes place (one large cluster and small clusters with only single actor).  
The average size of the largest cluster is still very large and contains about 90\% of all actors in the lattice.}
{Although the average number of clusters $\langle n_c\rangle$ is quite large, when compared with the average number of small clusters $\langle n_s\rangle$ it indicates the existence of one large cluster and single small clusters (the average number of clusters differs by approximately one from the average number of small clusters $|\langle n_c\rangle-\langle n_s\rangle|\approx 1$---see Figs.}~\ref{fig:rev2aveclunumK3} {and} \ref{fig:rev2aveclusizK3}).
{For $\alpha=3$ and $T=4$, despite the still high $\langle\mathcal{S}_{\max}\rangle$, the number of small clusters increases, which leads to a disorder phase for $T=5$.}
{The difference between average number of clusters $\langle n_c\rangle$ and average number of small clusters $\langle n_s\rangle$ is definitely greater than one cluster $|\langle n_c\rangle-\langle n_s\rangle|\gg 1$ (cf. Figs.}~\ref{fig:rev2SmaxK3}, \ref{fig:rev2aveclunumK3}).

{For $\alpha>3$ and all $T$ values, system disorder (many clusters of all opinions) is visible---Figs.}~\ref{fig:rev2K3xi} {and} \ref{fig:histK3}{, with the exception of $\alpha=4$ and $\alpha=5$, when a slight increase of noise ($T=2$) does induce more order than for $T=1$ (see Figs.}~\ref{fig:rev2K3} {and} \ref{fig:histK3}).
{As for $K=2$, in the case of three opinions, simulations for $L=21$ and $L=61$ were also carried out. 
The simulations for the smaller and larger network of agents showed similar results to the presented for networks with size $L=41$. 
Slight differences were observed for $\alpha=3$, and $T=1$ and $T=2$ after 1000 steps of simulation, where ordering in clusters is more visible  in small networks than in large ones.}

{To sum up, we can notice three main phases in the behaviour of the model for $K=3$:}
\begin{itemize}
\item {formation of single cluster, when all agents adopt one opinion and consensus takes place (for $\alpha \le 2$; $T=1$, 2, 3, 4, 5 and $\alpha = 3$, $T=3$),}
\item {greater orderliness---polarization of opinions in clusters (sometimes one cluster or ordering opinions in two or three clusters, where two or three opinions occur ($\alpha=3$, $T=1$, $T=2$ and for $\alpha=3$, 4, $T=2$),}
\item {formation of plenty clusters with all three opinions---disorder (for other values of $\alpha$ and $T$).}
\end{itemize}

{}

{}
{}
%%	\label{eq:histalpha1}
{}

{}

{}

{}

%% ###############################################################
\section{\label{S:conculsions}Summary and conclusions}
%% ###############################################################

In this paper, we are interested in how opinions are formed and how they spread in the community.
We were investigating how flow of information in the community and randomness of human behaviour influence formation of opinions, its spreading and its polarisation.
The community was presented as a square lattice of linear size $L$ with open boundary conditions, which is fully filled by {actors}.

The flow of information was controled by the parameter $\alpha$. 
This parameter reflects the effective impact of the neighbourhood on the opinion of the {actors}. 
In case of low values of this parameter, {actors} shape their opinion basing on a large number of {actors} (including distant neighbours). 
In our research, we also take into account the randomness in adopting opinions, which is expressed in the noise parameter $T$. 
The larger $T$, the more often {actors} adopt opinions which have no greatest impact on them

Each {actor} in our model is characterised, in addition to the opinion, by two parameters.
They are the intensity of persuasion ($p_i$) and the intensity of support ($s_i$).
The higher the value of persuasiveness $p_i$, the {actor} more easier convincing other {actors} to accept his/her opinion.
With bigger ($s_i$),  the agent convinces more strongly other {actors}.
These parameters therefore determine the effectiveness of which an individual may interact with or influence other individuals by changing or confirming their opinions.
In all performed simulations, we adopted random values of ($p_i$) and ($s_i$) parameters, which brings us closer to the social reality, in which we do not usually have data on the strength with which the unit affects other units.
Simulations have been carried out when {actors} have a choice of two or three opinions on a given topic.
First, the spatial distribution of opinions after {}{thousand} steps of simulation was analysed.
The simulations showed how clusters of opinion are formed depending on {(i)} the flow of information in the agents' network, {} {and (ii)} the randomness in forming the opinion.
{For both $K=2$ and $K=3$ we can see consensus, polarization of opinions and the formation of many clusters of available opinions.}
{}

{As it was shown in previous sections, the clustering of opinions is influenced by both the level of randomness in actors' decisions (noise) and the impact coming from neighbours.
Generally, the size of the largest cluster of opinions decreases with the increase of $\alpha$ (as can be seen by inspection of columns in Figs.}~\ref{fig:rev2K2xi} {and}~\ref{fig:rev2K3xi}).
{Furthermore, the number of clusters for both $K = 2$ and $K = 3$ increases with $\alpha$, i.e. the smaller effective range of actors' interaction the more difficult forming clusters of opinions. 
Intuitively, an increase in the number of clusters with an increase in noise level $T$ is expected. 
In fact, the number of clusters is growing, but for $\alpha = 3$ and $T=3$ we have one cluster and single agents with opposite opinions, as can be seen in Figs.}~\ref{fig:rev2K2xi} {and} \ref{fig:rev2K3xi}. 
{In this case, introducing of noise ($T$) leads to consensus with single representatives of the opposite opinion(s).
In addition, for $\alpha\ge 3$, a slight increase of noisa elevel ($T=2$) induces more order than for $T=1$ (with the exception for $\alpha=6$ when the interaction effectively takes place only among the nearest neighbors).}

{}

{}

{}

{}

{}

{}

%% \begin{table*}
%%\begin{center}
%%{TABLE X: The impact of both $T$ and $\alpha$ on the formation and spread of opinions.}
%%\end{center}
%% \caption{\label{tab:summary}{The impact of both $T$ and $\alpha$ on the formation and spread of opinions.}}
%%\begin{ruledtabular}
%%	\begin{tabular}{p{5cm}p{5cm}p{5cm}}
%%Impact on&	$T$&	$\alpha$\\ \hline
%%Spatial distribution of opinions &	The greater the $T$, there are more and more clusters&	The greater the $\alpha$, there are more and more clusters\\ \hline
%%		Probability of sustaining opinion&	The {greater} $T$, the less probability
%%		of sustaining the opinion& 	The {greater} $\alpha$, the chances of sustaining the current opinion of agents decrease\\ \hline
%%Spatial distribution of social impact&	---&	The relative impact $I_n/I$ increases with $\alpha$\\ \hline
%%Clustering of opinions&	The number of clusters increases with $T$&
%%	The number of clusters increases with $\alpha$.
%% The size of the largest cluster decreases with the increase of $\alpha$\\ \hline
%%		Summary&	Generally randomness hinders polarisation of opinions in groups&	The better the flow of information in the community, the easier it is to form and spread opinions, which can also lead to consensus\\
%%\end{tabular}
%%\end{ruledtabular}
%% \end{table*}

In summary, the simulations showed that opinion formation and spread is influenced by both: {} {efficiency of information flow among actors and noise level}.
{}  {}
{Better information flow, i.e. better contacts among actors facilitates the spread of opinion and its formation. 
In the case of small values of $\alpha$ (when information flow is very good) the unaminity of opinion is reached and consensus takes place, as in most sociophysical models of opinion dynamics} \cite{RevModPhys.81.591}, {for both, two and three opinions available in the system. 
For large values of $\alpha$---when effectively only the nearest neighbours excert impact on given actor---the polarisation of opinions is weak and there are many small groups of actors with the same opinion.}
{}

The lack of consensus in models is mainly caused by the introduction of noise \cite{Carro-2016} or anti-conformism \cite{Galam-2004}.
In the presented model there is no global agreement also for $T=0$ (when there is no noise).
{For $T=0$ and $\alpha\ge 3$} clusters of both opinions (or three for $K=3$) appear. {}
{In addition, in the presented model, noise for certain values of $\alpha$ promotes unanimity.
This situation occurs for $\alpha=3$ (both for $K=2$ and $K=3$), when the system from the frozen state, with increasing noise $T$, achieves the consensus state for $T=3$, before disordering for $T=5$.
}

As it was mentioned earlier, many studies indicate irrationality and unpredictability in the process of forming opinions \cite{Kowalska-Pyzalska-2014a,Kowalska-Pyzalska-2014b,Byrka-2016,Stadelmann-2013,Sobkowicz-2018}.
As our simulations have shown, this randomness in adopting opinions (noise) plays a {}{crucial} role.
{}
{A low level of noise (low $T$ values) results in less clusters of opinion than in the absence of noise ($T=0$). 
However, the most interesting is the fact, that the high noise levels ($T=2$, 3, 4) results in a more ordered system than for small values of noise $T$ (this is the case with $\alpha=3$). 
Thus, noise favours consensus and polarization of opinion in groups, but only when the influence of distant neighbours is significant. 
If the exchange of opinions takes place only with the nearest neighbors, this effect is not observed.}

In future research, we intend to take into account the impact of strong leaders on the opinion dynamics.
Also the influence of external sources of information (for instance the impact of mass media) is worth of investigation.

%% ---------------------------------------------------------------
\begin{acknowledgments}
We are grateful to anonymous Referee for his/her valuable comments which greatly improved the current version of manuscirpt.
This research was supported by the National Science Centre (NCN) in Poland (grant no. UMO-2014/15/B/HS4/04433) and PL-Grid infrastructure. 
\end{acknowledgments}
%% ---------------------------------------------------------------

\bibliography{this,opiniondynamics,percolation,ca,km}

\begin{thebibliography}{77}%
\makeatletter
\providecommand \@ifxundefined [1]{%
 \@ifx{#1\undefined}
}%
\providecommand \@ifnum [1]{%
 \ifnum #1\expandafter \@firstoftwo
 \else \expandafter \@secondoftwo
 \fi
}%
\providecommand \@ifx [1]{%
 \ifx #1\expandafter \@firstoftwo
 \else \expandafter \@secondoftwo
 \fi
}%
\providecommand \natexlab [1]{#1}%
\providecommand \enquote  [1]{``#1''}%
\providecommand \bibnamefont  [1]{#1}%
\providecommand \bibfnamefont [1]{#1}%
\providecommand \citenamefont [1]{#1}%
\providecommand \href@noop [0]{\@secondoftwo}%
\providecommand \href [0]{\begingroup \@sanitize@url \@href}%
\providecommand \@href[1]{\@@startlink{#1}\@@href}%
\providecommand \@@href[1]{\endgroup#1\@@endlink}%
\providecommand \@sanitize@url [0]{\catcode `\\12\catcode `\$12\catcode
  `\&12\catcode `\#12\catcode `\^12\catcode `\_12\catcode `\%12\relax}%
\providecommand \@@startlink[1]{}%
\providecommand \@@endlink[0]{}%
\providecommand \url  [0]{\begingroup\@sanitize@url \@url }%
\providecommand \@url [1]{\endgroup\@href {#1}{\urlprefix }}%
\providecommand \urlprefix  [0]{URL }%
\providecommand \Eprint [0]{\href }%
\providecommand \doibase [0]{http://dx.doi.org/}%
\providecommand \selectlanguage [0]{\@gobble}%
\providecommand \bibinfo  [0]{\@secondoftwo}%
\providecommand \bibfield  [0]{\@secondoftwo}%
\providecommand \translation [1]{[#1]}%
\providecommand \BibitemOpen [0]{}%
\providecommand \bibitemStop [0]{}%
\providecommand \bibitemNoStop [0]{.\EOS\space}%
\providecommand \EOS [0]{\spacefactor3000\relax}%
\providecommand \BibitemShut  [1]{\csname bibitem#1\endcsname}%
\let\auto@bib@innerbib\@empty
%</preamble>
\bibitem [{\citenamefont {Acemoglu}\ and\ \citenamefont
  {Ozdaglar}(2011)}]{Acemoglu-2011}%
  \BibitemOpen
  \bibfield  {author} {\bibinfo {author} {\bibfnamefont {D.}~\bibnamefont
  {Acemoglu}}\ and\ \bibinfo {author} {\bibfnamefont {A.}~\bibnamefont
  {Ozdaglar}},\ }\bibfield  {title} {\enquote {\bibinfo {title} {Opinion
  dynamics and learning in social networks},}\ }\href {\doibase
  10.1007/s13235-010-0004-1} {\bibfield  {journal} {\bibinfo  {journal}
  {Dynamic Games and Applications}\ }\textbf {\bibinfo {volume} {1}},\ \bibinfo
  {pages} {3--49} (\bibinfo {year} {2011})}\BibitemShut {NoStop}%
\bibitem [{\citenamefont {Jackson}\ and\ \citenamefont
  {Yariv}(2011)}]{Jackson-2011}%
  \BibitemOpen
  \bibfield  {author} {\bibinfo {author} {\bibfnamefont {M.~O.}\ \bibnamefont
  {Jackson}}\ and\ \bibinfo {author} {\bibfnamefont {L.}~\bibnamefont
  {Yariv}},\ }\bibfield  {title} {\enquote {\bibinfo {title} {Diffusion,
  strategic interaction, and social structure},}\ \ }(\bibinfo  {publisher}
  {North-Holland},\ \bibinfo {year} {2011})\ pp.\ \bibinfo {pages}
  {645--678}\BibitemShut {NoStop}%
\bibitem [{\citenamefont {Duncan}\ and\ \citenamefont
  {Moriarty}(1998)}]{Duncan-1998}%
  \BibitemOpen
  \bibfield  {author} {\bibinfo {author} {\bibfnamefont {T.}~\bibnamefont
  {Duncan}}\ and\ \bibinfo {author} {\bibfnamefont {S.~E.}\ \bibnamefont
  {Moriarty}},\ }\bibfield  {title} {\enquote {\bibinfo {title} {A
  communication-based marketing model for managing relationships},}\ }\href
  {\doibase 10.2307/1252157} {\bibfield  {journal} {\bibinfo  {journal}
  {Journal of Marketing}\ }\textbf {\bibinfo {volume} {62}},\ \bibinfo {pages}
  {1--13} (\bibinfo {year} {1998})}\BibitemShut {NoStop}%
\bibitem [{\citenamefont {Simon}(1955)}]{Simon-1955}%
  \BibitemOpen
  \bibfield  {author} {\bibinfo {author} {\bibfnamefont {H.~A.}\ \bibnamefont
  {Simon}},\ }\bibfield  {title} {\enquote {\bibinfo {title} {{A behavioral
  model of rational choice}},}\ }\href {\doibase 10.2307/1884852} {\bibfield
  {journal} {\bibinfo  {journal} {The Quarterly Journal of Economics}\ }\textbf
  {\bibinfo {volume} {69}},\ \bibinfo {pages} {99--118} (\bibinfo {year}
  {1955})}\BibitemShut {NoStop}%
\bibitem [{\citenamefont {Bentley}\ \emph {et~al.}(2011)\citenamefont
  {Bentley}, \citenamefont {Ormerod},\ and\ \citenamefont
  {Batty}}]{Bentley-2011}%
  \BibitemOpen
  \bibfield  {author} {\bibinfo {author} {\bibfnamefont {R.~A.}\ \bibnamefont
  {Bentley}}, \bibinfo {author} {\bibfnamefont {P.}~\bibnamefont {Ormerod}}, \
  and\ \bibinfo {author} {\bibfnamefont {M.}~\bibnamefont {Batty}},\ }\bibfield
   {title} {\enquote {\bibinfo {title} {Evolving social influence in large
  populations},}\ }\href {\doibase 10.1007/s00265-010-1102-1} {\bibfield
  {journal} {\bibinfo  {journal} {Behavioral Ecology and Sociobiology}\
  }\textbf {\bibinfo {volume} {65}},\ \bibinfo {pages} {537--546} (\bibinfo
  {year} {2011})}\BibitemShut {NoStop}%
\bibitem [{\citenamefont {Ku{\l}akowski}\ \emph {et~al.}(2016)\citenamefont
  {Ku{\l}akowski}, \citenamefont {Kulczycki}, \citenamefont {Misztal},
  \citenamefont {Dydejczyk}, \citenamefont {Gronek},\ and\ \citenamefont
  {Krawczyk}}]{Kulakowski-2018}%
  \BibitemOpen
  \bibfield  {author} {\bibinfo {author} {\bibfnamefont {K.}~\bibnamefont
  {Ku{\l}akowski}}, \bibinfo {author} {\bibfnamefont {P.}~\bibnamefont
  {Kulczycki}}, \bibinfo {author} {\bibfnamefont {K.}~\bibnamefont {Misztal}},
  \bibinfo {author} {\bibfnamefont {A.}~\bibnamefont {Dydejczyk}}, \bibinfo
  {author} {\bibfnamefont {P.}~\bibnamefont {Gronek}}, \ and\ \bibinfo {author}
  {\bibfnamefont {M.~J.}\ \bibnamefont {Krawczyk}},\ }\bibfield  {title}
  {\enquote {\bibinfo {title} {Naming boys after {U.S.} presidents in 20th
  century},}\ }\href {\doibase 10.12693/APhysPolA.129.1038} {\bibfield
  {journal} {\bibinfo  {journal} {Acta Physica Polonica A}\ }\textbf {\bibinfo
  {volume} {129}},\ \bibinfo {pages} {1038--1044} (\bibinfo {year}
  {2016})}\BibitemShut {NoStop}%
\bibitem [{\citenamefont {Krawczyk}\ \emph {et~al.}(2014)\citenamefont
  {Krawczyk}, \citenamefont {Dydejczyk},\ and\ \citenamefont
  {Ku{\l}akowski}}]{Krawczyk-2014}%
  \BibitemOpen
  \bibfield  {author} {\bibinfo {author} {\bibfnamefont {M.~J.}\ \bibnamefont
  {Krawczyk}}, \bibinfo {author} {\bibfnamefont {A.}~\bibnamefont {Dydejczyk}},
  \ and\ \bibinfo {author} {\bibfnamefont {K.}~\bibnamefont {Ku{\l}akowski}},\
  }\bibfield  {title} {\enquote {\bibinfo {title} {The {S}immel effect and
  babies' names},}\ }\href {\doibase 10.1016/j.physa.2013.10.018} {\bibfield
  {journal} {\bibinfo  {journal} {Physica A}\ }\textbf {\bibinfo {volume}
  {395}},\ \bibinfo {pages} {384--391} (\bibinfo {year} {2014})}\BibitemShut
  {NoStop}%
\bibitem [{\citenamefont {Guffy}\ \emph {et~al.}(2005)\citenamefont {Guffy},
  \citenamefont {Rhoddes},\ and\ \citenamefont {Rogin}}]{Guffy-2005}%
  \BibitemOpen
  \bibfield  {author} {\bibinfo {author} {\bibfnamefont {M.~E.}\ \bibnamefont
  {Guffy}}, \bibinfo {author} {\bibfnamefont {K.}~\bibnamefont {Rhoddes}}, \
  and\ \bibinfo {author} {\bibfnamefont {P.}~\bibnamefont {Rogin}},\
  }\href@noop {} {\emph {\bibinfo {title} {Business Communication}}}\ (\bibinfo
   {publisher} {South-Western},\ \bibinfo {address} {Toronto},\ \bibinfo {year}
  {2005})\BibitemShut {NoStop}%
\bibitem [{\citenamefont {{Kowalska-Pyzalska}}\ \emph
  {et~al.}(2014)\citenamefont {{Kowalska-Pyzalska}}, \citenamefont
  {{Maciejowska}}, \citenamefont {{Sznajd-Weron}},\ and\ \citenamefont
  {{Weron}}}]{Kowalska-Pyzalska-2014a}%
  \BibitemOpen
  \bibfield  {author} {\bibinfo {author} {\bibfnamefont {A.}~\bibnamefont
  {{Kowalska-Pyzalska}}}, \bibinfo {author} {\bibfnamefont {K.}~\bibnamefont
  {{Maciejowska}}}, \bibinfo {author} {\bibfnamefont {K.}~\bibnamefont
  {{Sznajd-Weron}}}, \ and\ \bibinfo {author} {\bibfnamefont {R.}~\bibnamefont
  {{Weron}}},\ }\bibfield  {title} {\enquote {\bibinfo {title} {Modeling
  consumer opinions towards dynamic pricing: {A}n agent-based approach},}\ }in\
  \href {\doibase 10.1109/EEM.2014.6861272} {\emph {\bibinfo {booktitle} {11th
  International Conference on the European Energy Market}}}\ (\bibinfo {year}
  {2014})\ pp.\ \bibinfo {pages} {1--5}\BibitemShut {NoStop}%
\bibitem [{\citenamefont {Kowalska-Pyzalska}\ \emph {et~al.}(2014)\citenamefont
  {Kowalska-Pyzalska}, \citenamefont {Maciejowska}, \citenamefont
  {Suszczy\'nski}, \citenamefont {Sznajd-Weron},\ and\ \citenamefont
  {Weron}}]{Kowalska-Pyzalska-2014b}%
  \BibitemOpen
  \bibfield  {author} {\bibinfo {author} {\bibfnamefont {A.}~\bibnamefont
  {Kowalska-Pyzalska}}, \bibinfo {author} {\bibfnamefont {K.}~\bibnamefont
  {Maciejowska}}, \bibinfo {author} {\bibfnamefont {K.}~\bibnamefont
  {Suszczy\'nski}}, \bibinfo {author} {\bibfnamefont {K.}~\bibnamefont
  {Sznajd-Weron}}, \ and\ \bibinfo {author} {\bibfnamefont {R.}~\bibnamefont
  {Weron}},\ }\bibfield  {title} {\enquote {\bibinfo {title} {Turning green:
  {A}gent-based modeling of the adoption of dynamic electricity tariffs},}\
  }\href {\doibase 10.1016/j.enpol.2014.04.021} {\bibfield  {journal} {\bibinfo
   {journal} {Energy Policy}\ }\textbf {\bibinfo {volume} {72}},\ \bibinfo
  {pages} {164--174} (\bibinfo {year} {2014})}\BibitemShut {NoStop}%
\bibitem [{\citenamefont {Byrka}\ \emph {et~al.}(2016)\citenamefont {Byrka},
  \citenamefont {J\k{e}drzejewski}, \citenamefont {Sznajd-Weron},\ and\
  \citenamefont {Weron}}]{Byrka-2016}%
  \BibitemOpen
  \bibfield  {author} {\bibinfo {author} {\bibfnamefont {K.}~\bibnamefont
  {Byrka}}, \bibinfo {author} {\bibfnamefont {A.}~\bibnamefont
  {J\k{e}drzejewski}}, \bibinfo {author} {\bibfnamefont {K.}~\bibnamefont
  {Sznajd-Weron}}, \ and\ \bibinfo {author} {\bibfnamefont {R.}~\bibnamefont
  {Weron}},\ }\bibfield  {title} {\enquote {\bibinfo {title} {Difficulty is
  critical: The importance of social factors in modeling diffusion of green
  products and practices},}\ }\href {\doibase 10.1016/j.rser.2016.04.063}
  {\bibfield  {journal} {\bibinfo  {journal} {Renewable and Sustainable Energy
  Reviews}\ }\textbf {\bibinfo {volume} {62}},\ \bibinfo {pages} {723--735}
  (\bibinfo {year} {2016})}\BibitemShut {NoStop}%
\bibitem [{\citenamefont {Stadelmann}\ and\ \citenamefont
  {Torgler}(2013)}]{Stadelmann-2013}%
  \BibitemOpen
  \bibfield  {author} {\bibinfo {author} {\bibfnamefont {D.}~\bibnamefont
  {Stadelmann}}\ and\ \bibinfo {author} {\bibfnamefont {B.}~\bibnamefont
  {Torgler}},\ }\bibfield  {title} {\enquote {\bibinfo {title} {Bounded
  rationality and voting decisions over 160 years: {V}oter behavior and
  increasing complexity in decision-making},}\ }\href {\doibase
  10.1371/journal.pone.0084078} {\bibfield  {journal} {\bibinfo  {journal}
  {PLoS ONE}\ }\textbf {\bibinfo {volume} {8}},\ \bibinfo {pages} {e84078}
  (\bibinfo {year} {2013})}\BibitemShut {NoStop}%
\bibitem [{\citenamefont {Sobkowicz}(2018)}]{Sobkowicz-2018}%
  \BibitemOpen
  \bibfield  {author} {\bibinfo {author} {\bibfnamefont {P.}~\bibnamefont
  {Sobkowicz}},\ }\bibfield  {title} {\enquote {\bibinfo {title} {Opinion
  dynamics model based on cognitive biases of complex agents},}\ }\href
  {\doibase 10.18564/jasss.3867} {\bibfield  {journal} {\bibinfo  {journal}
  {JASSS---the Journal of Artificial Societies and Social Simulation}\ }\textbf
  {\bibinfo {volume} {21}},\ \bibinfo {pages} {(4)8} (\bibinfo {year}
  {2018})}\BibitemShut {NoStop}%
\bibitem [{\citenamefont {Apolloni}\ and\ \citenamefont
  {Gargiulo}(2011)}]{Apolloni-2011}%
  \BibitemOpen
  \bibfield  {author} {\bibinfo {author} {\bibfnamefont {A.}~\bibnamefont
  {Apolloni}}\ and\ \bibinfo {author} {\bibfnamefont {F.}~\bibnamefont
  {Gargiulo}},\ }\bibfield  {title} {\enquote {\bibinfo {title} {Diffusion
  processes through social groups' dynamics},}\ }\href {\doibase
  10.1142/S0219525911003037} {\bibfield  {journal} {\bibinfo  {journal}
  {Advances in Complex Systems}\ }\textbf {\bibinfo {volume} {14}},\ \bibinfo
  {pages} {151--167} (\bibinfo {year} {2011})}\BibitemShut {NoStop}%
\bibitem [{\citenamefont {Latan\'e}(1981)}]{Latane-1981}%
  \BibitemOpen
  \bibfield  {author} {\bibinfo {author} {\bibfnamefont {B.}~\bibnamefont
  {Latan\'e}},\ }\bibfield  {title} {\enquote {\bibinfo {title} {The psychology
  of social impact},}\ }\href {\doibase 10.1037/0003-066X.36.4.343} {\bibfield
  {journal} {\bibinfo  {journal} {American Psychologist}\ }\textbf {\bibinfo
  {volume} {36}},\ \bibinfo {pages} {343--356} (\bibinfo {year}
  {1981})}\BibitemShut {NoStop}%
\bibitem [{\citenamefont {Ba\'ncerowski}\ and\ \citenamefont
  {Malarz}(2019)}]{1902.03454}%
  \BibitemOpen
  \bibfield  {author} {\bibinfo {author} {\bibfnamefont {P.}~\bibnamefont
  {Ba\'ncerowski}}\ and\ \bibinfo {author} {\bibfnamefont {K.}~\bibnamefont
  {Malarz}},\ }\bibfield  {title} {\enquote {\bibinfo {title} {Multi-choice
  opinion dynamics model based on {L}atan\'e theory},}\ }\href {\doibase
  10.1140/epjb/e2019-90533-0} {\bibfield  {journal} {\bibinfo  {journal}
  {European Physical Journal B}\ }\textbf {\bibinfo {volume} {92}},\ \bibinfo
  {pages} {219} (\bibinfo {year} {2019})}\BibitemShut {NoStop}%
\bibitem [{\citenamefont
  {Ba\'ncerowski}(2017{\natexlab{a}})}]{ThesisBancerowski}%
  \BibitemOpen
  \bibfield  {author} {\bibinfo {author} {\bibfnamefont {P.}~\bibnamefont
  {Ba\'ncerowski}},\ }\href@noop {} {Master's thesis},\ \bibinfo  {school}
  {{AGH} University of Science and Technology}, \bibinfo {address} {Krak\'ow}
  (\bibinfo {year} {2017}{\natexlab{a}}),\ \bibinfo {note} {in
  Polish}\BibitemShut {NoStop}%
\bibitem [{\citenamefont {Latan{\'e}}\ and\ \citenamefont
  {Harkins}(1976)}]{Latane-1976}%
  \BibitemOpen
  \bibfield  {author} {\bibinfo {author} {\bibfnamefont {B.}~\bibnamefont
  {Latan{\'e}}}\ and\ \bibinfo {author} {\bibfnamefont {S.}~\bibnamefont
  {Harkins}},\ }\bibfield  {title} {\enquote {\bibinfo {title} {Cross-modality
  matches suggest anticipated stage fright a multiplicative power function of
  audience size and status},}\ }\href {\doibase 10.3758/BF03208286} {\bibfield
  {journal} {\bibinfo  {journal} {Perception {\&} Psychophysics}\ }\textbf
  {\bibinfo {volume} {20}},\ \bibinfo {pages} {482--488} (\bibinfo {year}
  {1976})}\BibitemShut {NoStop}%
\bibitem [{\citenamefont {Darley}\ and\ \citenamefont
  {Latan\'e}({1968})}]{Darley1968}%
  \BibitemOpen
  \bibfield  {author} {\bibinfo {author} {\bibfnamefont {J.~M.}\ \bibnamefont
  {Darley}}\ and\ \bibinfo {author} {\bibfnamefont {B.}~\bibnamefont
  {Latan\'e}},\ }\bibfield  {title} {\enquote {\bibinfo {title} {Bystander
  intervention in emergencies---{D}iffusion of responsibility},}\ }\href
  {\doibase 10.1037/h0025589} {\bibfield  {journal} {\bibinfo  {journal}
  {Journal of Personality and Social Psychology}\ }\textbf {\bibinfo {volume}
  {{8}}},\ \bibinfo {pages} {377--383} (\bibinfo {year} {{1968}})}\BibitemShut
  {NoStop}%
\bibitem [{\citenamefont {Latan\'e}\ and\ \citenamefont
  {Nida}(1981)}]{Latane-1981a}%
  \BibitemOpen
  \bibfield  {author} {\bibinfo {author} {\bibfnamefont {B.}~\bibnamefont
  {Latan\'e}}\ and\ \bibinfo {author} {\bibfnamefont {S.}~\bibnamefont
  {Nida}},\ }\bibfield  {title} {\enquote {\bibinfo {title} {Ten years of
  research on group size and helping},}\ }\href {\doibase
  10.1037/0033-2909.89.2.308} {\bibfield  {journal} {\bibinfo  {journal}
  {Psychological Bulletin}\ }\textbf {\bibinfo {volume} {89}},\ \bibinfo
  {pages} {308--324} (\bibinfo {year} {1981})}\BibitemShut {NoStop}%
\bibitem [{\citenamefont {Nowak}\ \emph {et~al.}(1990)\citenamefont {Nowak},
  \citenamefont {Szamrej},\ and\ \citenamefont {Latan\'e}}]{Nowak-1990}%
  \BibitemOpen
  \bibfield  {author} {\bibinfo {author} {\bibfnamefont {A.}~\bibnamefont
  {Nowak}}, \bibinfo {author} {\bibfnamefont {J.}~\bibnamefont {Szamrej}}, \
  and\ \bibinfo {author} {\bibfnamefont {B.}~\bibnamefont {Latan\'e}},\
  }\bibfield  {title} {\enquote {\bibinfo {title} {From private attitude to
  public opinion: A dynamic theory of social impact},}\ }\href {\doibase
  10.1037/0033-295X.97.3.362} {\bibfield  {journal} {\bibinfo  {journal}
  {Psychological Review}\ }\textbf {\bibinfo {volume} {97}},\ \bibinfo {pages}
  {362--376} (\bibinfo {year} {1990})}\BibitemShut {NoStop}%
\bibitem [{\citenamefont {Burgos}\ \emph {et~al.}(2015)\citenamefont {Burgos},
  \citenamefont {Hern\'andez}, \citenamefont {Ceva},\ and\ \citenamefont
  {Perazzo}}]{PhysRevE.91.032808}%
  \BibitemOpen
  \bibfield  {author} {\bibinfo {author} {\bibfnamefont {E.}~\bibnamefont
  {Burgos}}, \bibinfo {author} {\bibfnamefont {L.}~\bibnamefont {Hern\'andez}},
  \bibinfo {author} {\bibfnamefont {H.}~\bibnamefont {Ceva}}, \ and\ \bibinfo
  {author} {\bibfnamefont {R.~P.~J.}\ \bibnamefont {Perazzo}},\ }\bibfield
  {title} {\enquote {\bibinfo {title} {Entropic determination of the phase
  transition in a coevolving opinion-formation model},}\ }\href {\doibase
  10.1103/PhysRevE.91.032808} {\bibfield  {journal} {\bibinfo  {journal}
  {Physical Review E}\ }\textbf {\bibinfo {volume} {91}},\ \bibinfo {pages}
  {032808} (\bibinfo {year} {2015})}\BibitemShut {NoStop}%
\bibitem [{\citenamefont {Holme}\ and\ \citenamefont
  {Newman}(2006)}]{PhysRevE.74.056108}%
  \BibitemOpen
  \bibfield  {author} {\bibinfo {author} {\bibfnamefont {P.}~\bibnamefont
  {Holme}}\ and\ \bibinfo {author} {\bibfnamefont {M.~E.~J.}\ \bibnamefont
  {Newman}},\ }\bibfield  {title} {\enquote {\bibinfo {title} {Nonequilibrium
  phase transition in the coevolution of networks and opinions},}\ }\href
  {\doibase 10.1103/PhysRevE.74.056108} {\bibfield  {journal} {\bibinfo
  {journal} {Physical Review E}\ }\textbf {\bibinfo {volume} {74}},\ \bibinfo
  {pages} {056108} (\bibinfo {year} {2006})}\BibitemShut {NoStop}%
\bibitem [{\citenamefont {Martins}(2019)}]{Martins_2019}%
  \BibitemOpen
  \bibfield  {author} {\bibinfo {author} {\bibfnamefont {A.~C.~R.}\
  \bibnamefont {Martins}},\ }\href@noop {} {\enquote {\bibinfo {title}
  {{Discrete opinion dynamics with $M$ choices}},}\ } (\bibinfo {year}
  {2019}),\ \Eprint {http://arxiv.org/abs/1905.10878} {arXiv:1905.10878
  [physics.soc-ph]} \BibitemShut {NoStop}%
\bibitem [{\citenamefont {Wu}\ and\ \citenamefont
  {Szeto}({2018})}]{000432967700004}%
  \BibitemOpen
  \bibfield  {author} {\bibinfo {author} {\bibfnamefont {Degang}\ \bibnamefont
  {Wu}}\ and\ \bibinfo {author} {\bibfnamefont {Kwok~Yip}\ \bibnamefont
  {Szeto}},\ }\bibfield  {title} {\enquote {\bibinfo {title} {Analysis of
  timescale to consensus in voting dynamics with more than two options},}\
  }\href {\doibase 10.1103/PhysRevE.97.042320} {\bibfield  {journal} {\bibinfo
  {journal} {Physical Review E}\ }\textbf {\bibinfo {volume} {{97}}},\ \bibinfo
  {pages} {042320} (\bibinfo {year} {{2018}})}\BibitemShut {NoStop}%
\bibitem [{\citenamefont {Galam}(2013)}]{000316891200004}%
  \BibitemOpen
  \bibfield  {author} {\bibinfo {author} {\bibfnamefont {S.}~\bibnamefont
  {Galam}},\ }\bibfield  {title} {\enquote {\bibinfo {title} {The drastic
  outcomes from voting alliances in three-party democratic voting
  (1990--2013)},}\ }\href {\doibase 10.1007/s10955-012-0641-4} {\bibfield
  {journal} {\bibinfo  {journal} {Journal of Statistical Physics}\ }\textbf
  {\bibinfo {volume} {151}},\ \bibinfo {pages} {46--68} (\bibinfo {year}
  {2013})}\BibitemShut {NoStop}%
\bibitem [{\citenamefont {Malarz}\ and\ \citenamefont
  {Ku{\l}akowski}(2010)}]{Kulakowski2010}%
  \BibitemOpen
  \bibfield  {author} {\bibinfo {author} {\bibfnamefont {K.}~\bibnamefont
  {Malarz}}\ and\ \bibinfo {author} {\bibfnamefont {K.}~\bibnamefont
  {Ku{\l}akowski}},\ }\bibfield  {title} {\enquote {\bibinfo {title}
  {Indifferents as an interface between contra and pro},}\ }\href {\doibase
  10.12693/APhysPolA.117.695} {\bibfield  {journal} {\bibinfo  {journal} {Acta
  Physica Polonica A}\ }\textbf {\bibinfo {volume} {117}},\ \bibinfo {pages}
  {695--699} (\bibinfo {year} {2010})}\BibitemShut {NoStop}%
\bibitem [{\citenamefont {Gekle}\ \emph {et~al.}(2005)\citenamefont {Gekle},
  \citenamefont {Peliti},\ and\ \citenamefont {Galam}}]{Gekle-2005}%
  \BibitemOpen
  \bibfield  {author} {\bibinfo {author} {\bibfnamefont {S.}~\bibnamefont
  {Gekle}}, \bibinfo {author} {\bibfnamefont {L.}~\bibnamefont {Peliti}}, \
  and\ \bibinfo {author} {\bibfnamefont {S.}~\bibnamefont {Galam}},\ }\bibfield
   {title} {\enquote {\bibinfo {title} {Opinion dynamics in a three-choice
  system},}\ }\href {\doibase 10.1140/epjb/e2005-00215-3} {\bibfield  {journal}
  {\bibinfo  {journal} {European Physical Journal B}\ }\textbf {\bibinfo
  {volume} {45}},\ \bibinfo {pages} {569--575} (\bibinfo {year}
  {2005})}\BibitemShut {NoStop}%
\bibitem [{\citenamefont {de~la Lama}\ \emph {et~al.}(2006)\citenamefont {de~la
  Lama}, \citenamefont {Szendro}, \citenamefont {Iglesias},\ and\ \citenamefont
  {Wio}}]{000238502800018}%
  \BibitemOpen
  \bibfield  {author} {\bibinfo {author} {\bibfnamefont {M.~S.}\ \bibnamefont
  {de~la Lama}}, \bibinfo {author} {\bibfnamefont {I.~G.}\ \bibnamefont
  {Szendro}}, \bibinfo {author} {\bibfnamefont {J.~R.}\ \bibnamefont
  {Iglesias}}, \ and\ \bibinfo {author} {\bibfnamefont {H.~S.}\ \bibnamefont
  {Wio}},\ }\bibfield  {title} {\enquote {\bibinfo {title} {Van {K}ampen's
  expansion approach in an opinion formation model},}\ }\href {\doibase
  10.1140/epjb/e2006-00232-8} {\bibfield  {journal} {\bibinfo  {journal}
  {European Physical Journal B}\ }\textbf {\bibinfo {volume} {51}},\ \bibinfo
  {pages} {435--442} (\bibinfo {year} {2006})}\BibitemShut {NoStop}%
\bibitem [{\citenamefont {Vazquez}\ and\ \citenamefont
  {Redner}(2004)}]{000223811700008}%
  \BibitemOpen
  \bibfield  {author} {\bibinfo {author} {\bibfnamefont {F.}~\bibnamefont
  {Vazquez}}\ and\ \bibinfo {author} {\bibfnamefont {S.}~\bibnamefont
  {Redner}},\ }\bibfield  {title} {\enquote {\bibinfo {title} {Ultimate fate of
  constrained voters},}\ }\href {\doibase 10.1088/0305-4470/37/35/006}
  {\bibfield  {journal} {\bibinfo  {journal} {Journal of Physics
  A---Mathematical and General}\ }\textbf {\bibinfo {volume} {37}},\ \bibinfo
  {pages} {8479--8494} (\bibinfo {year} {2004})}\BibitemShut {NoStop}%
\bibitem [{\citenamefont {Galam}(1991)}]{A1991FU10100002}%
  \BibitemOpen
  \bibfield  {author} {\bibinfo {author} {\bibfnamefont {S.}~\bibnamefont
  {Galam}},\ }\bibfield  {title} {\enquote {\bibinfo {title} {Political
  paradoxes of majority-rule voting and hierarchical systems},}\ }\href
  {\doibase 10.1080/03081079108935145} {\bibfield  {journal} {\bibinfo
  {journal} {International Journal of General Systems}\ }\textbf {\bibinfo
  {volume} {18}},\ \bibinfo {pages} {191--200} (\bibinfo {year}
  {1991})}\BibitemShut {NoStop}%
\bibitem [{\citenamefont {Galam}(1990)}]{A1990EP96200025}%
  \BibitemOpen
  \bibfield  {author} {\bibinfo {author} {\bibfnamefont {S.}~\bibnamefont
  {Galam}},\ }\bibfield  {title} {\enquote {\bibinfo {title} {Social paradoxes
  of majority-rule voting and renormalization-group},}\ }\href {\doibase
  10.1007/BF01027314} {\bibfield  {journal} {\bibinfo  {journal} {Journal of
  Statistical Physics}\ }\textbf {\bibinfo {volume} {61}},\ \bibinfo {pages}
  {943--951} (\bibinfo {year} {1990})}\BibitemShut {NoStop}%
\bibitem [{\citenamefont {Axelrod}(1997)}]{Axelrode-1997}%
  \BibitemOpen
  \bibfield  {author} {\bibinfo {author} {\bibfnamefont {R.}~\bibnamefont
  {Axelrod}},\ }\bibfield  {title} {\enquote {\bibinfo {title} {The
  dissemination of culture: {A} model with local convergence and global
  polarization},}\ }\href {\doibase 10.1177/0022002797041002001} {\bibfield
  {journal} {\bibinfo  {journal} {Journal of Conflict Resolution}\ }\textbf
  {\bibinfo {volume} {41}},\ \bibinfo {pages} {203--226} (\bibinfo {year}
  {1997})}\BibitemShut {NoStop}%
\bibitem [{\citenamefont {Weimer}\ \emph {et~al.}(2019)\citenamefont {Weimer},
  \citenamefont {Miller}, \citenamefont {Hill},\ and\ \citenamefont
  {Hodson}}]{Weimer2019}%
  \BibitemOpen
  \bibfield  {author} {\bibinfo {author} {\bibfnamefont {C.}~\bibnamefont
  {Weimer}}, \bibinfo {author} {\bibfnamefont {J.~O.}\ \bibnamefont {Miller}},
  \bibinfo {author} {\bibfnamefont {R.}~\bibnamefont {Hill}}, \ and\ \bibinfo
  {author} {\bibfnamefont {D.}~\bibnamefont {Hodson}},\ }\bibfield  {title}
  {\enquote {\bibinfo {title} {Agent scheduling in opinion dynamics: {A}
  taxonomy and comparison using generalized models},}\ }\href {\doibase
  10.18564/jasss.4065} {\bibfield  {journal} {\bibinfo  {journal} {JASSS---the
  Journal of Artificial Societies and Social Simulation}\ }\textbf {\bibinfo
  {volume} {22}},\ \bibinfo {pages} {(4)5} (\bibinfo {year}
  {2019})}\BibitemShut {NoStop}%
\bibitem [{\citenamefont {Krawczyk}\ and\ \citenamefont
  {Ku{\l}akowski}(2013)}]{ISI:000317452800014}%
  \BibitemOpen
  \bibfield  {author} {\bibinfo {author} {\bibfnamefont {M.~J.}\ \bibnamefont
  {Krawczyk}}\ and\ \bibinfo {author} {\bibfnamefont {K.}~\bibnamefont
  {Ku{\l}akowski}},\ }\bibfield  {title} {\enquote {\bibinfo {title} {On a
  combinatorial aspect of fashion},}\ }\href {\doibase
  10.12693/APhysPolA.123.560} {\bibfield  {journal} {\bibinfo  {journal} {Acta
  Physica Polonica A}\ }\textbf {\bibinfo {volume} {123}},\ \bibinfo {pages}
  {560--563} (\bibinfo {year} {2013})}\BibitemShut {NoStop}%
\bibitem [{\citenamefont {Sznajd-Weron}\ and\ \citenamefont
  {Sznajd}(2005)}]{Sznajd-2005a}%
  \BibitemOpen
  \bibfield  {author} {\bibinfo {author} {\bibfnamefont {K.}~\bibnamefont
  {Sznajd-Weron}}\ and\ \bibinfo {author} {\bibfnamefont {J.}~\bibnamefont
  {Sznajd}},\ }\bibfield  {title} {\enquote {\bibinfo {title} {Who is left, who
  is right?}}\ }\href {\doibase 10.1016/j.physa.2004.12.038} {\bibfield
  {journal} {\bibinfo  {journal} {Physica A}\ }\textbf {\bibinfo {volume}
  {351}},\ \bibinfo {pages} {593--604} (\bibinfo {year} {2005})}\BibitemShut
  {NoStop}%
\bibitem [{\citenamefont {S{\^i}rbu}\ \emph {et~al.}(2017)\citenamefont
  {S{\^i}rbu}, \citenamefont {Loreto}, \citenamefont {Servedio},\ and\
  \citenamefont {Tria}}]{Sirbu2017}%
  \BibitemOpen
  \bibfield  {author} {\bibinfo {author} {\bibfnamefont {A.}~\bibnamefont
  {S{\^i}rbu}}, \bibinfo {author} {\bibfnamefont {V.}~\bibnamefont {Loreto}},
  \bibinfo {author} {\bibfnamefont {V.~D.~P.}\ \bibnamefont {Servedio}}, \ and\
  \bibinfo {author} {\bibfnamefont {F.}~\bibnamefont {Tria}},\ }\enquote
  {\bibinfo {title} {Opinion dynamics: {M}odels, extensions and external
  effects},}\ in\ \href {\doibase 10.1007/978-3-319-25658-0_17} {\emph
  {\bibinfo {booktitle} {Participatory Sensing, Opinions and Collective
  Awareness}}},\ \bibinfo {editor} {edited by\ \bibinfo {editor} {\bibfnamefont
  {V.}~\bibnamefont {Loreto}}, \bibinfo {editor} {\bibfnamefont
  {M.}~\bibnamefont {Haklay}}, \bibinfo {editor} {\bibfnamefont
  {A.}~\bibnamefont {Hotho}}, \bibinfo {editor} {\bibfnamefont {V.~D.~P.}\
  \bibnamefont {Servedio}}, \bibinfo {editor} {\bibfnamefont {G.}~\bibnamefont
  {Stumme}}, \bibinfo {editor} {\bibfnamefont {J.}~\bibnamefont {Theunis}}, \
  and\ \bibinfo {editor} {\bibfnamefont {F.}~\bibnamefont {Tria}}}\ (\bibinfo
  {publisher} {Springer International Publishing},\ \bibinfo {address} {Cham},\
  \bibinfo {year} {2017})\ pp.\ \bibinfo {pages} {363--401}\BibitemShut
  {NoStop}%
\bibitem [{\citenamefont {Castellano}\ \emph {et~al.}(2009)\citenamefont
  {Castellano}, \citenamefont {Fortunato},\ and\ \citenamefont
  {Loreto}}]{RevModPhys.81.591}%
  \BibitemOpen
  \bibfield  {author} {\bibinfo {author} {\bibfnamefont {C.}~\bibnamefont
  {Castellano}}, \bibinfo {author} {\bibfnamefont {S.}~\bibnamefont
  {Fortunato}}, \ and\ \bibinfo {author} {\bibfnamefont {V.}~\bibnamefont
  {Loreto}},\ }\bibfield  {title} {\enquote {\bibinfo {title} {Statistical
  physics of social dynamics},}\ }\href {\doibase 10.1103/RevModPhys.81.591}
  {\bibfield  {journal} {\bibinfo  {journal} {Reviews of Modern Physics}\
  }\textbf {\bibinfo {volume} {81}},\ \bibinfo {pages} {591--646} (\bibinfo
  {year} {2009})}\BibitemShut {NoStop}%
\bibitem [{\citenamefont {Stauffer}(2009)}]{Stauffer2009}%
  \BibitemOpen
  \bibfield  {author} {\bibinfo {author} {\bibfnamefont {D.}~\bibnamefont
  {Stauffer}},\ }\enquote {\bibinfo {title} {Opinion dynamics and
  sociophysics},}\ in\ \href {\doibase 10.1007/978-0-387-30440-3_376} {\emph
  {\bibinfo {booktitle} {Encyclopedia of Complexity and Systems Science}}},\
  \bibinfo {editor} {edited by\ \bibinfo {editor} {\bibfnamefont {R.~A.}\
  \bibnamefont {Meyers}}}\ (\bibinfo  {publisher} {Springer},\ \bibinfo
  {address} {New York, NY},\ \bibinfo {year} {2009})\ pp.\ \bibinfo {pages}
  {6380--6388}\BibitemShut {NoStop}%
\bibitem [{\citenamefont {Anderson}\ \emph {et~al.}(1992)\citenamefont
  {Anderson}, \citenamefont {De~Palma},\ and\ \citenamefont
  {Thisse}}]{Anderson1992}%
  \BibitemOpen
  \bibfield  {author} {\bibinfo {author} {\bibfnamefont {S.~P.}\ \bibnamefont
  {Anderson}}, \bibinfo {author} {\bibfnamefont {A.}~\bibnamefont {De~Palma}},
  \ and\ \bibinfo {author} {\bibfnamefont {J.~F.}\ \bibnamefont {Thisse}},\
  }\href@noop {} {\emph {\bibinfo {title} {Discrete Choice Theory of Product
  Differentiation}}}\ (\bibinfo  {publisher} {MIT Press},\ \bibinfo {address}
  {Cambridge, MA},\ \bibinfo {year} {1992})\BibitemShut {NoStop}%
\bibitem [{\citenamefont {Galam}(2008)}]{GalamReview}%
  \BibitemOpen
  \bibfield  {author} {\bibinfo {author} {\bibfnamefont {S.}~\bibnamefont
  {Galam}},\ }\bibfield  {title} {\enquote {\bibinfo {title} {Sociophysics: {A}
  review of {G}alam models},}\ }\href {\doibase 10.1142/S0129183108012297}
  {\bibfield  {journal} {\bibinfo  {journal} {International Journal of Modern
  Physics C}\ }\textbf {\bibinfo {volume} {19}},\ \bibinfo {pages} {409--440}
  (\bibinfo {year} {2008})}\BibitemShut {NoStop}%
\bibitem [{\citenamefont {Malarz}\ and\ \citenamefont
  {Ku{\l}akowski}(2008)}]{Kulakowski2008}%
  \BibitemOpen
  \bibfield  {author} {\bibinfo {author} {\bibfnamefont {K.}~\bibnamefont
  {Malarz}}\ and\ \bibinfo {author} {\bibfnamefont {K.}~\bibnamefont
  {Ku{\l}akowski}},\ }\bibfield  {title} {\enquote {\bibinfo {title} {The
  {S}znajd dynamics on a directed clustered network},}\ }\href {\doibase
  10.12693/APhysPolA.114.581} {\bibfield  {journal} {\bibinfo  {journal} {Acta
  Physica Polonica A}\ }\textbf {\bibinfo {volume} {114}},\ \bibinfo {pages}
  {581--588} (\bibinfo {year} {2008})}\BibitemShut {NoStop}%
\bibitem [{\citenamefont {Slanina}\ \emph {et~al.}(2008)\citenamefont
  {Slanina}, \citenamefont {Sznajd-Weron},\ and\ \citenamefont
  {Przyby{\l}a}}]{Slanina-2008}%
  \BibitemOpen
  \bibfield  {author} {\bibinfo {author} {\bibfnamefont {F.}~\bibnamefont
  {Slanina}}, \bibinfo {author} {\bibfnamefont {K.}~\bibnamefont
  {Sznajd-Weron}}, \ and\ \bibinfo {author} {\bibfnamefont {P.}~\bibnamefont
  {Przyby{\l}a}},\ }\bibfield  {title} {\enquote {\bibinfo {title} {Some new
  results on one-dimensional outflow dynamics},}\ }\href {\doibase
  10.1209/0295-5075/82/18006} {\bibfield  {journal} {\bibinfo  {journal} {EPL}\
  }\textbf {\bibinfo {volume} {82}},\ \bibinfo {pages} {18006} (\bibinfo {year}
  {2008})}\BibitemShut {NoStop}%
\bibitem [{\citenamefont {Sznajd-Weron}(2005)}]{Sznajd-2005}%
  \BibitemOpen
  \bibfield  {author} {\bibinfo {author} {\bibfnamefont {K.}~\bibnamefont
  {Sznajd-Weron}},\ }\bibfield  {title} {\enquote {\bibinfo {title} {Sznajd
  model and its applications},}\ }\href
  {http://www.actaphys.uj.edu.pl/fulltext?series=Reg&vol=36&page=2537}
  {\bibfield  {journal} {\bibinfo  {journal} {Acta Physica Polonica B}\
  }\textbf {\bibinfo {volume} {36}},\ \bibinfo {pages} {2537--2547} (\bibinfo
  {year} {2005})}\BibitemShut {NoStop}%
\bibitem [{\citenamefont {Sznajd-Weron}\ and\ \citenamefont
  {Sznajd}(2000)}]{Sznajd-2000}%
  \BibitemOpen
  \bibfield  {author} {\bibinfo {author} {\bibfnamefont {K.}~\bibnamefont
  {Sznajd-Weron}}\ and\ \bibinfo {author} {\bibfnamefont {J.}~\bibnamefont
  {Sznajd}},\ }\bibfield  {title} {\enquote {\bibinfo {title} {Opinion
  evolution in closed community},}\ }\href {\doibase 10.1142/S0129183100000936}
  {\bibfield  {journal} {\bibinfo  {journal} {International Journal of Modern
  Physics C}\ }\textbf {\bibinfo {volume} {11}},\ \bibinfo {pages} {1157--1165}
  (\bibinfo {year} {2000})}\BibitemShut {NoStop}%
\bibitem [{\citenamefont {Gargiulo}\ and\ \citenamefont
  {Gandica}(2017)}]{Gargiulo-2017}%
  \BibitemOpen
  \bibfield  {author} {\bibinfo {author} {\bibfnamefont {F.}~\bibnamefont
  {Gargiulo}}\ and\ \bibinfo {author} {\bibfnamefont {Y.}~\bibnamefont
  {Gandica}},\ }\bibfield  {title} {\enquote {\bibinfo {title} {The role of
  homophily in the emergence of opinion controversies},}\ }\href {\doibase
  10.18564/jasss.3448} {\bibfield  {journal} {\bibinfo  {journal} {JASSS---the
  Journal of Artificial Societies and Social Simulation}\ }\textbf {\bibinfo
  {volume} {20}},\ \bibinfo {pages} {(3)8} (\bibinfo {year}
  {2017})}\BibitemShut {NoStop}%
\bibitem [{\citenamefont {Mathias}\ \emph {et~al.}(2016)\citenamefont
  {Mathias}, \citenamefont {Huet},\ and\ \citenamefont
  {Deffuant}}]{Mathias-2016}%
  \BibitemOpen
  \bibfield  {author} {\bibinfo {author} {\bibfnamefont {J.-D.}\ \bibnamefont
  {Mathias}}, \bibinfo {author} {\bibfnamefont {S.}~\bibnamefont {Huet}}, \
  and\ \bibinfo {author} {\bibfnamefont {G.}~\bibnamefont {Deffuant}},\
  }\bibfield  {title} {\enquote {\bibinfo {title} {Bounded confidence model
  with fixed uncertainties and extremists: The opinions can keep fluctuating
  indefinitely},}\ }\href {\doibase 10.18564/jasss.2967} {\bibfield  {journal}
  {\bibinfo  {journal} {JASSS---the Journal of Artificial Societies and Social
  Simulation}\ }\textbf {\bibinfo {volume} {19}},\ \bibinfo {pages} {(1)6}
  (\bibinfo {year} {2016})}\BibitemShut {NoStop}%
\bibitem [{\citenamefont {Malarz}\ and\ \citenamefont
  {Ku{\l}akowski}(2014)}]{Kulakowski2014}%
  \BibitemOpen
  \bibfield  {author} {\bibinfo {author} {\bibfnamefont {K.}~\bibnamefont
  {Malarz}}\ and\ \bibinfo {author} {\bibfnamefont {K.}~\bibnamefont
  {Ku{\l}akowski}},\ }\bibfield  {title} {\enquote {\bibinfo {title} {Mental
  ability and common sense in an artificial society},}\ }\href {\doibase
  10.1051/epn/2014402} {\bibfield  {journal} {\bibinfo  {journal} {Europhysics
  News}\ }\textbf {\bibinfo {volume} {45}},\ \bibinfo {pages} {21--23}
  (\bibinfo {year} {2014})}\BibitemShut {NoStop}%
\bibitem [{\citenamefont {Malarz}\ and\ \citenamefont
  {Ku{\l}akowski}(2012)}]{Kulakowski2012}%
  \BibitemOpen
  \bibfield  {author} {\bibinfo {author} {\bibfnamefont {K.}~\bibnamefont
  {Malarz}}\ and\ \bibinfo {author} {\bibfnamefont {K.}~\bibnamefont
  {Ku{\l}akowski}},\ }\bibfield  {title} {\enquote {\bibinfo {title} {Bounded
  confidence model: addressed information maintain diversity of opinions},}\
  }\href {\doibase 10.12693/APhysPolA.121.B-86} {\bibfield  {journal} {\bibinfo
   {journal} {Acta Physica Polonica A}\ }\textbf {\bibinfo {volume} {121}},\
  \bibinfo {pages} {B86--B88} (\bibinfo {year} {2012})}\BibitemShut {NoStop}%
\bibitem [{\citenamefont {Malarz}\ \emph {et~al.}(2011)\citenamefont {Malarz},
  \citenamefont {Gronek},\ and\ \citenamefont {Ku{\l}akowski}}]{Gronek2011}%
  \BibitemOpen
  \bibfield  {author} {\bibinfo {author} {\bibfnamefont {K.}~\bibnamefont
  {Malarz}}, \bibinfo {author} {\bibfnamefont {P.}~\bibnamefont {Gronek}}, \
  and\ \bibinfo {author} {\bibfnamefont {K.}~\bibnamefont {Ku{\l}akowski}},\
  }\bibfield  {title} {\enquote {\bibinfo {title} {{Z}aller--{D}effuant model
  of mass opinion},}\ }\href {\doibase 10.18564/jasss.1719} {\bibfield
  {journal} {\bibinfo  {journal} {JASSS---the Journal of Artificial Societies
  and Social Simulation}\ }\textbf {\bibinfo {volume} {14}},\ \bibinfo {pages}
  {(1)2} (\bibinfo {year} {2011})}\BibitemShut {NoStop}%
\bibitem [{\citenamefont {Ku{\l}akowski}(2009)}]{Kulakowski2009469}%
  \BibitemOpen
  \bibfield  {author} {\bibinfo {author} {\bibfnamefont {K.}~\bibnamefont
  {Ku{\l}akowski}},\ }\bibfield  {title} {\enquote {\bibinfo {title} {Opinion
  polarization in the receipt--accept--sample model},}\ }\href {\doibase
  10.1016/j.physa.2008.10.037} {\bibfield  {journal} {\bibinfo  {journal}
  {Physica A}\ }\textbf {\bibinfo {volume} {388}},\ \bibinfo {pages} {469--476}
  (\bibinfo {year} {2009})}\BibitemShut {NoStop}%
\bibitem [{\citenamefont {Deffuant}(2006)}]{Deffuant-2006}%
  \BibitemOpen
  \bibfield  {author} {\bibinfo {author} {\bibfnamefont {G.}~\bibnamefont
  {Deffuant}},\ }\bibfield  {title} {\enquote {\bibinfo {title} {Comparing
  extremism propagation patterns in continuous opinion models},}\ }\href
  {http://jasss.soc.surrey.ac.uk/9/3/8.html} {\bibfield  {journal} {\bibinfo
  {journal} {JASSS---the Journal of Artificial Societies and Social
  Simulation}\ }\textbf {\bibinfo {volume} {9}},\ \bibinfo {pages} {(3)8}
  (\bibinfo {year} {2006})}\BibitemShut {NoStop}%
\bibitem [{\citenamefont {Hegselmann}\ and\ \citenamefont
  {Krause}(2002)}]{Hegselmann-2002}%
  \BibitemOpen
  \bibfield  {author} {\bibinfo {author} {\bibfnamefont {R.}~\bibnamefont
  {Hegselmann}}\ and\ \bibinfo {author} {\bibfnamefont {U.}~\bibnamefont
  {Krause}},\ }\bibfield  {title} {\enquote {\bibinfo {title} {Opinion dynamics
  and bounded confidence: {M}odels, analysis and simulation},}\ }\href
  {http://jasss.soc.surrey.ac.uk/5/3/2.html} {\bibfield  {journal} {\bibinfo
  {journal} {JASSS---the Journal of Artificial Societies and Social
  Simulation}\ }\textbf {\bibinfo {volume} {5}},\ \bibinfo {pages} {(3)2}
  (\bibinfo {year} {2002})}\BibitemShut {NoStop}%
\bibitem [{\citenamefont {Deffuant}\ \emph {et~al.}(2000)\citenamefont
  {Deffuant}, \citenamefont {Neau}, \citenamefont {Amblard},\ and\
  \citenamefont {Weisbuch}}]{Deffuant-2000}%
  \BibitemOpen
  \bibfield  {author} {\bibinfo {author} {\bibfnamefont {G.}~\bibnamefont
  {Deffuant}}, \bibinfo {author} {\bibfnamefont {D.}~\bibnamefont {Neau}},
  \bibinfo {author} {\bibfnamefont {F.}~\bibnamefont {Amblard}}, \ and\
  \bibinfo {author} {\bibfnamefont {G.}~\bibnamefont {Weisbuch}},\ }\bibfield
  {title} {\enquote {\bibinfo {title} {Mixing beliefs among interacting
  agents},}\ }\href {\doibase 10.1142/S0219525900000078} {\bibfield  {journal}
  {\bibinfo  {journal} {Advances in Complex Systems}\ }\textbf {\bibinfo
  {volume} {3}},\ \bibinfo {pages} {87} (\bibinfo {year} {2000})}\BibitemShut
  {NoStop}%
\bibitem [{\citenamefont {Lima}(2017)}]{000411951900001}%
  \BibitemOpen
  \bibfield  {author} {\bibinfo {author} {\bibfnamefont {F.~W.~S.}\
  \bibnamefont {Lima}},\ }\bibfield  {title} {\enquote {\bibinfo {title}
  {Kinetic continuous opinion dynamics model on two types of {A}rchimedean
  lattices},}\ }\href {\doibase 10.3389/fphy.2017.00047} {\bibfield  {journal}
  {\bibinfo  {journal} {Frontiers in Physics}\ }\textbf {\bibinfo {volume}
  {5}},\ \bibinfo {pages} {47} (\bibinfo {year} {2017})}\BibitemShut {NoStop}%
\bibitem [{\citenamefont {Malarz}(2006)}]{Malarz2006b}%
  \BibitemOpen
  \bibfield  {author} {\bibinfo {author} {\bibfnamefont {K.}~\bibnamefont
  {Malarz}},\ }\bibfield  {title} {\enquote {\bibinfo {title} {Truth seekers in
  opinion dynamics models},}\ }\href {\doibase 10.1142/S0129183106009850}
  {\bibfield  {journal} {\bibinfo  {journal} {International Journal of Modern
  Physics C}\ }\textbf {\bibinfo {volume} {17}},\ \bibinfo {pages} {1521--1524}
  (\bibinfo {year} {2006})}\BibitemShut {NoStop}%
\bibitem [{\citenamefont {Baccelli}\ \emph {et~al.}(2017)\citenamefont
  {Baccelli}, \citenamefont {Chatterjee},\ and\ \citenamefont
  {Vishwanath}}]{000413837700014}%
  \BibitemOpen
  \bibfield  {author} {\bibinfo {author} {\bibfnamefont {F.}~\bibnamefont
  {Baccelli}}, \bibinfo {author} {\bibfnamefont {A.}~\bibnamefont
  {Chatterjee}}, \ and\ \bibinfo {author} {\bibfnamefont {S.}~\bibnamefont
  {Vishwanath}},\ }\bibfield  {title} {\enquote {\bibinfo {title} {Pairwise
  stochastic bounded confidence opinion dynamics: Heavy tails and stability},}\
  }\href {\doibase 10.1109/TAC.2017.2691312} {\bibfield  {journal} {\bibinfo
  {journal} {IEEE Transactions on Automatic Control}\ }\textbf {\bibinfo
  {volume} {62}},\ \bibinfo {pages} {5678--5693} (\bibinfo {year}
  {2017})}\BibitemShut {NoStop}%
\bibitem [{\citenamefont {Su}\ \emph {et~al.}(2017)\citenamefont {Su},
  \citenamefont {Chen},\ and\ \citenamefont {Hong}}]{000414818100049}%
  \BibitemOpen
  \bibfield  {author} {\bibinfo {author} {\bibfnamefont {W.}~\bibnamefont
  {Su}}, \bibinfo {author} {\bibfnamefont {G.}~\bibnamefont {Chen}}, \ and\
  \bibinfo {author} {\bibfnamefont {Y.}~\bibnamefont {Hong}},\ }\bibfield
  {title} {\enquote {\bibinfo {title} {Noise leads to quasi-consensus of
  {H}egselmann--{K}rause opinion dynamics},}\ }\href {\doibase
  10.1016/j.automatica.2017.08.008} {\bibfield  {journal} {\bibinfo  {journal}
  {Automatica}\ }\textbf {\bibinfo {volume} {85}},\ \bibinfo {pages} {448--454}
  (\bibinfo {year} {2017})}\BibitemShut {NoStop}%
\bibitem [{\citenamefont {Zhu}\ \emph {et~al.}(2017)\citenamefont {Zhu},
  \citenamefont {Wang}, \citenamefont {Li},\ and\ \citenamefont
  {Cai}}]{000409101900001}%
  \BibitemOpen
  \bibfield  {author} {\bibinfo {author} {\bibfnamefont {Y.}~\bibnamefont
  {Zhu}}, \bibinfo {author} {\bibfnamefont {Q.~A.}\ \bibnamefont {Wang}},
  \bibinfo {author} {\bibfnamefont {W.}~\bibnamefont {Li}}, \ and\ \bibinfo
  {author} {\bibfnamefont {X.}~\bibnamefont {Cai}},\ }\bibfield  {title}
  {\enquote {\bibinfo {title} {The formation of continuous opinion dynamics
  based on a gambling mechanism and its sensitivity analysis},}\ }\href
  {\doibase 10.1088/1742-5468/aa7df1} {\bibfield  {journal} {\bibinfo
  {journal} {Journal of Statistical Mechanics---Theory and Experiment}\
  }\textbf {\bibinfo {volume} {2017}},\ \bibinfo {pages} {093401} (\bibinfo
  {year} {2017})}\BibitemShut {NoStop}%
\bibitem [{\citenamefont {Anteneodo}\ and\ \citenamefont
  {Crokidakis}(2017)}]{000399951700004}%
  \BibitemOpen
  \bibfield  {author} {\bibinfo {author} {\bibfnamefont {C.}~\bibnamefont
  {Anteneodo}}\ and\ \bibinfo {author} {\bibfnamefont {N.}~\bibnamefont
  {Crokidakis}},\ }\bibfield  {title} {\enquote {\bibinfo {title} {Symmetry
  breaking by heating in a continuous opinion model},}\ }\href {\doibase
  10.1103/PhysRevE.95.042308} {\bibfield  {journal} {\bibinfo  {journal}
  {Physical Review E}\ }\textbf {\bibinfo {volume} {95}},\ \bibinfo {pages}
  {042308} (\bibinfo {year} {2017})}\BibitemShut {NoStop}%
\bibitem [{\citenamefont {Chen}\ \emph {et~al.}(2017)\citenamefont {Chen},
  \citenamefont {Cheng}, \citenamefont {Huang}, \citenamefont {Han},
  \citenamefont {Dai}, \citenamefont {Li},\ and\ \citenamefont
  {Yang}}]{000399951000005}%
  \BibitemOpen
  \bibfield  {author} {\bibinfo {author} {\bibfnamefont {G.}~\bibnamefont
  {Chen}}, \bibinfo {author} {\bibfnamefont {H.}~\bibnamefont {Cheng}},
  \bibinfo {author} {\bibfnamefont {C.}~\bibnamefont {Huang}}, \bibinfo
  {author} {\bibfnamefont {W.}~\bibnamefont {Han}}, \bibinfo {author}
  {\bibfnamefont {Q.}~\bibnamefont {Dai}}, \bibinfo {author} {\bibfnamefont
  {H.}~\bibnamefont {Li}}, \ and\ \bibinfo {author} {\bibfnamefont
  {J.}~\bibnamefont {Yang}},\ }\bibfield  {title} {\enquote {\bibinfo {title}
  {Deffuant model on a ring with repelling mechanism and circular opinions},}\
  }\href {\doibase 10.1103/PhysRevE.95.042118} {\bibfield  {journal} {\bibinfo
  {journal} {Physical Review E}\ }\textbf {\bibinfo {volume} {95}},\ \bibinfo
  {pages} {042118} (\bibinfo {year} {2017})}\BibitemShut {NoStop}%
\bibitem [{\citenamefont {Zhang}\ \emph {et~al.}(2017)\citenamefont {Zhang},
  \citenamefont {Liu},\ and\ \citenamefont {Zhang}}]{000396054300030}%
  \BibitemOpen
  \bibfield  {author} {\bibinfo {author} {\bibfnamefont {Y.}~\bibnamefont
  {Zhang}}, \bibinfo {author} {\bibfnamefont {Q.}~\bibnamefont {Liu}}, \ and\
  \bibinfo {author} {\bibfnamefont {S.}~\bibnamefont {Zhang}},\ }\bibfield
  {title} {\enquote {\bibinfo {title} {Opinion formation with time-varying
  bounded confidence},}\ }\href {\doibase 10.1371/journal.pone.0172982}
  {\bibfield  {journal} {\bibinfo  {journal} {PLoS ONE}\ }\textbf {\bibinfo
  {volume} {12}},\ \bibinfo {pages} {e0172982} (\bibinfo {year}
  {2017})}\BibitemShut {NoStop}%
\bibitem [{\citenamefont {De~Sanctis}\ and\ \citenamefont
  {Galla}(2009)}]{PhysRevE.79.046108}%
  \BibitemOpen
  \bibfield  {author} {\bibinfo {author} {\bibfnamefont {Luca}\ \bibnamefont
  {De~Sanctis}}\ and\ \bibinfo {author} {\bibfnamefont {Tobias}\ \bibnamefont
  {Galla}},\ }\bibfield  {title} {\enquote {\bibinfo {title} {Effects of noise
  and confidence thresholds in nominal and metric axelrod dynamics of social
  influence},}\ }\href {\doibase 10.1103/PhysRevE.79.046108} {\bibfield
  {journal} {\bibinfo  {journal} {Physical Review E}\ }\textbf {\bibinfo
  {volume} {79}},\ \bibinfo {pages} {046108} (\bibinfo {year}
  {2009})}\BibitemShut {NoStop}%
\bibitem [{\citenamefont {Ren}\ \emph {et~al.}(2007)\citenamefont {Ren},
  \citenamefont {Wang},\ and\ \citenamefont {Qi}}]{PhysRevE.75.045101}%
  \BibitemOpen
  \bibfield  {author} {\bibinfo {author} {\bibfnamefont {Jie}\ \bibnamefont
  {Ren}}, \bibinfo {author} {\bibfnamefont {Wen-Xu}\ \bibnamefont {Wang}}, \
  and\ \bibinfo {author} {\bibfnamefont {Feng}\ \bibnamefont {Qi}},\ }\bibfield
   {title} {\enquote {\bibinfo {title} {Randomness enhances cooperation: A
  resonance-type phenomenon in evolutionary games},}\ }\href {\doibase
  10.1103/PhysRevE.75.045101} {\bibfield  {journal} {\bibinfo  {journal}
  {Physical Review E}\ }\textbf {\bibinfo {volume} {75}},\ \bibinfo {pages}
  {045101} (\bibinfo {year} {2007})}\BibitemShut {NoStop}%
\bibitem [{\citenamefont {Biondo}\ \emph {et~al.}(2013)\citenamefont {Biondo},
  \citenamefont {Pluchino},\ and\ \citenamefont {Rapisarda}}]{Biondo2013}%
  \BibitemOpen
  \bibfield  {author} {\bibinfo {author} {\bibfnamefont {Alessio~Emanuele}\
  \bibnamefont {Biondo}}, \bibinfo {author} {\bibfnamefont {Alessandro}\
  \bibnamefont {Pluchino}}, \ and\ \bibinfo {author} {\bibfnamefont {Andrea}\
  \bibnamefont {Rapisarda}},\ }\bibfield  {title} {\enquote {\bibinfo {title}
  {The beneficial role of random strategies in social and financial systems},}\
  }\href {\doibase 10.1007/s10955-013-0691-2} {\bibfield  {journal} {\bibinfo
  {journal} {Journal of Statistical Physics}\ }\textbf {\bibinfo {volume}
  {151}},\ \bibinfo {pages} {607--622} (\bibinfo {year} {2013})}\BibitemShut
  {NoStop}%
\bibitem [{\citenamefont {Shirado}\ and\ \citenamefont
  {Christakis}(2017)}]{Shirado2017}%
  \BibitemOpen
  \bibfield  {author} {\bibinfo {author} {\bibfnamefont {Hirokazu}\
  \bibnamefont {Shirado}}\ and\ \bibinfo {author} {\bibfnamefont {Nicholas~A.}\
  \bibnamefont {Christakis}},\ }\bibfield  {title} {\enquote {\bibinfo {title}
  {Locally noisy autonomous agents improve global human coordination in network
  experiments},}\ }\href {\doibase 10.1038/nature22332} {\bibfield  {journal}
  {\bibinfo  {journal} {Nature}\ }\textbf {\bibinfo {volume} {545}},\ \bibinfo
  {pages} {370--374} (\bibinfo {year} {2017})}\BibitemShut {NoStop}%
\bibitem [{\citenamefont {Ho{\l}yst}\ \emph {et~al.}(2011)\citenamefont
  {Ho{\l}yst}, \citenamefont {Kacperski},\ and\ \citenamefont
  {Schweitzer}}]{ARCPIX253}%
  \BibitemOpen
  \bibfield  {author} {\bibinfo {author} {\bibfnamefont {J.~A.}\ \bibnamefont
  {Ho{\l}yst}}, \bibinfo {author} {\bibfnamefont {K.}~\bibnamefont
  {Kacperski}}, \ and\ \bibinfo {author} {\bibfnamefont {F.}~\bibnamefont
  {Schweitzer}},\ }\enquote {\bibinfo {title} {Social impact models of opinion
  dynamics},}\ in\ \href {\doibase 10.1142/9789812811578_0005} {\emph {\bibinfo
  {booktitle} {Annual Reviews of Computational Physics IX}}},\ \bibinfo
  {editor} {edited by\ \bibinfo {editor} {\bibfnamefont {D.}~\bibnamefont
  {Stauffer}}}\ (\bibinfo  {publisher} {World Scientific},\ \bibinfo {address}
  {Singapore},\ \bibinfo {year} {2011})\ pp.\ \bibinfo {pages}
  {253--273}\BibitemShut {NoStop}%
\bibitem [{\citenamefont {Gehrke}(1996)}]{Fortran95}%
  \BibitemOpen
  \bibfield  {author} {\bibinfo {author} {\bibfnamefont {W.}~\bibnamefont
  {Gehrke}},\ }\href {\doibase 10.1007/978-1-4471-1025-5} {\emph {\bibinfo
  {title} {Fortran 95 Language Guide}}}\ (\bibinfo  {publisher}
  {Springer-Verlag},\ \bibinfo {address} {London},\ \bibinfo {year}
  {1996})\BibitemShut {NoStop}%
\bibitem [{\citenamefont
  {Ba\'ncerowski}(2017{\natexlab{b}})}]{Bancerowski-app}%
  \BibitemOpen
  \bibfield  {author} {\bibinfo {author} {\bibfnamefont {P.}~\bibnamefont
  {Ba\'ncerowski}},\ }\href
  {http://www.zis.agh.edu.pl/app/MSc/Przemyslaw_Bancerowski/} {\enquote
  {\bibinfo {title} {{www.zis.agh.edu.pl/app/MSc/Przemyslaw\_Bancerowski/}},}\
  } (\bibinfo {year} {2017}{\natexlab{b}})\BibitemShut {NoStop}%
\bibitem [{\citenamefont {Hoshen}\ and\ \citenamefont
  {Kopelman}(1976)}]{Hoshen1976a}%
  \BibitemOpen
  \bibfield  {author} {\bibinfo {author} {\bibfnamefont {J.}~\bibnamefont
  {Hoshen}}\ and\ \bibinfo {author} {\bibfnamefont {R.}~\bibnamefont
  {Kopelman}},\ }\bibfield  {title} {\enquote {\bibinfo {title} {Percolation
  and cluster distribution. 1. {C}luster multiple labeling technique and
  critical concentration algorithm},}\ }\href {\doibase
  10.1103/PhysRevB.14.3438} {\bibfield  {journal} {\bibinfo  {journal}
  {Physical Review B}\ }\textbf {\bibinfo {volume} {14}},\ \bibinfo {pages}
  {3438--3445} (\bibinfo {year} {1976})}\BibitemShut {NoStop}%
\bibitem [{\citenamefont {Kotwica}\ \emph {et~al.}(2019)\citenamefont
  {Kotwica}, \citenamefont {Gronek},\ and\ \citenamefont
  {Malarz}}]{1803.09504}%
  \BibitemOpen
  \bibfield  {author} {\bibinfo {author} {\bibfnamefont {M.}~\bibnamefont
  {Kotwica}}, \bibinfo {author} {\bibfnamefont {P.}~\bibnamefont {Gronek}}, \
  and\ \bibinfo {author} {\bibfnamefont {K.}~\bibnamefont {Malarz}},\
  }\bibfield  {title} {\enquote {\bibinfo {title} {Efficient space
  virtualisation for {H}oshen--{K}opelman algorithm},}\ }\href {\doibase
  10.1142/S0129183119500554} {\bibfield  {journal} {\bibinfo  {journal}
  {International Journal of Modern Physics C}\ }\textbf {\bibinfo {volume}
  {30}},\ \bibinfo {pages} {1950055} (\bibinfo {year} {2019})}\BibitemShut
  {NoStop}%
\bibitem [{\citenamefont {Malarz}(2015)}]{Malarz2015}%
  \BibitemOpen
  \bibfield  {author} {\bibinfo {author} {\bibfnamefont {K.}~\bibnamefont
  {Malarz}},\ }\bibfield  {title} {\enquote {\bibinfo {title} {Simple cubic
  random-site percolation thresholds for neighborhoods containing
  fourth-nearest neighbors},}\ }\href {\doibase 10.1103/PhysRevE.91.043301}
  {\bibfield  {journal} {\bibinfo  {journal} {Physical Review E}\ }\textbf
  {\bibinfo {volume} {91}},\ \bibinfo {pages} {043301} (\bibinfo {year}
  {2015})}\BibitemShut {NoStop}%
\bibitem [{\citenamefont {Kurzawski}\ and\ \citenamefont
  {Malarz}(2012)}]{Kurzawski2012}%
  \BibitemOpen
  \bibfield  {author} {\bibinfo {author} {\bibfnamefont {{\L}.}~\bibnamefont
  {Kurzawski}}\ and\ \bibinfo {author} {\bibfnamefont {K.}~\bibnamefont
  {Malarz}},\ }\bibfield  {title} {\enquote {\bibinfo {title} {Simple cubic
  random-site percolation thresholds for complex neighbourhoods},}\ }\href
  {\doibase 10.1016/S0034-4877(12)60036-6} {\bibfield  {journal} {\bibinfo
  {journal} {Reports on Mathematical Physics}\ }\textbf {\bibinfo {volume}
  {70}},\ \bibinfo {pages} {163--169} (\bibinfo {year} {2012})}\BibitemShut
  {NoStop}%
\bibitem [{\citenamefont {Majewski}\ and\ \citenamefont
  {Malarz}(2007)}]{Majewski2007}%
  \BibitemOpen
  \bibfield  {author} {\bibinfo {author} {\bibfnamefont {M.}~\bibnamefont
  {Majewski}}\ and\ \bibinfo {author} {\bibfnamefont {K.}~\bibnamefont
  {Malarz}},\ }\bibfield  {title} {\enquote {\bibinfo {title} {Square lattice
  site percolation thresholds for complex neighbourhoods},}\ }\href
  {http://www.actaphys.uj.edu.pl/fulltext?series=Reg&vol=38&page=2191}
  {\bibfield  {journal} {\bibinfo  {journal} {Acta Physica Polonica B}\
  }\textbf {\bibinfo {volume} {38}},\ \bibinfo {pages} {2191--2199} (\bibinfo
  {year} {2007})}\BibitemShut {NoStop}%
\bibitem [{\citenamefont {Malarz}\ and\ \citenamefont
  {Galam}(2005)}]{Galam2005a}%
  \BibitemOpen
  \bibfield  {author} {\bibinfo {author} {\bibfnamefont {K.}~\bibnamefont
  {Malarz}}\ and\ \bibinfo {author} {\bibfnamefont {S.}~\bibnamefont {Galam}},\
  }\bibfield  {title} {\enquote {\bibinfo {title} {Square-lattice site
  percolation at increasing ranges of neighbor bonds},}\ }\href {\doibase
  10.1103/PhysRevE.71.016125} {\bibfield  {journal} {\bibinfo  {journal}
  {Physical Review E}\ }\textbf {\bibinfo {volume} {71}},\ \bibinfo {pages}
  {016125} (\bibinfo {year} {2005})}\BibitemShut {NoStop}%
\bibitem [{\citenamefont {Carro}\ \emph {et~al.}(2016)\citenamefont {Carro},
  \citenamefont {Toral},\ and\ \citenamefont {San~Miguel}}]{Carro-2016}%
  \BibitemOpen
  \bibfield  {author} {\bibinfo {author} {\bibfnamefont {A.}~\bibnamefont
  {Carro}}, \bibinfo {author} {\bibfnamefont {R.}~\bibnamefont {Toral}}, \ and\
  \bibinfo {author} {\bibfnamefont {M.}~\bibnamefont {San~Miguel}},\ }\bibfield
   {title} {\enquote {\bibinfo {title} {The noisy voter model on complex
  networks},}\ }\href {\doibase 10.1038/srep24775} {\bibfield  {journal}
  {\bibinfo  {journal} {Scientific Reports}\ }\textbf {\bibinfo {volume} {6}},\
  \bibinfo {pages} {24775} (\bibinfo {year} {2016})}\BibitemShut {NoStop}%
\bibitem [{\citenamefont {Galam}(2004)}]{Galam-2004}%
  \BibitemOpen
  \bibfield  {author} {\bibinfo {author} {\bibfnamefont {S.}~\bibnamefont
  {Galam}},\ }\bibfield  {title} {\enquote {\bibinfo {title} {Contrarian
  deterministic effects on opinion dynamics: `{T}he hung elections
  scenario'},}\ }\href {\doibase 10.1016/j.physa.2003.10.041} {\bibfield
  {journal} {\bibinfo  {journal} {Physica A}\ }\textbf {\bibinfo {volume}
  {333}},\ \bibinfo {pages} {453--460} (\bibinfo {year} {2004})}\BibitemShut
  {NoStop}%
\end{thebibliography}%

\appendix

%% ###############################################################
\section{\label{A:small}Example of small system evolution ($L=3$, $K=3$)}
%% ###############################################################

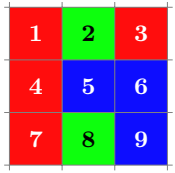
\begin{figure}[!htbp]
\centering
\begin{tikzpicture}[scale=.7]
\tkzDefPoint(0,0){A1}
\tkzDefPoint(1,0){A2}
\tkzDefPoint(1,3){A3}
\tkzDefPoint(0,3){A4}
\tkzFillPolygon[red,opacity=0.95](A1,A2,A3,A4)
\tkzDefPoint(2,2){B1}
\tkzDefPoint(3,2){B2}
\tkzDefPoint(3,3){B3}
\tkzDefPoint(2,3){B4}
\tkzFillPolygon[red,opacity=0.95](B1,B2,B3,B4)
\tkzDefPoint(1,2){C1}
\tkzFillPolygon[green,opacity=0.95](C1,B1,B4,A3)
\tkzDefPoint(2,0){D1}
\tkzDefPoint(2,1){D2}
\tkzDefPoint(1,1){D3}
\tkzFillPolygon[green,opacity=0.95](D1,D2,D3,A2)
\tkzDefPoint(3,0){E1}
\tkzFillPolygon[blue,opacity=0.95](D1,E1,B2,C1,D3,D2)
\node[] at (0.5,2.5) {\color{white} {\bf 1}};
\node[] at (1.5,2.5) {\bf 2};
\node[] at (2.5,2.5) {\color{white} {\bf 3}};
\node[] at (0.5,1.5) {\color{white} {\bf 4}};
\node[] at (1.5,1.5) {\color{white} {\bf 5}};
\node[] at (2.5,1.5) {\color{white} {\bf 6}};
\node[] at (0.5,0.5) {\color{white} {\bf 7}};
\node[] at (1.5,0.5) {\bf 8};
\node[] at (2.5,0.5) {\color{white} {\bf 9}};
\draw[step=10mm,gray,very thin] (-0.1,-0.1) grid (3.1,3.1);
\end{tikzpicture}
\caption{\label{fig:small} (Colour online) Example of small lattice with nine actors and three opinions. The numbers are actors labels $i$. The colours correspond to various actors opinions (`red'---$\Xi_1$, `blue'---$\Xi_2$ and `green'---$\Xi_3$).}
\end{figure}

To better explain the model rules we calculate social impact on single actor for case of small lattice ($L=3$).
We assume $K=3$ opinions available in the system marked as `red' ($\Xi_1$), `blue' ($\Xi_2$) and `green' ($\Xi_3$).
We will calculate the impact exerting by nine actors on the actors labelled as `5' and `9' in Fig.~\ref{fig:small}.
We assume the supportiveness $s_i=i/10$ and persuasiveness $p_i=1-i/10$.

According to Eq.~\eqref{eq:wplyw} to evaluate the opinion $\xi_5(t+1)$ in the next time step we have to calculated $K=3$ impacts exerted on actor $i=5$ for three opinions available in the system.

As $\xi_5(t)=\Xi_2$ (`blue') we use Eq.~\eqref{eq:wplyw_ta_sama} to calculate impact
%% ---------------------------------------------------------------
\begin{equation}
I_{5,\text{blue}}(t)=4\mathcal{J}_s \left(\dfrac{q(s_5)}{g(d_{5,5})} + \dfrac{q(s_6)}{g(d_{5,6})} + \dfrac{q(s_9)}{g(d_{5,9})} \right),
\end{equation}
%% ---------------------------------------------------------------
from all actors with `blue' opinions (i.e. for $i=6,9$), including actor $i=5$ himself/herself. The impacts from actors with `red' and `green' opinions are calculated basing on Eq.~\eqref{eq:wplyw_inna}:
%% ---------------------------------------------------------------
\begin{equation}
I_{5,\text{red}}(t) = 4\mathcal{J}_p\left(
	\dfrac{q(p_1)}{g(d_{5,1})}
       +\dfrac{q(p_3)}{g(d_{5,3})}
       +\dfrac{q(p_4)}{g(d_{5,4})}
       +\dfrac{q(p_7)}{g(d_{5,7})}
	\right),
\end{equation}
%% ---------------------------------------------------------------
\begin{equation}
I_{5,\text{green}}(t) = 4\mathcal{J}_p\left(
	\dfrac{q(p_2)}{g(d_{5,2})}
       +\dfrac{q(p_8)}{g(d_{5,8})}
	\right),
\end{equation}
%% ---------------------------------------------------------------

%% ---------------------------------------------------------------
%% ---------------------------------------------------------------

We assume identity function for scaling functions $\mathcal{J}_S(x)\equiv x$, $\mathcal{J}_P(x)\equiv x$, $q(x)\equiv x$ and the distance scaling function $g(x)=1+x^\alpha$, with $\alpha=2$.
These assumptions yield
%% ---------------------------------------------------------------
\begin{equation}
I_{5,\text{blue}}(t)= 4\left(
	\dfrac{s_5}{1+d_{5,5}^2} 
       +\dfrac{s_6}{1+d_{5,6}^2} 
       +\dfrac{s_9}{1+d_{5,9}^2}
	\right)
	= 4\left(
        \dfrac{0.5}{1+0^2}
       +\dfrac{0.6}{1+1^2}
	+\dfrac{0.9}{1+\sqrt{2}^2}
        \right)= 4.4,
\end{equation}
%% ---------------------------------------------------------------
\begin{multline}
I_{5,\text{red}}(t) 
       = 4\left(
        \dfrac{p_1}{1+d_{5,1}^2}
       +\dfrac{p_3}{1+d_{5,3}^2}
       +\dfrac{p_4}{1+d_{5,4}^2}
       +\dfrac{p_7}{1+d_{5,7}^2}
        \right)\\
	= 4\left(
	\dfrac{0.9}{1+\sqrt{2}^2}
       +\dfrac{0.7}{1+\sqrt{2}^2}
       +\dfrac{0.6}{1+1^2}
       +\dfrac{0.3}{1+\sqrt{2}^2}
	\right)=7.(3),
\end{multline}
%% ---------------------------------------------------------------
\begin{equation}
I_{5,\text{green}}(t)
      = 4\left(
        \dfrac{p_2}{1+d_{5,2}^2}
       +\dfrac{p_8}{1+d_{5,8}^2}
        \right)\\
       =4\left(
        \dfrac{0.8}{1+1^2}
       +\dfrac{0.2}{1+1^2}
        \right)=2.
\end{equation}
%% ---------------------------------------------------------------

For $T=0$ the largest impact on actor $i=5$ is excreted by `red' actors and thus---according to Eq.~\eqref{eq:T=0}---actor $i=5$ in the next time step {\bf will change} his/her opinion from `blue' ($\xi_5(t)=\Xi_2$) to `red' ($\xi_5(t+1)=\Xi_1$).

For $T>0$ we calculate probabilities $P_{5,\text{blue}}$, $P_{5,\text{red}}$ and $P_{5,\text{green}}$ of choosing opinion by actor $i=5$ (see Eqs.~\eqref{eq:E_ik}--\eqref{eq:P_ik}).
For example, for $T=1$ these probabilities are
\begin{equation}
\begin{split}
	P_{5,\text{blue}} &= \dfrac{\exp(I_{5,\text{blue}}/1)}{P_1},\\
	P_{5,\text{red}}  &= \dfrac{\exp(I_{5,\text{red}}/1)}{P_1}, \\ 
	P_{5,\text{green}}&= \dfrac{\exp(I_{5,\text{green}}/1)}{P_1},
\end{split}
\end{equation}
while for $T=10$ we have
\begin{equation}
\begin{split}
	P_{5,\text{blue}}  &= \dfrac{\exp(I_{5,\text{blue}}/10)}{P_{10}},\\
	P_{5,\text{red}}   &= \dfrac{\exp(I_{5,\text{red}}/10)}{P_{10}}, \\
	P_{5,\text{green}} &= \dfrac{\exp(I_{5,\text{green}}/10)}{P_{10}},
\end{split}
\end{equation}
where normalisation constants are 
\[ P_1=\exp(I_{5,\text{blue}}/1)+\exp(I_{5,\text{red}}/1)+\exp(I_{5,\text{green}}/1) \]
and 
\[ P_{10}=\exp(I_{5,\text{blue}}/10)+\exp(I_{5,\text{red}}/10)+\exp(I_{5,\text{green}}/10). \] 

The calculated probabilities for $T=1$ are
\begin{equation}
\begin{split}
	P_{5,\text{blue}} &= \dfrac{\exp(4.4/1)}{e^{4.4}+e^{7.(3)}+e^2}   \approx 0.050,  \\
	P_{5,\text{red}}  &= \dfrac{\exp(7.(3)/1)}{e^{4.4}+e^{7.(3)}+e^2} \approx 0.945,   \\
	P_{5,\text{green}}&= \dfrac{\exp(2/1)}{e^{4.4}+e^{7.(3)}+e^2}     \approx 0.005,
\end{split}
\end{equation}
while for $T=10$ we have
\begin{equation}
\begin{split}
	P_{5,\text{blue}} &= \dfrac{\exp(4.4/10)}{e^{0.44}+e^{0.7(3)}+e^{0.2}}  \approx 0.320, \\
	P_{5,\text{red}}  &= \dfrac{\exp(7.(3)/10)}{e^{0.44}+e^{0.7(3)}+e^{0.2}}\approx 0.429, \\
	P_{5,\text{green}}&= \dfrac{\exp(2/10)}{e^{0.44}+e^{0.7(3)}+e^{0.2}}    \approx 0.251.
\end{split}
\end{equation}

For non-deterministic version of algorithm (i.e. for $T>0$) still the most probably state $\xi_5(t+1)$ is $\Xi_1$ (`red').
But probability of such evolution for actor $i=5$ decreases from 100\% for $T=0$ to 94.5\% for $T=1$ and to 42.9\% for $T=10$ to become 33.3\%$=1/K$ for $T\to\infty$.

Let us repeat these calculation for actor $i=9$:
%% ---------------------------------------------------------------
\begin{equation}
I_{9,\text{blue}}(t)=4\mathcal{J}_s \left(\dfrac{q(s_5)}{g(d_{9,5})} + \dfrac{q(s_6)}{g(d_{9,6})} + \dfrac{q(s_9)}{g(d_{9,9})} \right),
\end{equation}
%% ---------------------------------------------------------------
\begin{equation}
I_{9,\text{red}}(t) = 4\mathcal{J}_p\left(
	\dfrac{q(p_1)}{g(d_{9,1})}
       +\dfrac{q(p_3)}{g(d_{9,3})}
       +\dfrac{q(p_4)}{g(d_{9,4})}
       +\dfrac{q(p_7)}{g(d_{9,7})}
	\right),
\end{equation}
%% ---------------------------------------------------------------
\begin{equation}
I_{9,\text{green}}(t) = 4\mathcal{J}_p\left(
	\dfrac{q(p_2)}{g(d_{9,2})}
       +\dfrac{q(p_8)}{g(d_{9,8})}
	\right),
\end{equation}
%% ---------------------------------------------------------------
\begin{equation}
I_{9,\text{blue}}(t)= 4\left(
	\dfrac{s_5}{1+d_{9,5}^2} 
       +\dfrac{s_6}{1+d_{9,6}^2} 
       +\dfrac{s_9}{1+d_{9,9}^2}
	\right)
	= 4\left(
	\dfrac{0.5}{1+\sqrt{2}^2}
       +\dfrac{0.6}{1+1^2}
	+\dfrac{0.9}{1+0^2}
	\right)= 5.4(6),
\end{equation}
%% ---------------------------------------------------------------
\begin{multline}
I_{9,\text{red}}(t) 
       = 4\left(
        \dfrac{p_1}{1+d_{9,1}^2}
       +\dfrac{p_3}{1+d_{9,3}^2}
       +\dfrac{p_4}{1+d_{9,4}^2}
       +\dfrac{p_7}{1+d_{9,7}^2}
        \right)\\
	= 4\left(
	\dfrac{0.9}{1+(2\sqrt{2})^2}
       +\dfrac{0.7}{1+2^2}
	+\dfrac{0.6}{1+\sqrt{5}^2}
       +\dfrac{0.3}{1+2^2}
	\right)=1.6,
\end{multline}
%% ---------------------------------------------------------------
\begin{equation}
I_{9,\text{green}}(t)
      = 4\left(
        \dfrac{p_2}{1+d_{9,2}^2}
       +\dfrac{p_8}{1+d_{9,8}^2}
        \right)\\
       =4\left(
	\dfrac{0.8}{1+\sqrt{5}^2}
       +\dfrac{0.2}{1+1^2}
	\right)=0.9(3).
\end{equation}
%% ---------------------------------------------------------------

For $T=0$ the largest impact on actor $i=9$ is excreted by `blue' actors and thus---according to Eq.~\eqref{eq:T=0}---actor $i=9$ in the next time step {\bf will sustain} his/her `blue' opinion ($\xi_9(t+1)=\xi_9(t)=\Xi_2$).
Two factors influence the difference in actors $i=5$ and $i=9$ opinion in time $(t+1)$.
Namely, the difference in supportiveness of these two actors and their distance to `red' actors: {}{actor} $i=5$ has moderate supportiveness ($s_5=0.5$) and his/her distance to `red' actors is no longer than $\sqrt 2$. In contrast, actor $i=9$ has very high supportiveness ($s_9=0.9$) and distance to `red' actors no shorter than 2. 
Please note however, that ultimate fate of the system is the state with the unanimity of opinions.
As we have shown above, in the next time step at least the actor in the middle of the system ($i=5$) will convert his{/her} opinion to the `red' one. 
The same presumably will occur for actor $i=2$ who has low supportiveness ($s_2=0.2$) and who has only a single supporter.
Thus in time $(t+3)$ all actors will convert to the supporters of the `red' opinion.

For $T>0$ we calculate probabilities $P_{9,\text{blue}}$, $P_{9,\text{red}}$ and $P_{9,\text{green}}$ of choosing opinion by actor $i=9$ (see Eqs.~\eqref{eq:E_ik}--\eqref{eq:P_ik}).
For example, for $T=1$ these probabilities are
\begin{equation}
\begin{split}
	P_{9,\text{blue}} &= \dfrac{\exp(I_{9,\text{blue}}/1)}{P_1},\\
	P_{9,\text{red}}  &= \dfrac{\exp(I_{9,\text{red}}/1)}{P_1}, \\ 
	P_{9,\text{green}}&= \dfrac{\exp(I_{9,\text{green}}/1)}{P_1},
\end{split}
\end{equation}
while for $T=10$ we have
\begin{equation}
\begin{split}
	P_{9,\text{blue}}  &= \dfrac{\exp(I_{9,\text{blue}}/10)}{P_{10}},\\
	P_{9,\text{red}}   &= \dfrac{\exp(I_{9,\text{red}}/10)}{P_{10}}, \\
	P_{9,\text{green}} &= \dfrac{\exp(I_{9,\text{green}}/10)}{P_{10}},
\end{split}
\end{equation}
where normalisation constants are 
\[ P_1=\exp(I_{9,\text{blue}}/1)+\exp(I_{9,\text{red}}/1)+\exp(I_{9,\text{green}}/1) \]
and 
\[ P_{10}=\exp(I_{9,\text{blue}}/10)+\exp(I_{9,\text{red}}/10)+\exp(I_{9,\text{green}}/10). \] 

The calculated probabilities for $T=1$ are
\begin{equation}
\begin{split}
	P_{9,\text{blue}} &= \dfrac{\exp(5.4(6)/1)}{e^{5.4(6)}+e^{1.6}+e^{0.9(3)}} \approx 0.969, \\
	P_{9,\text{red}}  &= \dfrac{\exp(1.6/1)}   {e^{5.4(6)}+e^{1.6}+e^{0.9(3)}} \approx 0.020, \\
	P_{9,\text{green}}&= \dfrac{\exp(0.9(3)/1)}{e^{5.4(6)}+e^{1.6}+e^{0.9(3)}} \approx 0.011, 
\end{split}
\end{equation}
while for $T=10$ we have
\begin{equation}
\begin{split}
	P_{9,\text{blue}} &= \dfrac{\exp(5.4(6)/10)}{e^{0.54(6)}+e^{0.16}+e^{0.09(3)}} \approx 0.432, \\
	P_{9,\text{red}}  &= \dfrac{\exp(1.6/10)}{e^{0.54(6)}+e^{0.16}+e^{0.09(3)}}    \approx 0.293, \\
	P_{9,\text{green}}&= \dfrac{\exp(0.9(3)/10)}{e^{0.54(6)}+e^{0.16}+e^{0.09(3)}} \approx 0.275.
\end{split}
\end{equation}
Similarly to the actor $i=5$, the increase of the social temperature reduces chance of keeping initial opinion for actor $i=9$.
For $T=10$ these probabilities do not differ from $1/K$ for more than 0.1.

%% ###############################################################
\section{\label{B:small}{Small example of clustering ($L=10$, $K=3$)}}
%% ###############################################################

\begin{figure}[htbp]
\centering
\begin{subfigure}[b]{.44\textwidth}
\caption{\label{fig:samllcluster1}{}}
\begin{tikzpicture}[scale=.7]
\tkzDefPoint(0,0){A1}
\tkzDefPoint(0,3){B1}
\tkzDefPoint(1,3){C1}
\tkzDefPoint(1,2){D1}
\tkzDefPoint(2,2){E1}
\tkzDefPoint(2,0){F1}
\tkzFillPolygon[red,opacity=0.95](A1,B1,C1,D1,E1,F1)
\tkzDefPoint(9,0){A2}
\tkzDefPoint(10,0){B2}
\tkzDefPoint(10,1){C2}
\tkzDefPoint(9,1){D2}
\tkzFillPolygon[red,opacity=0.95](A2,B2,C2,D2)
\tkzDefPoint(9,4){A3}
\tkzDefPoint(10,4){B3}
\tkzDefPoint(10,5){C3}
\tkzDefPoint(9,5){D3}
\tkzFillPolygon[red,opacity=0.95](A3,B3,C3,D3)
\tkzDefPoint(4,3){A4}
\tkzDefPoint(6,3){B4}
\tkzDefPoint(6,4){C4}
\tkzDefPoint(5,4){D4}
\tkzDefPoint(5,5){E4}
\tkzDefPoint(4,5){F4}
\tkzFillPolygon[red,opacity=0.95](A4,B4,C4,D4,E4,F4)
\tkzDefPoint(10,8){A5}
\tkzDefPoint(10,9){B5}
\tkzDefPoint(8,9){C5}
\tkzDefPoint(8,10){D5}
\tkzDefPoint(4,10){E5}
\tkzDefPoint(4,8){F5}
\tkzDefPoint(5,8){I5}
\tkzDefPoint(5,9){J5}
\tkzDefPoint(7,9){G5}
\tkzDefPoint(7,8){H5}
\tkzFillPolygon[red,opacity=0.95](A5,B5,C5,D5,E5,F5,I5,J5,G5,H5)
\tkzDefPoint(10,10){A6}
%% \tkzFillPolygon[green,opacity=0.95](B5,C5,D5,A6)
\tkzDefPoint(10,6){A7}
\tkzDefPoint(10,7){B7}
\tkzDefPoint(9,7){C7}
\tkzDefPoint(9,6){D7}
%% \tkzFillPolygon[green,opacity=0.95](A7,B7,C7,D7)
\tkzDefPoint(5,0){A8}
\tkzDefPoint(8,7){B8}
\tkzDefPoint(7,7){C8}
\tkzDefPoint(7,3){D8}
\tkzDefPoint(5,3){E8}
\tkzDefPoint(8,8){F8}
\tkzFillPolygon[blue,opacity=0.95](A8,A2,D2,C2,B3,A3,D3,C3,A7,D7,C7,B7,A5,F8,B8,C8,D8,E8)
\tkzDefPoint(0,10){A9}
\tkzDefPoint(4,7){B9}
\tkzDefPoint(3,7){C9}
\tkzDefPoint(3,6){D9}
\tkzDefPoint(6,6){E9}
\tkzDefPoint(6,5){F9}
\tkzDefPoint(2,5){G9}
\tkzDefPoint(2,3){H9}
\tkzFillPolygon[blue,opacity=0.95](B1,A9,E5,B9,C9,D9,E9,F9,G9,H9)
%% \tkzFillPolygon[green,opacity=0.95](B4,C4,D4,E4,F9,E9,D9,C9,B9,F5,I5,J5,G5,H5,F8,B8,C8,D8)
%% \tkzFillPolygon[green,opacity=0.95]()
\draw[step=10mm,gray,very thin] (-0.1,-0.1) grid (10.1,10.1);
\node[] at (2,8)      {\color{white} {\bf 1}};
\node[] at (7.5,9)    {\color{white} {\bf 2}};
\node[] at (9.5,9.5)                 {\bf 3};
\node[] at (6.5,7.5)                 {\bf 4};
\node[] at (8,4)      {\color{white} {\bf 5}};
\node[] at (9.5,6.5)                 {\bf 6};
\node[] at (3.5,1.5)                 {\bf 7};
\node[] at (4.5,3.5)  {\color{white} {\bf 8}};
\node[] at (9.5,4.5)  {\color{white} {\bf 9}};
\node[] at (1,1)     {\color{white} {\bf 10}};
\node[] at (9.5,0.5) {\color{white} {\bf 11}};
\end{tikzpicture}
\end{subfigure}
%% ---------------------------------------------------------------
\begin{subfigure}[b]{.44\textwidth}
\caption{\label{fig:samllcluster2}{}}
\begin{tikzpicture}[scale=.7]
\tkzDefPoint(0,0){A1}
\tkzDefPoint(0,8){B1}
\tkzDefPoint(1,8){C1}
\tkzDefPoint(1,9){D1}
\tkzDefPoint(2,9){E1}
\tkzDefPoint(2,10){F1}
\tkzDefPoint(5,10){G1}
\tkzDefPoint(5,5){H1}
\tkzDefPoint(4,5){I1}
\tkzDefPoint(4,0){J1}
\tkzFillPolygon[red,opacity=0.95](A1,B1,C1,D1,E1,F1,G1,H1,I1,J1)
\tkzDefPoint(9,0){A2}
\tkzDefPoint(9,1){B2}
\tkzDefPoint(10,1){C2}
\tkzDefPoint(10,0){D2}
\tkzFillPolygon[blue,opacity=0.95](A2,B2,C2,D2)
\node[] at (0.5,9.5) {\bf 12};
\node[] at (2.5,5.5) {\color{white} {\bf 13}};
\node[] at (7.5,5.5) {\bf 14};
\node[] at (9.5,0.5) {\color{white} {\bf 15}};
\draw[step=10mm,gray,very thin] (-0.1,-0.1) grid (10.1,10.1);
\end{tikzpicture}
\end{subfigure}
%% ---------------------------------------------------------------
\caption{\label{fig:samllcluster}{Example of sites labelling for $K=3$ and $L=10$ {and two lattice realizations}.}}
\end{figure}
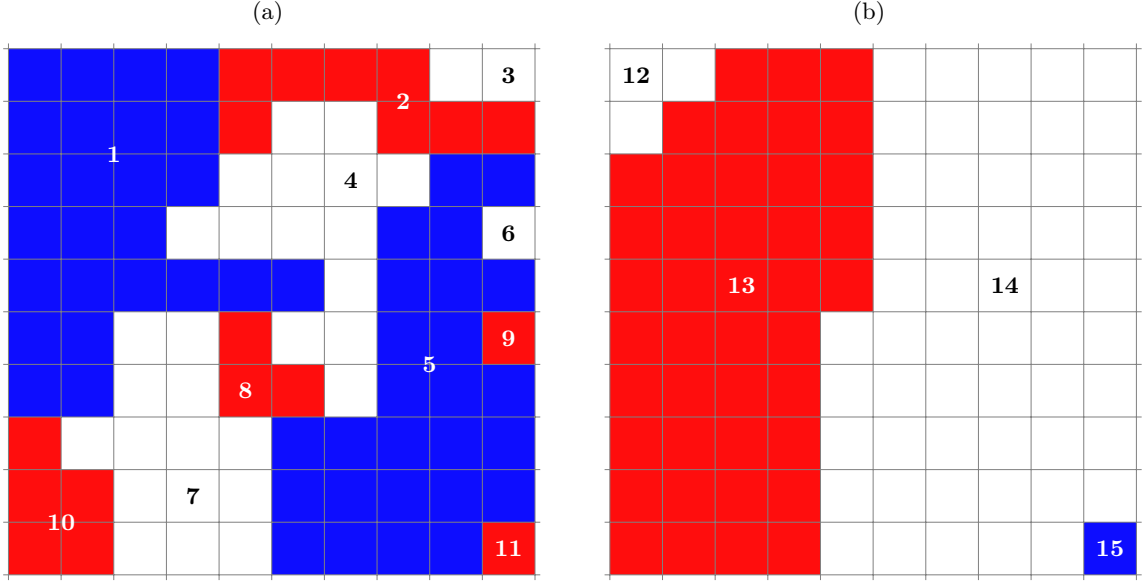

Two sites are in the same cluster if they are adjacent (in von Neumann neighbourhood, Fig. \ref{F:vN}) to each other and simultaneously actors at these sites share the same opinion.
The Hoshen--Kopelman algorithm allows for sites labelling in such way that sites in the same cluster have the same labels and sites in different cluster have different labels.
Example{s} of sites labelling for $L=10$ and $K=3$ {} {are} presented in Fig{s}.~\ref{fig:samllcluster1} {and} \ref{fig:samllcluster2}, where $n_c=11$ {and $n_c=4$} clusters have been identified{, respectively}. 
{}
{The average number of cluster for these two lattice realization is $\langle n_c\rangle=(11+4)/2=7.5$.
The number of sites in each cluster defines its size $\mathcal{S}$.
For these two lattice ralizations the largest culsters are labelled as 1 and 5 (Fig.}~\ref{fig:samllcluster1}) {and as 14 (Fig.}~\ref{fig:samllcluster2}) {and their sizes are $\mathcal{S}_{\max}=25$ and $\mathcal{S}_{\max}=54$, respectively.}
{Thus average largest cluster size is $\langle\mathcal{S}_{\max}\rangle=(25+54)/2=39.5$.}
{In given example histogram $H(\mathcal{S})$ of clusters sizes is presented in Table}~\ref{tab:hist}.
{Basing on Table}~\ref{tab:hist} {we evaluate number of small clusters (with $\mathcal{S}\le 5$) as 4+1+2+1=8.
As this sum comes from merging results of two lattice realization the average number of small clusters is $\langle n_s\rangle=8/2=4$.}

%% ---------------------------------------------------------------
%% ---------------------------------------------------------------

%% ---------------------------------------------------------------
\begin{table}
\begin{ruledtabular}
	\caption{\label{tab:hist}{Histogram of {} cluster sizes $\mathcal{S}$ for lattice{s} presented in Fig.}~\ref{fig:samllcluster}.}
\begin{tabular}{lrrrrrrrrr}
labels $i$:       & 6, 9, 11, {15} & 3 & 8, {12}   & 10 & 2 & 4, 7 & 1, 5 & {13} & {14} \\ \hline
$\mathcal{S}$:    & 1                 & 2 & 3            &  5 & 8 &   14 &   25 & {42} & {54} \\ \hline
$n(\mathcal{S})$: & {}{4}      & 1 & {}{2} &  1 & 1 &    2 &    2 & {1}  & {1}  \\
\end{tabular}
\end{ruledtabular}
\end{table}
%% ---------------------------------------------------------------

%% ###############################################################
\section{\label{A:code}Source codes}
%% ###############################################################

In Listing{s} \ref{lst:T0} {and} \ref{lst:TT} the Fortran 95 code{s} allowing for reproductions of data for Figs.~\ref{fig:histK2}, \ref{fig:histK3}, \ref{fig:rev2K2}, \ref{fig:rev2K3} (for {both, noiseless and} non-deterministic version of simulations{}) {} are presented.

The module {\tt settings} provides model parameters including lattice size $L$ {({\tt Xmax} and {\tt Ymax})}, number of opinions $K$ {({\tt Kmax})}, {} number of time steps $t_{\max}$ { ({\tt Tmax}) and number of lattice realizations ({\tt Run})} {}.

In module {\tt utils} the scaling functions $g(x)$ and $q(x)$ as well as the Euclidean distance $d(x,y)$ are defined.
Also the {\tt reclassify} function for Hoshen--Kopelman algorithm is defined there.

The main program starts in line 52.
The actors supportiveness ($s_i$) and persuasiveness ($p_i$) are initialised randomly in lines {} {85--90}, while initial actors opinions ($\xi_i$) are given in lines {93--98}{}.
Loop 88 provides time evolution of the system.
Loop 77 realises Hoshen--Kopelman algorithm of sites (actors) labelling {for $t=0$}.
{}  {}  {}
{}
{}
{}
In loop 99 the system characteristics after the system time evolution is completed are calculated.
Loop 777 realises averaging procedure over independent runnnigs for various initial conditions.

\definecolor{mygray}{rgb}{0.92,0.97,0.92}
\lstset{backgroundcolor=\color{mygray},commentstyle=\itshape\color{blue},stringstyle=\color{teal}}

%% ===============================================================
\subsection{\label{A:codeT0} $T=0$}
%% ===============================================================
{An input data ($\alpha$ parameter) is read in line 71.}
{In lines 272--275 histograms $H(\mathcal{S})$ of clusters sizes $\mathcal{S}$ are printed.}
{In line 278 values of $\langle n_c\rangle$, $\langle S\rangle$ (not presented in this paper) and $\langle\mathcal{S}_{\max}\rangle$ are printed.}

\begin{widetext}
\begin{lstlisting}[language={[95]Fortran},frame=single,numbers=left,numberstyle=\tiny,basicstyle=\footnotesize,stepnumber=1,breaklines=true,caption={Fortan95 code implementing Eq.~\eqref{eq:T=0} i.e. for $T=0$},label=lst:T0]
!!! Latane-Nowak-Szamrej model + Hoshen-Kopelman algorithm
!!! K. Malarz
!!! created: Tue, 21 May 2019, 13:06:13 CEST
!!! revised: Sun, 05 Apr 2020, 14:43:19 CEST

!!! ================================================================
module settings
!!! ================================================================
implicit none

integer, parameter :: Xmax=41,Ymax=41,Tmax=1000,Kmax=2,L2=(Xmax+1)*(Ymax+1),Run=100
real*8, parameter :: T=0.d0
real*8 :: alpha
end module settings

!!! ================================================================
module utils
!!! ================================================================
use settings
implicit none
contains

real*8 function g(x)
        real*8 :: x
        g=1.0d0+x**alpha
end function

real*8 function q(x)
        real*8 :: x
        q=x
end function

real*8 function d(x1,y1,x2,y2)
        integer :: x1,y1,x2,y2
        d=dsqrt((1.d0*x1-1.d0*x2)**2 + (1.d0*y1-1.d0*y2)**2)
end function

integer function reclassify(ix)
integer :: ix
integer, dimension (0:Xmax,0:Ymax) :: label
integer, dimension (L2) :: iclass
common/block/ label, iclass

reclassify=iclass(ix)
90 if(iclass(reclassify).eq.reclassify) return
reclassify=iclass(reclassify)
goto 90
end function

end module utils

!!! ################################################################
program Latane_Hoshen_Kopelman
!!! ################################################################
use settings
use utils
implicit none
integer :: x,y,it, xx,yy,k,kk,strongest_k,irun,maxlabel,Smax,largestS
real :: r
real*8 :: sump,strongest_I,avenc,aveS,aveSmax

integer, dimension (0:Xmax,0:Ymax) :: label
integer, dimension (L2) :: iclass
integer, dimension (0:Xmax,0:Ymax) :: xi
integer, dimension (0:Xmax*Ymax) :: isize, histograminrun,histogram
integer, dimension (0:Xmax*Ymax,Kmax) :: histogramK
real*8, dimension (Xmax,Ymax) :: p, s
real*8, dimension (Xmax,Ymax,Kmax) :: I, prob
common/block/ label, iclass

  read *,alpha
  print '(A3,7A11)','###','Xmax','Ymax','K','alpha','T','tmax','run'
  print '(A3,3I11,2F11.3,2I11)','###',Xmax,Ymax,Kmax,alpha,T,Tmax,Run

  histogram=0
  aveSmax=0.d0
  aveS=0.d0
  avenc=0.d0

  do 777 irun=1,Run
  histograminrun=0
!  print *,'#########'
!  print *,'# irun=',irun

  do x=1,Xmax !! initial state
  do y=1,Ymax
    s(x,y)=rand()
    p(x,y)=rand()
  enddo
  enddo

  it=0
  xi=0
  do x=1,Xmax
  do y=1,Ymax
    xi(x,y)=1+Kmax*rand()
  enddo
  enddo

  histogramK=0

  do 77 k=1,Kmax
    isize=0

    label=L2
    do kk=1,L2
      iclass(kk)=kk
    enddo
    maxlabel=0

    do x=1,Xmax
    do y=1,Ymax
      if(xi(x,y).eq.k) then !! labeling clusters
        if(xi(x-1,y).eq.k .or. xi(x,y-1).eq.k) then
          !! reclassifing neighbouring sites
          if(xi(x-1,y).eq.k) label(x-1,y)=reclassify(label(x-1,y))
          if(xi(x,y-1).eq.k) label(x,y-1)=reclassify(label(x,y-1))
          label(x,y)=min(label(x-1,y),label(x,y-1))
          iclass(label(x-1,y))=label(x,y)
          iclass(label(x,y-1))=label(x,y)
        else
          maxlabel=maxlabel+1
          label(x,y)=maxlabel
        endif
      endif
    enddo
    enddo
    ! reclassifing all occupied sites
    do x=1,Xmax
    do y=1,Ymax
      if((xi(x,y).eq.k) .and. (label(x,y).gt.iclass(label(x,y)))) label(x,y)=reclassify(label(x,y))
    enddo
    enddo

    do x=1,Xmax
    do y=1,Ymax
      if(xi(x,y).eq.k) isize(label(x,y))=isize(label(x,y))+1
    enddo
    enddo

    do kk=1,Xmax*Ymax
      histogramK(isize(kk),k)=histogramK(isize(kk),k)+1
    enddo

77 enddo

   do 88 it=1,Tmax !!! time evolution
     I=0.0d0
     do x=1,Xmax
     do y=1,Ymax
       do xx=1,Xmax
       do yy=1,Ymax
         if(xi(x,y).eq.xi(xx,yy)) then
           I(x,y,xi(xx,yy))=I(x,y,xi(xx,yy))+q(s(xx,yy))/g(d(x,y,xx,yy))
         else
           I(x,y,xi(xx,yy))=I(x,y,xi(xx,yy))+q(p(xx,yy))/g(d(x,y,xx,yy))
         endif
       enddo
       enddo
     enddo
     enddo
  
     do x=1,Xmax
     do y=1,Ymax
     do k=1,Kmax
       I(x,y,k)=4.0d0*I(x,y,k)
     enddo
     enddo
     enddo

     do x=1,Xmax
     do y=1,Ymax
       strongest_I=I(x,y,1)
       strongest_k=1
       do k=2,Kmax
         if(I(x,y,k).gt.strongest_I) then
           strongest_I=I(x,y,k)
           strongest_k=k
         endif
       enddo
       xi(x,y)=strongest_k
     enddo
     enddo
88 enddo !!! time evolution

!   print *,'# it=',it,'xi:'
!   do x=1,Xmax
!     print '(41I5)',(xi(x,y),y=1,Ymax)
!   enddo
  histogramK=0
  Smax=0

  do 99 k=1,Kmax
    isize=0

!    print *,"# k=",k
    label=L2
    do kk=1,L2
      iclass(kk)=kk
    enddo
    maxlabel=0

    do x=1,Xmax
    do y=1,Ymax
      if(xi(x,y).eq.k) then ! labeling clusters
        if(xi(x-1,y).eq.k .or. xi(x,y-1).eq.k) then
          ! reclassifing neighbouring sites
          if(xi(x-1,y).eq.k) label(x-1,y)=reclassify(label(x-1,y))
          if(xi(x,y-1).eq.k) label(x,y-1)=reclassify(label(x,y-1))
          label(x,y)=min(label(x-1,y),label(x,y-1))
          iclass(label(x-1,y))=label(x,y)
          iclass(label(x,y-1))=label(x,y)
        else
          maxlabel=maxlabel+1
          label(x,y)=maxlabel
        endif
      endif
    enddo
    enddo
    ! reclassifing all occupied sites
    do x=1,Xmax
    do y=1,Ymax
      if((xi(x,y).eq.k) .and. (label(x,y).gt.iclass(label(x,y)))) label(x,y)=reclassify(label(x,y))
    enddo
    enddo

!    do x=1,Xmax
!      print '(41I5)',(label(x,y),y=1,Ymax)
!    enddo

    do x=1,Xmax
    do y=1,Ymax
      if(xi(x,y).eq.k) isize(label(x,y))=isize(label(x,y))+1
    enddo
    enddo

    do kk=1,Xmax*Ymax
      histogramK(isize(kk),k)=histogramK(isize(kk),k)+1
    enddo

!    print *,"# histogram, irun=",irun," k=",k
    do kk=1,Xmax*Ymax
!      if(histogramK(kk,k).gt.0) print *,kk,histogramK(kk,k)
      histograminrun(kk)=histograminrun(kk)+histogramK(kk,k)
    enddo

    do kk=Xmax*Ymax,1,-1
      if(histogramK(kk,k).gt.0) then
        largestS=kk
        goto 33
      endif
    enddo
33  Smax=max(Smax,largestS)
!   print *,"# largest S=",largestS
!   print *,"# Smax=",Smax

99  enddo

!   do k=1,Xmax*Ymax
!     if(histograminrun(k).gt.0) print *,k,histograminrun(k)
!   enddo
!   print *,"# nc=",sum(histograminrun)
    avenc=avenc+1.d0*sum(histograminrun)
    aveSmax=aveSmax+1.d0*Smax
    do k=1,Xmax*Ymax
      aveS=aveS+(1.d0*k*histograminrun(k))/(1.d0*sum(histograminrun))
      histogram(k)=histogram(k)+histograminrun(k)
    enddo

777 enddo

  print *,"# total histogram:"
  do k=1,Xmax*Ymax
    if(histogram(k).gt.0) print *,k,histogram(k)
  enddo

  print '(A2,A3,5A9)',"#","K","T","alpha","<nc>","<S>","<Smax>"
  print '(A2,I3,5F9.3)',"#",Kmax,T,alpha,avenc/(1.d0*Run),aveS/(1.d0*Run),aveSmax/(1.d0*Run)

end program Latane_Hoshen_Kopelman
\end{lstlisting}

%% ===============================================================
\subsection{\label{A:codeTT} $T>0$}
%% ===============================================================
{An input data ($\alpha$ and $T$ parameters) are read in line 71.}
{In lines 284--287 histograms $H(\mathcal{S})$ of clusters size $\mathcal{S}$ are printed.}
{In line 290 values of $\langle n_c\rangle$, $\langle S\rangle$ (not presented in this paper), $\langle\mathcal{S}_{\max}\rangle$ are printed.}

\begin{lstlisting}[language={[95]Fortran},frame=single,numbers=left,numberstyle=\tiny,basicstyle=\footnotesize,stepnumber=1,breaklines=true,caption={Fortan95 code implementing Eq.~\eqref{eq:T=T} i.e. for $T>0$},label=lst:TT]
!!! Latane-Nowak-Szamrej model + Hoshen-Kopelman algorithm
!!! K. Malarz
!!! created: Tue, 21 May 2019, 13:06:13 CEST
!!! revised: Sun, 05 Apr 2020, 14:43:19 CEST

!!! ================================================================
module settings
!!! ================================================================
implicit none

integer, parameter :: Xmax=41,Ymax=41,Tmax=1000,Kmax=2,L2=(Xmax+1)*(Ymax+1),Run=100
real*8 :: T
real*8 :: alpha
end module settings

!!! ================================================================
module utils
!!! ================================================================
use settings
implicit none
contains

real*8 function g(x)
        real*8 :: x
        g=1.0d0+x**alpha
end function

real*8 function q(x)
        real*8 :: x
        q=x
end function

real*8 function d(x1,y1,x2,y2)
        integer :: x1,y1,x2,y2
        d=dsqrt((1.d0*x1-1.d0*x2)**2 + (1.d0*y1-1.d0*y2)**2)
end function

integer function reclassify(ix)
integer :: ix
integer, dimension (0:Xmax,0:Ymax) :: label
integer, dimension (L2) :: iclass
common/block/ label, iclass

reclassify=iclass(ix)
90 if(iclass(reclassify).eq.reclassify) return
reclassify=iclass(reclassify)
goto 90
end function

end module utils

!!! ################################################################
program Latane_Hoshen_Kopelman
!!! ################################################################
use settings
use utils
implicit none
integer :: x,y,it, xx,yy,k,kk,strongest_k,irun,maxlabel,Smax,largestS
real :: r
real*8 :: sump,avenc,aveS,aveSmax

integer, dimension (0:Xmax,0:Ymax) :: label
integer, dimension (L2) :: iclass
integer, dimension (0:Xmax,0:Ymax) :: xi
integer, dimension (0:Xmax*Ymax) :: isize, histograminrun,histogram
integer, dimension (0:Xmax*Ymax,Kmax) :: histogramK
real*8, dimension (Xmax,Ymax) :: p, s
real*8, dimension (Xmax,Ymax,Kmax) :: I, prob
common/block/ label, iclass

  read *,T,alpha
  print '(A3,7A11)','###','Xmax','Ymax','K','alpha','T','tmax','run'
  print '(A3,3I11,2F11.3,2I11)','###',Xmax,Ymax,Kmax,alpha,T,Tmax,Run

  histogram=0
  aveSmax=0.d0
  aveS=0.d0
  avenc=0.d0

  do 777 irun=1,Run
  histograminrun=0
!  print *,'#########'
!  print *,'# irun=',irun

  do x=1,Xmax !! initial state
  do y=1,Ymax
    s(x,y)=rand()
    p(x,y)=rand()
  enddo
  enddo

  it=0
  xi=0
  do x=1,Xmax
  do y=1,Ymax
    xi(x,y)=1+Kmax*rand()
  enddo
  enddo

  histogramK=0

  do 77 k=1,Kmax
    isize=0

    label=L2
    do kk=1,L2
      iclass(kk)=kk
    enddo
    maxlabel=0

    do x=1,Xmax
    do y=1,Ymax
      if(xi(x,y).eq.k) then !! labeling clusters
        if(xi(x-1,y).eq.k .or. xi(x,y-1).eq.k) then
          !! reclassifing neighbouring sites
          if(xi(x-1,y).eq.k) label(x-1,y)=reclassify(label(x-1,y))
          if(xi(x,y-1).eq.k) label(x,y-1)=reclassify(label(x,y-1))
          label(x,y)=min(label(x-1,y),label(x,y-1))
          iclass(label(x-1,y))=label(x,y)
          iclass(label(x,y-1))=label(x,y)
        else
          maxlabel=maxlabel+1
          label(x,y)=maxlabel
        endif
      endif
    enddo
    enddo
    ! reclassifing all occupied sites
    do x=1,Xmax
    do y=1,Ymax
      if((xi(x,y).eq.k) .and. (label(x,y).gt.iclass(label(x,y)))) label(x,y)=reclassify(label(x,y))
    enddo
    enddo

    do x=1,Xmax
    do y=1,Ymax
      if(xi(x,y).eq.k) isize(label(x,y))=isize(label(x,y))+1
    enddo
    enddo

    do kk=1,Xmax*Ymax
      histogramK(isize(kk),k)=histogramK(isize(kk),k)+1
    enddo

77 enddo

   do 88 it=1,Tmax !!! time evolution
     I=0.0d0
     do x=1,Xmax
     do y=1,Ymax
       do xx=1,Xmax
       do yy=1,Ymax
         if(xi(x,y).eq.xi(xx,yy)) then
           I(x,y,xi(xx,yy))=I(x,y,xi(xx,yy))+q(s(xx,yy))/g(d(x,y,xx,yy))
         else
           I(x,y,xi(xx,yy))=I(x,y,xi(xx,yy))+q(p(xx,yy))/g(d(x,y,xx,yy))
         endif
       enddo
       enddo
     enddo
     enddo

     do x=1,Xmax
     do y=1,Ymax
     do k=1,Kmax
       I(x,y,k)=4.0d0*I(x,y,k)
     enddo
     enddo
     enddo
  
     do x=1,Xmax
     do y=1,Ymax
       sump=0.0d0
       do k=1,Kmax
         prob(x,y,k)=dexp(I(x,y,k)/T) 
         sump=sump+prob(x,y,k)
       enddo
       do k=1,Kmax
         prob(x,y,k)=prob(x,y,k)/sump
       enddo
     enddo
     enddo

     do x=1,Xmax
     do y=1,Ymax
       r=rand()
       sump=0.0d0
       do k=1,Kmax
         sump=sump+prob(x,y,k)
         if(r.lt.sump) goto 666
       enddo
666    xi(x,y)=k
     enddo
     enddo

88 enddo !!! time evolution
 
!   print *,'# it=',it,'xi:'
!   do x=1,Xmax
!     print '(41I5)',(xi(x,y),y=1,Ymax)
!   enddo
  histogramK=0
  Smax=0

  do 99 k=1,Kmax
    isize=0

!    print *,"# k=",k
    label=L2
    do kk=1,L2
      iclass(kk)=kk
    enddo
    maxlabel=0

    do x=1,Xmax
    do y=1,Ymax
      if(xi(x,y).eq.k) then ! labeling clusters
        if(xi(x-1,y).eq.k .or. xi(x,y-1).eq.k) then
          ! reclassifing neighbouring sites
          if(xi(x-1,y).eq.k) label(x-1,y)=reclassify(label(x-1,y))
          if(xi(x,y-1).eq.k) label(x,y-1)=reclassify(label(x,y-1))
          label(x,y)=min(label(x-1,y),label(x,y-1))
          iclass(label(x-1,y))=label(x,y)
          iclass(label(x,y-1))=label(x,y)
        else
          maxlabel=maxlabel+1
          label(x,y)=maxlabel
        endif
      endif
    enddo
    enddo
    ! reclassifing all occupied sites
    do x=1,Xmax
    do y=1,Ymax
      if((xi(x,y).eq.k) .and. (label(x,y).gt.iclass(label(x,y)))) label(x,y)=reclassify(label(x,y))
    enddo
    enddo

!    do x=1,Xmax
!      print '(41I5)',(label(x,y),y=1,Ymax)
!    enddo

    do x=1,Xmax
    do y=1,Ymax
      if(xi(x,y).eq.k) isize(label(x,y))=isize(label(x,y))+1
    enddo
    enddo

    do kk=1,Xmax*Ymax
      histogramK(isize(kk),k)=histogramK(isize(kk),k)+1
    enddo

!    print *,"# histogram, irun=",irun," k=",k
    do kk=1,Xmax*Ymax
!      if(histogramK(kk,k).gt.0) print *,kk,histogramK(kk,k)
      histograminrun(kk)=histograminrun(kk)+histogramK(kk,k)
    enddo

    do kk=Xmax*Ymax,1,-1
      if(histogramK(kk,k).gt.0) then
        largestS=kk
        goto 33
      endif
    enddo
33  Smax=max(Smax,largestS)
!   print *,"# largest S=",largestS
!   print *,"# Smax=",Smax

99  enddo

!   do k=1,Xmax*Ymax
!     if(histograminrun(k).gt.0) print *,k,histograminrun(k)
!   enddo
!   print *,"# nc=",sum(histograminrun)
    avenc=avenc+1.d0*sum(histograminrun)
    aveSmax=aveSmax+1.d0*Smax
    do k=1,Xmax*Ymax
      aveS=aveS+(1.d0*k*histograminrun(k))/(1.d0*sum(histograminrun))
      histogram(k)=histogram(k)+histograminrun(k)
    enddo

777 enddo

  print *,"# total histogram:"
  do k=1,Xmax*Ymax
    if(histogram(k).gt.0) print *,k,histogram(k)
  enddo

  print '(A2,A3,5A9)',"#","K","T","alpha","<nc>","<S>","<Smax>"
  print '(A2,I3,5F9.3)',"#",Kmax,T,alpha,avenc/(1.d0*Run),aveS/(1.d0*Run),aveSmax/(1.d0*Run)

end program Latane_Hoshen_Kopelman
\end{lstlisting}
\end{widetext}

\end{document}